\documentclass[11pt]{article}
\usepackage[utf8]{inputenc}
\usepackage{enumitem}

\usepackage{amsmath,amssymb,array,setspace,fullpage}
\usepackage{algorithm, algcompatible}
\algnewcommand\INPUT{\item[\textbf{Input:}]}%
\algnewcommand\OUTPUT{\item[\textbf{Output:}]}%
\usepackage{mathrsfs}
\usepackage{caption}
\usepackage{booktabs}
\usepackage{float}
\usepackage{lscape}
\usepackage{dcolumn}
\usepackage{ulem}
\usepackage{multicol,multirow}
\usepackage[all]{xy}
\usepackage[dvips]{graphicx}
\usepackage{url}
\usepackage[table]{xcolor}
\usepackage{colortbl}
\usepackage{wrapfig}
\usepackage{comment}
\graphicspath{{}}
\usepackage{natbib}
 
\newcommand{\nonpara}{model-free}
\newcommand{\Nonpara}{Model-free}
\newcommand{\degcor}{degree-corrected}
\newcommand{\avgden}{E[W_t]}

\title{The Value of Summary Statistics for Anomaly Detection in Temporally-Evolving Networks: A Performance Evaluation Study}
\author{Lata Kodali\footnote{Contact Email: latak215@vt.edu} , Srijan Sengupta, Leanna House, William H. Woodall}
\date{}

\begin{document}
\maketitle
 
\begin{abstract}
Analysis of network data has emerged as an active research area in statistics. Much of the focus of ongoing research has been on static networks that represent a single snapshot or aggregated historical data unchanging over time. 
However, most networks result from temporally-evolving systems that exhibit intrinsic dynamic behavior. Monitoring such temporally-varying networks to detect anomalous changes has applications in both social and physical sciences. In this work, we perform an evaluation study of the use of summary statistics for anomaly detection in temporally-evolving networks by incorporating principles from statistical process monitoring. 
In contrast to previous studies, we deliberately incorporate temporal auto-correlation in our study. Other considerations in our comprehensive assessment include types and duration of anomaly, model type, and sparsity in temporally-evolving networks. We conclude that the use of summary statistics can be valuable tools for network monitoring and often perform better than more complicated statistics.
\end{abstract}

\section{Introduction}


Recent decades have witnessed an explosion of data in the form of
networks, representing important systems in various fields, e.g.,
physical infrastructure \citep{huberman1999internet,pagani2013power},
social interaction \citep{milgram1967small,adamic2005political},
and biological systems \citep{bassett2006small,lynall2010functional}.
Consequently, statistical modeling and analysis have become fundamental
tools for studying physical and virtual networked systems, and are
poised to become even more critical in the near future. 

Traditionally, most research has focused on static modeling of
networks, in which either a single snapshot or aggregated historical data of a system are available. However, usually
networks result from time-evolving systems that exhibit intrinsic dynamic behavior. The terms dynamic, temporally-evolving, time-evolving, or temporally dependent are all used interchangeably to describe such networks that span over time. For example, often the relationships between members of a social network evolve over time due to finding new friends, collaborating with new colleagues, moving to another department, etc. Recent studies have focused on analyzing dynamic networks where a network is represented by a statistical/probabilistic model that is adaptively updated over time. However, many of these models have been developed on the premise that a system either is stationary or has smooth dynamics, and it does not experience abrupt changes.

In applications, the occurrence of sudden large changes and shocks in time-varying networks is very common. For example, resignation of a key employee or occurrence of a conflict in an organization may cause a significant change in the professional network of the employees. As another example, the occurrence of a change in the communications network of a terrorist group may indicate a high possibility of a
terrorist attack that could be prevented if the change is detected quickly. Similarly, significant changes in the brain connectome network \citep{xia2013brainnet} of an individual can indicate the onset of a neurological disorder like Alzheimer's disease or epilepsy. Monitoring, change detection, and accurate estimation of the change time are crucial for effective decision-making and for taking necessary actions
in a timely manner. Moreover, abrupt changes often affect a network locally. That is, only a subset of nodes and their corresponding links are altered by an event. Consequently, diagnosis, defined as identifying affected sub-networks, plays an important role in root-cause determination and action planning. For example, it is crucial to determine the group
of people who might be involved in a terrorist plot, or the parts of the brain involved in a particular disease, by identifying the set of nodes that caused the change in the corresponding network.

A recent review paper by \cite{woodall2017overview} provides an assessment of  monitoring methods that may detect anomalies in time-evolving networks.   Specifically, they reviewed statistical process monitoring methods for social, dynamic networks which fall into five broad categories: hypothesis testing (signals based on likelihood ratio tests), Bayesian methods (control limits are calculated using a Bayesian predictive distribution), scan methods (signals from a moving window based monitoring statistic), time series models (signals from large residuals), and changes in community structures/membership.  While summarizing the categories, \cite{woodall2017overview} highlight differences among available methods for specifying tuning parameters, such as specifying the size of moving average window, defining control limits based on baseline data, and approaches for removing seasonal effects in data.    
A notable point made by \cite{woodall2017overview}, which was re-iterated in \cite{sengupta2018discussion}, is that even with variation in parameter settings comparable relevant work in the literature are not available for detecting abrupt changes in a stream of time-varying networks.  Similarly, several papers contain studies of network monitoring under specific parametric modeling frameworks \citep{wilson2016modeling,yu2018detecting,zhao2018performance,zhao2018aggregation}.
However, such monitoring methods work under the assumption of a specific network model, and cannot be extended to the general task of network monitoring without model assumptions.

In this paper, we assess the performance of model-free summary statistics in network monitoring, such as network density, maximum degree, and their linear combinations, in anomaly detection.  Such  summary statistics are simple to calculate and often used in practice, but little is understood about how these summary statistics behave under varying network conditions and over time.  In turn, the utility of common network summaries, such as  density and maximum degree, for anomaly detection is also currently unclear.  For our work,  we conduct a comprehensive simulation study to assess both the successes and failures of four monitoring statistics that are functions of network density and/or maximum degree in comparison to a well-studied scan-based moving window approach  \citep{priebe2005scan}. We measure success and failure for the methods based on  false alarm rates, anomaly detection rates, and Area Under Curve (AUC) calculations from receiver operating characteristic (ROC) curves. 

An important aspect of our work is in applying monitoring methods on temporally dependent network data to evaluate method performance in realistic scenarios. The data are simulated from well-studied network models so that we may introduce various kinds of anomalies in a controlled manner to facilitate a principled comparative evaluation of network monitoring methods. 
As pointed out by several authors \citep{woodall2017overview, Savage201462, azarnoush2016monitoring}, compared to case studies, such controlled scenarios from synthetic networks provide a more principled testbed for performance assessment.

The remainder of the paper is as follows. The anomaly detection methods utilized, mathematical formulas, and considerations for such methods utilized are given in Section \ref{sec:methods}. In order to better understand what types of anomalies can be detected, the network monitoring methods were applied to data generated from two popular latent variable models: the dynamic latent space model (DLSM) \citep{SewellChen2015} and the dynamic \degcor\ stochastic block model (DDCSBM) \citep{wilson2016modeling,matias2017statistical}. Brief overviews of these models are given in Section \ref{sec:models}.  We reiterate that no model fitting occurred and models were only used to generate data of temporally-evolving networks. A performance evaluation of network monitoring methods on summary statistics is accomplished using a comprehensive simulation study. Settings for the simulation study as well as planted anomalies are discussed in Section \ref{sec:simstudy}. Lastly, we summarize our findings and discuss future work in Section \ref{sec:conclusion}.

\section{Summary Statistics and Network Monitoring Methods} 
\label{sec:methods}

For anomaly detection, we discuss which statistics are monitored and methods by which to monitor these statistics. The summary statistics calculated from network data and interpretations of such quantities in a network are described in Section \ref{subsec:methods:stats}. How common techniques used in statistical process monitoring are applied to network data is described in Section \ref{subsec:methods:monitoring}. Threshold decisions for the selected monitoring approaches are further discussed in the evaluation of monitoring approaches in Section \ref{sec:simstudy}.

\subsection{Summary Statistics} \label{subsec:methods:stats}

We first define our mathematical notation.
Let $n$ represent the number of nodes in a network at time $t$, where $t$ evolves discretely until time $T$. Let $\boldsymbol{Y}_t$ represent an adjacency matrix at time $t \in \{1, 2, \ldots, T \}$ and $y_{ijt}$ represent an edge weight at time $t$ between nodes $i$ and $j$, for $i,j \in \{1, 2, \ldots, n\}$. 

\textbf{Density} is defined as the sum of edges in a network divided by its total number of possible edges at time $t$. For $n$ nodes, the total number of possible edges is $\binom{n}{2} = \frac{n(n-1)}{2}$, and for $Y_t$, the sum of all edges at time $t$ is $\sum_{i=1}^n \sum_{j=1; j \ne i}^n y_{ijt}$. If we let $W_t$ represent density at time $t$, then 
\begin{equation}
    W_t = \dfrac{2}{n(n-1)} \sum_{i=1}^n \sum_{j=1; j \ne i}^n y_{ijt}.
\end{equation}
Density is considered to be a global measure of a network; a measure that describes the entire network.   

A local measure of a network might summarize individual nodes, such as degree.  Let $D_{it}$ represent  the degree of node $i$ within a network at time $t$.   In a directed network,  $D_{it} = \sum_{j=1}^n (y_{ijt} + y_{jit})$; $D_{it}$ is the sum of in- and out-degrees within directed networks.  For undirected networks, one divides $D_{it}$ in half.  To form a global measure from the local node degrees, \textbf{maximum degree} is often reported.  Let $D_{t}$ represent the maximum degree of a network at time $t$; 
\begin{equation}
D_t = \max_i \{ D_{it} \}.
\end{equation}  

For this paper, we monitor network  density,  maximum degree, and \textbf{two linear combinations of density and maximum degree},  denoted $M_{t}^{-}$ and $M_{t}^{+}$.  We define  $M_{t}^{-}$ and $M_{t}^{+}$ as follows:
\begin{align}
M_{t}^{-} = \frac{1}{n} D_t - W_t\\
M_{t}^{+} = \frac{1}{n} D_t + W_t.
\end{align}

We compare the effectiveness of monitoring $W_t$, $D_{t}$, $M_{t}^{-}$, and $M_{t}^{+}$ to detect anomalies relative to each other and a scan statistic, $S_t^*$,  that was proposed by \cite{priebe2005scan}.   Calculations for $S_t^*$ result over moving windows of size $m$ (e.g., $m$=20) over time $t$ ($t \in \{2m+1,...,T\}$) and are based on the size of local neighborhoods of each node $i$. Neighborhoods within a network at time $t$ are determined from a pre-specified order $k$, e.g., $k=\{0,1,2\}$ in that each neighborhood $i$ ($i \in \{1,2,...,n\}$) is the set of all nodes (and edges between them) within $k$ edges of node $i$. Thus, the size of an order $k$ neighborhood is the number of edges contained in that neighborhood. Let $O^{(k)}_{i,t}$ denote the size of an order $k$ neighborhood of $i$ and time $t$. Note, when $k=0$, the size of an order 0 neighborhood is equivalent to degree. The calculated scan statistic for order $k$ with moving window $m$ involves a 2-step process on the sizes of order $k=\{0,1,2\}$ neighborhoods \citep{zhao2018performance}. We represent scan statistics based on order $k$ as $S_t^{*(k)}$, for  $k=\{0,1,2\}$  and define $S_t^* = \max\{S_t^{*(0)},S_t^{*(1)},S_t^{*(2)}\}$ when reporting results.   

We now overview this 2-step process as is explained in \cite{zhao2018performance} summarizing the work of \cite{priebe2005scan}. First step is to standardize $O^{(k)}_{i,t}$ using a previous window of size $m$. That is, the first standardization of $O^{*(k)}_{i,t}$ with $t > m$ is calculated by $$O^{*(k)}_{i,t} = \dfrac{O^{(k)}_{i,t} - \text{mean}(O^{(k)}_{i,t})}{ \max( \text{sd}(O^{(k)}_{i,t}), 1) },$$ with $$\text{mean}(O^{(k)}_{i,t}) = \frac{1}{m} \sum_{j=1}^m O^{(k)}_{i,t-j} ~~\text{ and }~~  \text{sd}(O^{(k)}_{i,t}) = \sqrt{\frac{1}{m-1} \sum_{j=1}^m [ O^{(k)}_{i,t-j} - \text{mean}(O^{(k)}_{i,t})]^2}$$ for $k=\{0,1,2\}$ and $i \in \{1, 2, \ldots, n\}$. The lower bound of 1 in the denominator of $O^{*(k)}_{i,t}$ prevents detection of relatively small (perhaps too small) changes in a network.  This first standardization of $O^{*(0)}_{i,t}, ~O^{*(1)}_{i,t},$ and $O^{*(2)}_{i,t}$ are done for all times $t > m$. The second step of the process involves standardizing the maxima of $O^{*(k)}_{i,t}$ across nodes $i$. Let $S^{(k)}_t = \max_{i} \{ O^{*(k)}_{1,t}, O^{*(k)}_{2,t}, \ldots, O^{*(k)}_{n,t} \}$ and calculate across nodes $i$ and time $t > 2m$, $S^{(0)}_t, ~S^{(1)}_t,$ and $S^{(2)}_t$. Then, the second standardization of $S^{*(k)}_t$ with $t > 2m$ is calculated by 
\begin{equation}
   S^{*(k)}_{t} = \dfrac{S^{(k)}_{t} - \text{mean}(S^{(k)}_{t})}{ \max( \text{sd}(S^{(k)}_{t}), 1) }, 
\end{equation}
 with 
$$\text{mean}(S^{(k)}_{t}) = \frac{1}{m} \sum_{j=1}^m S^{(k)}_{t-j} ~~\text{ and }~~  \text{sd}(S^{(k)}_{t}) = \sqrt{\frac{1}{m-1} \sum_{j=1}^m [ S^{(k)}_{t-j} - \text{mean}(S^{(k)}_{t})]^2}$$ for $k=\{0,1,2\}$.

Finally, monitoring scan statistics, $S_t^{*{k}}$, begin at time $t = 2m + 1$ since the first $2m$ windows are needed to start the procedure. For more details on $S_t^{*{k}}$, see  both \cite{priebe2005scan} and \cite{zhao2018performance}.

Table \ref{tab:Methods} provides a list of statistics that we study in this paper for network monitoring.
We chose these statistics for their simplicity, popularity, flexibility, and relative meaning in dynamic networks. For example,  network densities over time are easy to calculate and have the potential to reveal both global or local changes in networks. Globally, several nodes may increase or decrease communication (even slightly) over set time(s) in a network, and locally, a few nodes may  significantly increase or decrease communication. Both such  global and local changes could be reflected in changes of network densities.  Note, ``communication'', refers to the number or weight of edges in nodes. Similarly, maximum degree may capture global and local changes based on all or a set node behaviors.   By combining measures of degree and density, e.g., with ``Difference'' ($M_t^{-}$) and ``Sum'' ($M_t^{+}$) statistics, there is additional opportunity to capture global and/or local changes in networks.   For example, with binary networks where $y_{ijt} \in \{ 0, 1\}$),  we know $D_t \leq n$ and  $D_t/n \leq 1$.   Thus, $M_t^{-} = (\frac{1}{n}D_t - W_t) \leq 1$, because $0 \leq W_t \leq 1$. When $M_t^{-}$ values are high (close to 1)  there is large discrepancy between network density and the average node degree.  This would suggest that individual nodes may be highly connected with other nodes (local behavior), while  the overall communication of other nodes is scarce (global behavior). Finally, the scan method $S_t^*$ is popular and well-cited, thus it makes sense to ground our analyses of four summary statistics by its performance.  However, the scan statistic is harder to calculate than the other statistics that we study.  
Furthermore, for large-scale networks (e.g., social media) and real-time monitoring, computing the scan statistic can be computationally expensive and/or infeasible.
Thus, in addition to comparing the effectiveness of five common network summaries, we may form opinions on the trade-offs between complex and simple monitoring approaches.  

\begin{table}[H] \centering
\begin{tabular}{c|c} \toprule
Name (at time $t$) & Notation\\ \hline
Density & $W_t$\\
Max Degree & $D_{t}$ \\
Difference & $M_t^{-}$\\
Sum & $M_t^{+}$\\
Scan Statistic & $S_t^{*}$\\ \bottomrule
\end{tabular}
\caption{Summary statistics for dynamic networks: complexity increases from top to bottom.}
\label{tab:Methods}
\end{table}

\subsection{Statistical process monitoring for networks} \label{subsec:methods:monitoring}

Statistical process monitoring is a well-studied area, but applying these techniques to network data is not straightforward; e.g., it is unclear how to set control limits for network data. The general network monitoring approaches are discussed in Section \ref{subsec:methods:monitoring} while considerations for control limits are further discussed in Sections \ref{subsec:methods:SDcomp} and \ref{subsec:methods:Getq}.

The summary statistics calculated are monitored in two ways. One way utilizes a control chart described in Algorithm \ref{alg:Simstudy:Chart}, and the other utilizes a moving window approach over-viewed in Algorithm \ref{alg:Simstudy:Scan}. In the first approach, there are three main steps. Step one, establish $T_1$, the number of time-points required to determine baseline behavior of a network. We refer to data from $t=1$ through $t=T_1$  as Phase I data, where network snapshots are assumed to have no anomalies. Next, after establishing Phase I data, appropriate control limits, i.e., thresholds for expected non-anomalous behavior, must be determined. Third, established control limits from Phase I data are applied to remaining time-points, i.e., $t>T_1$, called Phase II data, in order to detect any anomalies or atypical behavior. 

Similar to the first approach, the second approach has three main steps. First, a moving window approach requires a window length, $m$, to be set, e.g. 20 time-points. Specifically for the scan statistic, in addition to the first set of $m$ time-points, yet another set of $m$ time-points is needed to start the method. Second, a threshold must be determined for acceptable behavior. Third, the remaining moving windows of length $m$ are observed using that threshold. In both approaches, whether or not an anomaly occurred is recorded using $\boldsymbol{A} = \{ A_1, A_2, \ldots, A_{T-T_1} \}$ such that $A_t = 1$ when an anomaly is detected and 0 otherwise. 

\begin{figure}[H] \centering
\includegraphics[width=2.75in,angle=270]{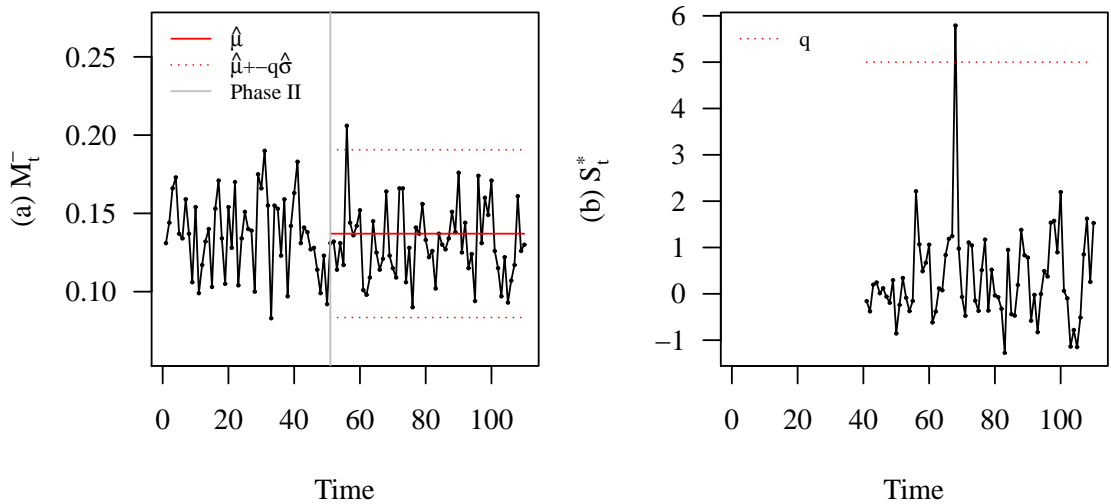}
\caption{Illustration of General Monitoring Approaches for (a) $M^{-}_t$ and (b) $S^{*}_t$. Plot (a) shows a Shewhart individuals control chart for $M^{-}_t$ where control limits are calculated using Phase I data, and plot (b) shows a moving window approach on $S^{*}_t$ using a pre-determined threshold. In this illustration, both methods signal around time-point 60.}
\label{fig:ExMonitor}
\end{figure}

We exemplify the two approaches on $M^{-}_{t}$ and $S^{*}_t$ in Figure \ref{fig:ExMonitor}. From Figure \ref{fig:ExMonitor} (a), we observe control limits, $\text{CL} = \hat{\mu} \pm q \hat{\sigma}$, are calculated using Phase I data, and likewise from Figure \ref{fig:ExMonitor} (b), we see the monitoring process begin at time 41 when using window size $m = 20$ with an acceptable threshold for the scan statistic. In this illustration, both methods signal around time-point 60. 

In the first monitoring approach, Shewhart individual control charts are utilized for a majority of our summary statistics. In any control chart, a statistic is observed over time with control limits indicating expected behavior of that statistic. Since the network data are \emph{correlated snapshots} over time, control limits should be adjusted for such correlation. Control limits of a Shewhart individuals control chart are $\overline{x} \pm (3 \overline{MR})/d_2$ where $d_2$  is an anti-biasing constant and set to 1.13; and 
$MR$ is a moving range, for a process $\{ x_{1}, x_{2}, \ldots, x_{n} \}$ so that $MR = |x_{i} - x_{i-1}| ~ (i > 1)$ \citep{montgomery2007introduction}. However, in the context of network data, these control limits may not be appropriate.  
In Section \ref{subsec:methods:SDcomp}, we assess  whether the standard deviation of our summary statistics is appropriately measured using $\overline{MR}/d_2$, and in Section \ref{subsec:methods:Getq}, we evaluate the utility of applying $3$ times the standard deviation as control limits.

\begin{algorithm}
    \caption{General Monitoring Approach for $W_t, ~D_t, ~ M_{t}^{-},$ and $M_{t}^{+}$.}
  \begin{algorithmic}[1]
  \INPUT Temporally-evolving network data.
  \OUTPUT $\boldsymbol{A} = \{ A_{T_1+1}, A_{T_1+2}, \ldots, A_{T} \}$ where $A_t = 1$ if anomaly is detected at time $t$ and 0 otherwise.
    \STATE Set number of time-points for Phase I data, $T_1$. Data from Phase I is assumed to be non-anomalous.
    \STATE Use Phase I data to determine control limits for a Shewhart individuals control chart. Control limits (CL) are of the form: $\text{CL} = \hat{\mu} \pm q \hat{\sigma}$.
    \STATE  \textbf{For $t$ in $(T_1 + 1):T$ \{}
    \begin{itemize}
        \item[] For a summary statistic at time $t$, denoted $x_t$, record if an anomaly occurred or not using: 
        $$A_t = \begin{cases} 1, & x_t < \hat{\mu} - q \hat{\sigma}\\ 1, & x_t > \hat{\mu} + q \hat{\sigma}\\ 0, & \text{otherwise} \end{cases}.$$
    \end{itemize}
    \textbf{ \} END}
  \end{algorithmic} \label{alg:Simstudy:Chart}
\end{algorithm}

\begin{algorithm}
    \caption{General Monitoring Approach for $S_t^{*}$.}
  \begin{algorithmic}[1]
  \INPUT Temporally-evolving network data.
  \OUTPUT $\boldsymbol{A} = \{ A_{T_1+1}, A_{T_1+2}, \ldots, A_{T} \}$ where $A_t = 1$ if anomaly is detected at time $t$ and 0 otherwise.
    \STATE Determine the length of moving window, i.e. set number of time-points for a window, $m$.
    \STATE Use $2m$ windows for starting the scan method. First set of $m$ windows is for maximizing across nodes. Second set of $m$ windows is for maximizing across time. 
    \STATE Set threshold, $q$, to determine if an anomaly occurred or not, where $q \in \{1, 2, 3, \ldots \}$.
    \STATE  \textbf{For $t$ in $(T_1 + 1):T$ \{}
    \begin{itemize}
        \item[] For scan statistic at time $t$, denoted $S^{*}_t$, record if an anomaly occurred or not using: 
        $$A_t = \begin{cases} 1, & S^{*}_t > q \\ 0, & \text{otherwise} \end{cases}.$$
    \end{itemize}
 \textbf{ \} END}
  \end{algorithmic} \label{alg:Simstudy:Scan}
\end{algorithm}

\subsubsection{Estimating Standard Deviation of Summary Statistics}
\label{subsec:methods:SDcomp}

To determine how to appropriately estimate the standard deviation of our summary statistics, we compare the standard $\overline{MR}/d_2$, i.e., average moving range (AMR), with two alternatives. The first alternative is to use the median of the moving ranges, median moving range (MMR). The second alternative is using a correlated data calculation for standard deviation (SD), $s = \sqrt{ \dfrac{s^2}{\gamma_1}}$ such that $\gamma_1 = 1 - \dfrac{2}{(n-1)} \displaystyle{\sum_{\kappa=1}^{n-1} \left( 1 - \dfrac{\kappa}{n} \right) \rho_\kappa}$ and $\rho_\kappa$ is autocorrelation at lag $\kappa$ \citep{box2008time}. These three measures of standard deviation are compared by the rate of false alarms above the upper control limit, $\hat{\mu} + q \cdot \hat{\sigma}$, where $\hat{\mu} = \overline{x}$, $q \in [2,4]$, and $\hat{\sigma}$ is $s$ for SD, $\overline{MR}/d_2$ for AMR, and median of $MR$ values multiplied by 1.047 for MMR \citep{montgomery2007introduction}. 

\begin{figure}[ht] \centering
\includegraphics[width=4.25in,angle=270]{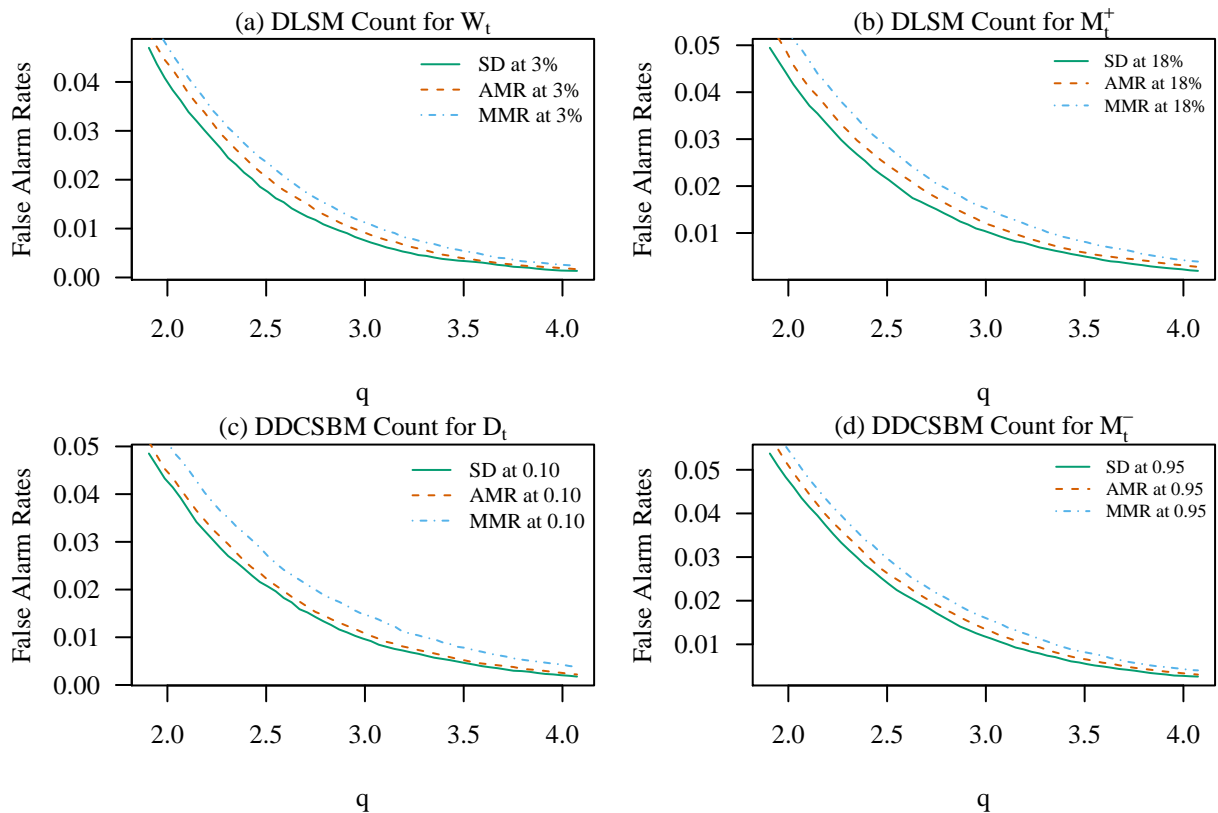}
\caption{False Alarm Rates of AMR, MMR, and SD as $\hat{\sigma}$ in Control Limits, $\overline{x} + q \cdot \hat{\sigma}$, with $q \in [2,4]$ across $W_t, ~M^{+}_t, ~D_t,$ and $M^{-}_t$. Plots (a) and (b) DLSM Count with Sparsity Levels and plots (c) and (d) DDCSBM Count with Correlation Levels. Dashed lines correspond to control limits using AMR, dotted lines are for MMR, and solid lines are for SD. False alarm rates are the lowest in all settings when using SD as $\hat{\sigma}$.}
\label{fig:SDComp}
\end{figure}

Results of 200 simulations were obtained using Phase I data with 100 nodes, $T_1=50$ time-periods for comparing metrics, and data generated from two common network models:  the Dynamic Latent Space Model (DSLM) and the Dynamic Degree-Corrected Stochastic Block Model (DDCSBM).  Details on models and model settings are described later in Section \ref{sec:models}.  Figure \ref{fig:SDComp} shows four examples of false alarm rates with Figures \ref{fig:SDComp} (a) and (b) from a DLSM count setting using $\avgden$ at 3\% and at 18\% respectively and with Figures \ref{fig:SDComp} (c) and (d) from a DDCSBM count setting using $\phi$ at 0.10 and at 0.95.  As seen in the figures, false alarm rates from control limits using SD and AMR perform similarly (solid and dashed lines in Figure \ref{fig:SDComp}), but MMR results in a larger number of false alarms when considering control limits between 2 to 4 standard deviations above the mean (dotted lines in Figure \ref{fig:SDComp}). Thus, to account for the correlation in monitoring dynamic network statistics and to obtain the smallest false alarm rates, the SD calculation is used as $\hat{\sigma}$ in calculating the control limits.

\subsubsection{Determining Appropriate Thresholds}
\label{subsec:methods:Getq}

We ascertain if $3 \sigma$ limits are appropriate in addition to suitable thresholds for monitoring the scan statistic, $S_t^{*}$. When monitoring the averages of batches via a Shewhart $\bar{X}$ chart and conditioning on Phase I data, there are several approaches to determine appropriate control limits \citep{jardim2019two}. Similar principles from the $\bar{X}$ chart can be used in our context for summary statistics. In general, we aim to fix the conditional false alarm probability for a statistic $x_t$ such that $Pr(x_t > \hat{\mu} + q\hat{\sigma}) = p$. When monitoring the scan statistic, the method signals when the scan statistic is above a threshold \citep{priebe2005scan}. The threshold recommended by \cite{priebe2005scan} is 5, while in \cite{zhao2018performance} the authors showed via simulation study that perhaps lowering the threshold with a tolerable rate of false alarms can improve the performance of the scan statistic. For the scan statistic, we have the conditional false alarm probability as $Pr(x_t > q) = p$. Ultimately, $p$ is determined by the practitioner, but in this work, we settle on $p=0.03$. After calculating $\overline{x}$, using the correlated standard deviation $s$, and setting $p$, we must calibrate to obtain an appropriate $q$.  

There are several ways to calibrate an appropriate $q$ in thresholds of network monitoring. One option is if $x_t$ reasonably follows a normal distribution (or some other well-known distribution), then $q$ can be set using one-tail probabilities of the normal distribution, $p$. 
However, this method can lead to errors when the normality (or standard distribution) assumption fails to hold. Another option is empirically obtaining false alarm rates, the rate of signals for data with no anomalies. False alarm rates can be calculated from Phase I data or generated Phase II data with no anomalies. In practice, one does not know whether there is any anomaly in Phase II, and therefore only Phase I data should be used for calibrating $q$.
However, there can be a difference between nominal false alarm rates (calibrated using Phase I data) and actual false alarm rates (resulting from Phase II data) in this approach.
This happens due to sampling variation as the distribution of Phase II data, even when non-anomalous, is unlikely to be an exact replicate of Phase I data.
In our case, we used synthetic network data for performance assessment. Since data are generated, Phase II data are generated from the same distribution as Phase I data. Therefore, established control limits from Phase I data can be applied in Phase II data. 
This allows an exact calibration of false alarm rates in Phase II, which allows us to accurately compare the anomaly detection performance of the various summary statistics and the scan statistic of \cite{priebe2005scan}.

\begin{figure}[H] \centering
\includegraphics[width=4.75in,angle=270]{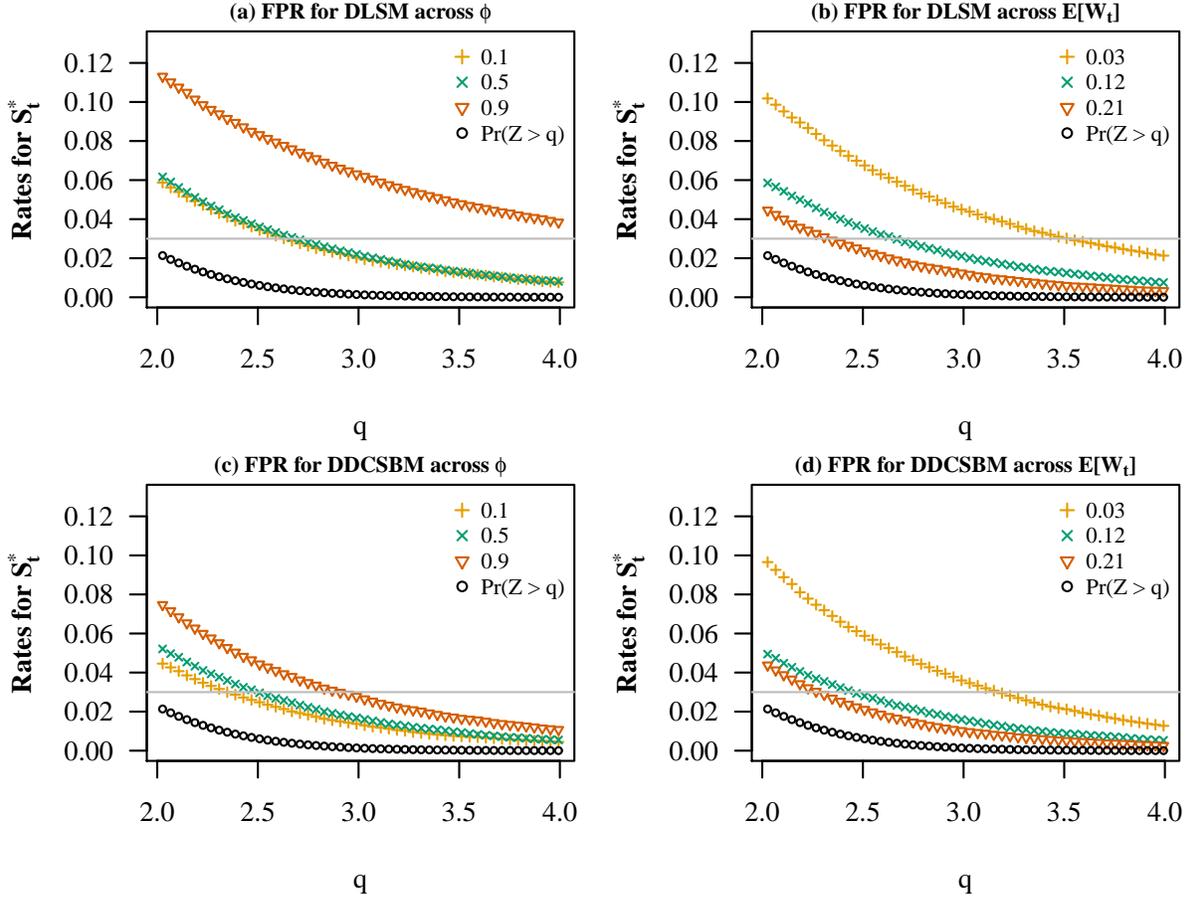}
\caption{Plot of False Alarm Rates Monitoring $S_t^{*}$ Across Varying Correlation [(a) and (c)] and Sparsity [(b) and (d)] in Binary DLSM and DDCSBM Settings.}
\label{fig:FPR:Bin:Scan}
\end{figure}

To determine if summary statistics approximately follow a normal distribution, empirical false alarm rates were compared to that based on the upper tail of a standard normal distribution, $Pr(Z > q)$. Here, false alarm rates are calculated using Phase II data with the same conditional false alarm probability as Phase I. Specifically in a Shewhart individuals control chart, the number of times $x_t$ exceeds $\overline{x} + q s$ is recorded, and for the scan statistic, the number of times $x_t$ exceeds $q$. Calibrating $q$ is determined empirically using false alarm rates across varying correlation and sparsity values. Correlation is denoted using $\phi$ and sparsity is average density denoted using $\avgden$. Further details on models, model settings, and correlation and sparsity values are discussed in Sections \ref{sec:methods} and \ref{sec:simstudy}. 
Monte Carlo simulations were performed a total of 1000 times using 100 nodes ($n=100$) and 110 time-points ($T=110$) in monitoring $W_t, ~D_t, ~M^{-}_t, ~M^{+}_t$ and $S^{*}_t$. For each simulation, all 110 time-points were simulated from the same model with Phase I data set at time-points up to time 50. Then, control limits are estimated from Phase I data, and the number of false alarms are calculated starting at time-point 51. To obtain false alarm rates, counts of false alarms across 1000 simulations are summed together and divided by 1000. An example in binary DLSM and DDCSBM settings is shown for monitoring the scan statistic in Figure \ref{fig:FPR:Bin:Scan}.  As can be seen in Figure \ref{fig:FPR:Bin:Scan}, false alarm rates are generally double or triple that of the area of the upper tail of a standard normal distribution for the scan statistic when varying the threshold of $q$ between $[2,4]$. False alarm rates tended to increase as correlation increases in the binary DLSM setting. 

Examples for binary or count networks are shown in Figures \ref{fig:FPR:Dens:Ct}- \ref{fig:FPR:Sum:Bin} in Appendix \ref{subsec:AppA:FPR} for $W_t, ~D_t, ~M^{-}_t,$ and $M^{+}_t$. In general, there is less of a pattern between correlation and false alarm rates as well as sparsity and false alarm rates in the DDCSBM setting. In the DLSM setting, higher correlation tends to increase false alarm rates when monitoring with these methods, while varying correlation and sparsity has little effect in the DDCSBM setting. Hence, $p=0.03$ is used as our false alarm threshold to accommodate those statistics which do not follow a normal distribution and have higher false alarm rates than that obtained using the upper one-sided tail of a standard normal distribution. Then, $q$ is calibrated from empirical false alarm rates as close as possible to $p=0.03$.

\section{Statistical Models for Dynamic Networks} \label{sec:models}

Our methods for anomaly detection do not rely on a parametric model.
Therefore, we neither have to estimate parameters of such a model nor determine if parameter estimates had changed significantly. 
Rather, we use statistical models to generate synthetic network data, and then implement the model-free methods on these synthetic networks.
We highlight this advantage of a \nonpara\ approach to parametric based methods by generating dynamic network data from two popular latent variable models. \cite{kim2018review} reviewed several latent variable models including latent space models and stochastic block models in both static and dynamic versions. Specifically, we focus on particular formulations of a dynamic latent space model (DLSM) and dynamic \degcor\ stochastic block models (DDCSBM).

\subsection{Dynamic Latent Space Model} \label{sec:models:DLSM}


Use of latent space models relies on defining a space of latent positions using either pairwise distances or projections \citep{kim2018review}. \cite{hoff2002latent} introduced both a distance based and projection based latent space model using Bayesian inference for parameter estimation via MCMC. The idea here is that the probability of forming an edge between nodes $i$ and $j$ depends on the latent positions $z_i$ and $z_j$. For example, $z_i$ and $z_j$ being relatively close together (via a distance metric) in the latent space suggests a higher probability of an edge between nodes $i$ and $j$ in the network. Hence, edges are conditionally independent given the latent positions. We focus on the formulation of \cite{SewellChen2015} who extended ideas of \cite{hoff2002latent} to a dynamic version of a latent space model. 

We now provide an overview of the DLSM in \cite{SewellChen2015}. More details can be found in their paper.  Let $\boldsymbol{Y}_t$ be an adjacency matrix at time $t \in \{1, 2, \ldots, T \}$ such that $y_{ijt} = 1$ when there is an edge between nodes $i$ and $j$ and 0 otherwise.  Let $\boldsymbol{\mathcal{X}}_t$ be a matrix of latent positions ($\boldsymbol{X}_{it} = (x_{it1}, x_{it2})$ for node $i \in \{ 1, 2, \ldots, n \}$) at time $t$. For ease in plotting and visualizations, two-dimensional coordinates are used. The probability of an edge between nodes $i$ and $j$ is given by

$$y_{ijt} ~|~ \boldsymbol{\Psi} = \{\boldsymbol{X}_{it}, \boldsymbol{X}_{jt}, \beta_{IN}, \beta_{OUT}, \boldsymbol{r}, \sigma^2 \} \sim \text{Bern} \left(p_{ijt} = \frac{\exp(\eta_{ijt})}{1 + \exp(\eta_{ijt})}\right) \text{ s.t. }$$ 
$$ \eta_{ijt} = \log \left( \frac{Pr(y_{ijt} =1 ~|~ \boldsymbol{\Psi})}{Pr(y_{ijt} = 0 ~|~ \boldsymbol{\Psi})}\right) = \beta_{IN} \left( 1 - \frac{d_{ijt}}{r_j} \right) + \beta_{OUT} \left( 1 - \frac{d_{ijt}}{r_i}  \right).$$

\noindent The parameters $\beta_{IN}$ and $\beta_{OUT}$ are global (system-scale) network characteristics describing popularity and social activity respectively. If $\beta_{IN} > \beta_{OUT}$ in a directed network, then this suggests the receiver is more important while the opposite scenario suggests the sender is more important. The radii in $\boldsymbol{r} = (r_{1},r_{2}, \ldots, r_{n})$, are node specific characteristic describing the radius of communication in the latent space such that the radii sum to 1, i.e., $\sum_{i=1}^n r_i = 1$. These radii are analogous to a scaled degree of a node. The pairwise distance between latent positions of nodes $i$ and $j$ is denoted as $d_{ijt} = \text{dist}(\boldsymbol{X}_{it}, \boldsymbol{X}_{jt})$, and the spread of the latent positions are controlled by $\sigma^2$. 

\begin{figure}[H]
\centering
\includegraphics[width=4in,angle=270]{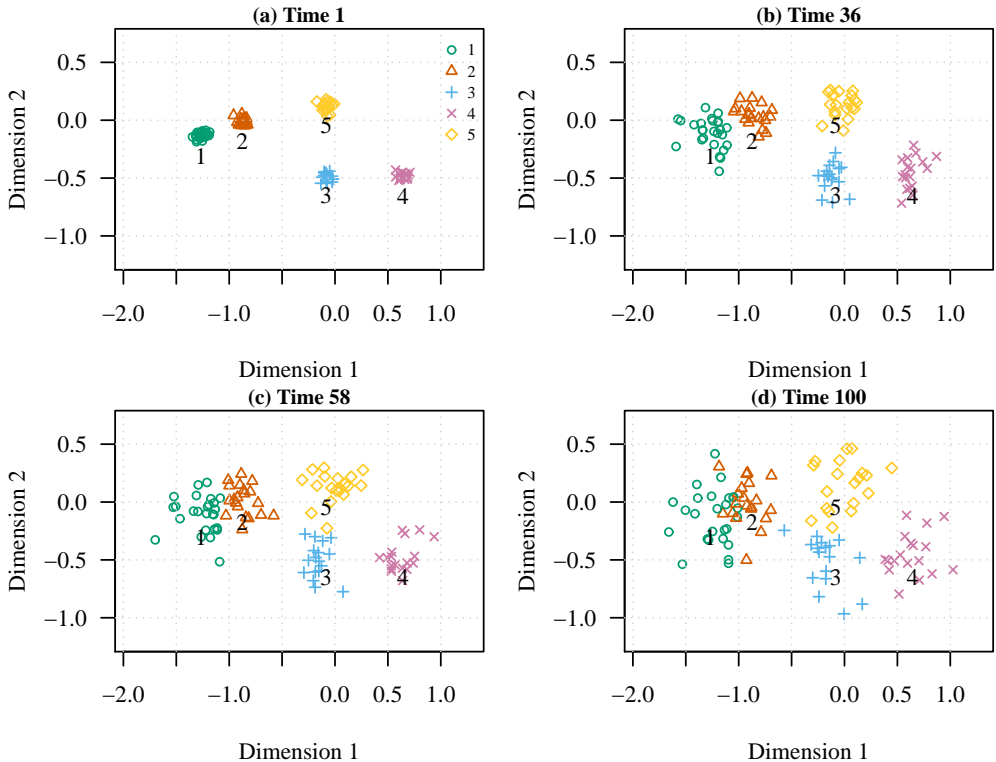}
\caption{Latent Positions at Times $\text{(a) } 2, \text{(b) } 26, \text{(c) } 58,$ and (d) $100$. For $\sigma^2 = 0.0003$ and $T = 100$, the spread of the latent positions grow over time when starting in tight pre-defined clusters.} 
\label{fig:1:origprior}
\end{figure}

In order to generate data from this model, we use the prior distribution on latent positions to get $\boldsymbol{\mathcal{X}}_{1:T}$ and subsequently $d_{ijt}$ and set values for the remaining parameters, $\beta_{IN}, ~\beta_{OUT}, ~\boldsymbol{r},$ and $\sigma^2$. The original prior in \cite{SewellChen2015} is $N(\boldsymbol{0}^T, \tau^2I_2)$ for the initial network and $N(\boldsymbol{X}_{i(t-1)},\sigma^2I_2)$ for all subsequent networks. Hence, a priori, latent positions may move around the space according to some random jump (dictated by $\tau^2$ or $\sigma^2$). For meaningful networks, the variances of the latent positions must be quite small in order to achieve reasonable edge probabilities. We demonstrate how latent positions move a priori in Figure \ref{fig:1:origprior} using $T=100, ~\sigma^2 = 0.0003$ and 5 clusters (defined by both color and shape). Latent positions are plotted at times $t = 2, 36, 58,$ and 100 in plots (a)-(d) respectively. As time increases, we observe a spread of latent positions from the initial tight clusters. Perhaps intermixing of clusters is desired a priori, but the spread of latent positions has an effect on edge probabilities, $p_{ijt}$.  By increasing pairwise distances of the latent positions, $d_{ijt}$, this decreases $p_{ijt}$ values over time, and in turn, decreases realizations of $p_{ijt}$ values via network density. An interpretation of Figure \ref{fig:1:origprior} is that people communicate less and less as over time, which may not make practical sense. Thus, generating latent positions from this distribution is a concern due to the spread in latent positions. 

Our solution is to scale the spread of the latent positions, $\sigma^2$, using a time series model, in particular, a vector auto-regression of order 1 (VAR(1)) model. Consider the following VAR(1) model:

$$ \boldsymbol{X}_{i(t+1)} = \phi \cdot \boldsymbol{X}_{it} + \boldsymbol{\epsilon}_{t+1}  \Rightarrow \begin{bmatrix} x_{i1(t+1)}\\ x_{i2(t+1)} \end{bmatrix}  = \phi * \begin{bmatrix} x_{i1t}\\ x_{i2t} \end{bmatrix} + \begin{bmatrix} \epsilon_{1(t+1)}\\ \epsilon_{2(t+1)} \end{bmatrix},$$
such that $|\phi| < 1$ and $\boldsymbol{\epsilon}_{t+1} \sim N((0,0)^T, \sigma^2I_2).$ Thus, a VAR(1) prior on latent positions has the following form:
\begin{align*}
\boldsymbol{\mathcal{X}}_1 ~|~ \phi, \sigma^2 & \sim \prod_{i=1}^n N \left(\textbf{0}, \tau^2 = \left( \frac{\sigma^2}{(1 - \phi^2)} \right) I_2 \right)\\
\boldsymbol{\mathcal{X}}_t ~|~ \boldsymbol{\mathcal{X}}_{t-1}, \phi, \sigma^2 & \sim \prod_{i=1}^n N \left( \boldsymbol{X}_{i(t-1)}, \tau^2 = \frac{\sigma^2}{(1- \phi^2)} I_2 \right), ~ \text{for } t \geq 2.
\end{align*}

\begin{figure}[H]
\centering
\includegraphics[width=2.5in,angle=270]{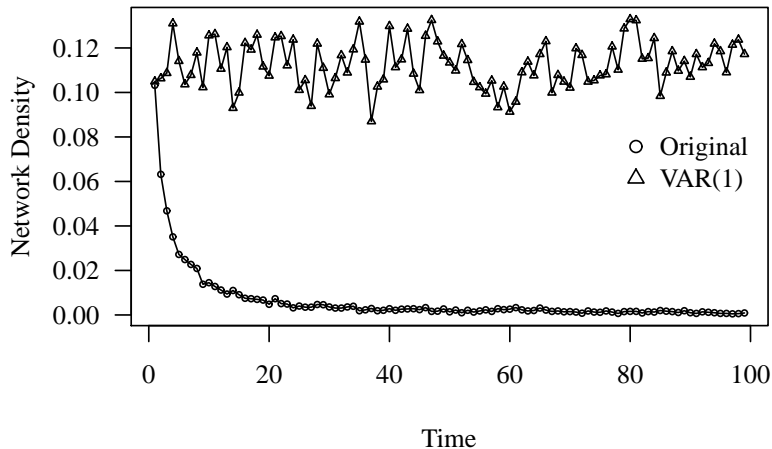}
\caption{Network Density using Original Prior and VAR(1) Prior on Latent Positions with $n=100, ~T=100, ~\beta_{IN} = \beta_{OUT} = 1,  ~\boldsymbol{r} = \{ 1/n \}, ~\phi = 0.3, $ and $\sigma^2 = 0.0003$.} 
\label{fig:2:density}
\end{figure}

\noindent We compare the network densities from the original prior to a VAR(1) prior when $n=100, ~T=100, ~\beta_{IN} = \beta_{OUT} = 1,  ~\boldsymbol{r} = \{ 1/n \}, ~\phi = 0.3, $ and $\sigma^2 = 0.0003$ in Figure \ref{fig:2:density}. Network density is plotted over time, and we observe a decay in density using the original prior whereas density using a VAR(1) prior is within a reasonable range around an average density of 11\%. Thus, using a VAR(1) prior appears able to control the spread of the latent positions a priori over time.

By default, the model is setup as a binary network, but we can make the following modification for a count network: $y_{ijt} ~|~ \boldsymbol{\Psi} \sim \text{Poisson} (p_{ijt} = \exp\{\eta_{ijt}\}) \text{ s.t. } \eta_{ijt} = \log (E[y_{ijt} ~|~ \boldsymbol{\Psi}])$ \citep{sewell2016weighed}. Default settings of parameters in the simulation study include $\sigma^2 = (1-\phi^2), ~\beta_{IN} = 1, ~\beta_{OUT} = 2, \text{ and } \boldsymbol{r} = \{ \frac{1}{n} \}_{i=1}^n$. An outline of the data generation from a DLSM is provided in Algorithm \ref{alg:DLSM}.

\begin{algorithm}
    \caption{Data Generation from a DLSM.}
  \begin{algorithmic}[1]
    \item Set $n$, $T$, $\phi$, $\sigma^2$, $\beta_{IN}$, $\beta_{OUT}$, $\boldsymbol{r}$ and choose binary or count network.
    \STATE Generate latent positions, $\boldsymbol{\mathcal{X}_{1:T}}$, using $\phi$ and $\sigma^2$.
    \begin{itemize}
        \item Set number of clusters and draw means for each cluster using $N(\boldsymbol{0}, (2/n)^2I_2)$.
        \item Assign cluster labels to $n$ nodes.
        \item Generate $\boldsymbol{\mathcal{X}_1}$ by drawing from a normal distribution with assigned cluster mean and variance $\sigma^2I$.
        \item Generate $\boldsymbol{\mathcal{X}_{2:T}}$ using $\boldsymbol{\mathcal{X}_t} = \phi \cdot \boldsymbol{\mathcal{X}_{t-1}} + N(0,\sigma^2)$. \emph{Note: For stationarity, generate an additional $T$ latent positions and keep last $T$ draws.}
    \end{itemize}
    \STATE Calculate pairwise distances of latent positions.
    \STATE  \textbf{For $t$ in $1:T$ \{}
    \begin{itemize}
        \item[] \textbf{For $i$ in $1:n$ \{}
        \begin{itemize}
        \item Get $\eta_{ijt} = \beta_{IN} \left( 1 - \frac{d_{ijt}}{r_j} \right) + \beta_{OUT} \left( 1 - \frac{d_{ijt}}{r_i}  \right)$.
        \item Draw $y_{ijt} \sim \text{Bern} \left(p_{ijt} = \frac{\exp(\eta_{ijt})}{1 + \exp(\eta_{ijt})}\right)$ (binary) or $y_{ijt} \sim \text{Poisson} (p_{ijt} = \exp\{\eta_{ijt}\})$ (count) .
        \end{itemize}
        \item[] \textbf{ \} END}
    \end{itemize}
     \textbf{ \} END}
  \end{algorithmic} \label{alg:DLSM}
\end{algorithm}

\subsection{Dynamic \degcor\ stochastic block model} \label{sec:models:DDCSBM}

Stochastic block models (SBM), on the other hand, utilize a latent community (block) assignment to assign edge probabilities \citep{snijders1997estimation}. That is, edge probabilities within a community should differ from edge probabilities between communities.  
\cite{matias2017statistical} extended an SBM model to a dynamic version which highlights evolving community memberships over time as well as any desired probability density measure on edge probabilities. A general issue with SBMs is when degree is heterogeneous among nodes \citep{karrer2011stochastic}. Many times, communities of high degree nodes and low degree nodes are formed, which may not be appropriate community assignments. \cite{karrer2011stochastic} developed a degree-corrected version of an SBM (DCSBM). \cite{wilson2016modeling} discussed a dynamic version of a DCSBM utilizing parameters to describe a propensity for a node to communicate. The DDCSBM model we use to generate data from adapts the dynamic SBM of \cite{matias2017statistical} to include degree heterogeneity and the dynamic DCSBM of \cite{wilson2016modeling} to include correlation over time in the propensities. Ultimately, we use a model that allows for movement within community assignment, degree heterogeneity, and correlation among propensities to communicate over time.

We now provide an overview of the adapted DDCSBM used for data generation. 
Let $\boldsymbol{Y}_t$ represent an adjacency matrix at time $t$ for $t \in \{1, 2, \ldots, T \}.$ Let $K$ represent the number of communities and the probability of an edge between nodes $i$ and $j$ is defined by: $$y_{ijt} \sim \text{Poisson}(p_{ijt} = \theta_{it}\theta_{jt}\omega_{Z_{it}Z_{jt}}).$$
Parameter $Z_{it}$ is a latent community assignment for node $i$ at time $t$. The initial assignment is found via $Z_{i0} \sim \text{Multinomial}(\boldsymbol{\alpha} = \{1/K\}_{k=1}^{K})$, and for all subsequent networks, community assignments can transition with probability $\boldsymbol{\pi}_{K \times K}$, where 
$$\boldsymbol{\pi}_{K \times K} = \begin{pmatrix} \boldsymbol{\pi}_{1}\\ \boldsymbol{\pi}_{2}\\ \vdots\\  \boldsymbol{\pi}_{K} \end{pmatrix} = \begin{pmatrix} \pi_{11} & \pi_{1K} & \cdots\ \pi_{1n}\\ \pi_{21} & \pi_{22} & \cdots\ \pi_{2K}\\ \vdots & \vdots & \ddots\ \vdots \\  \pi_{K1} & \pi_{K2} & \cdots\ \pi_{KK} \end{pmatrix}$$ such that if $Z_{it} = k$, $Z_{i(t+1)} \sim \text{Multinomial}(\boldsymbol{\pi}_{k})$ for $k \in \{ 1, 2, \ldots, K \}$. Next, we introduce propensity to communicate parameters, $\boldsymbol{\Theta} = \{ \boldsymbol{\theta}_1, \boldsymbol{\theta}_2, \ldots, \boldsymbol{\theta}_n \}$ such that $\boldsymbol{\theta}_i = \{ \theta_{i1}, \theta_{i2}, \ldots, \theta_{iT} \}$ and $\theta_{it} \in [1 - \delta, 1 + \delta]$ for some $\delta \in (0,1)$ \citep{wilson2016modeling}. A low propensity to communicate suggests the degree of that node will be relatively low. Parameter $\boldsymbol{\omega}_{K \times K}$ is community structure matrix such that $\omega_{Z_{it}Z_{jt}} \in (0,1)$ and $\text{diag}(\boldsymbol{\omega})$ are all distinct values. As noted in \cite{matias2017statistical}, if any intra-community communication values are the same, then distinguishing those communities remains unidentifiable. Community assignments affect propensities in the following way. Propensities of communication for a given community must be rescaled by the average propensity to preserve community structure. That is, if nodes transition between communities, then corresponding $\theta_{it}$ will be rescaled as will the average propensity for those communities at time $t$. 

\begin{figure}[H] \centering
\includegraphics[width=2.75in,angle=270]{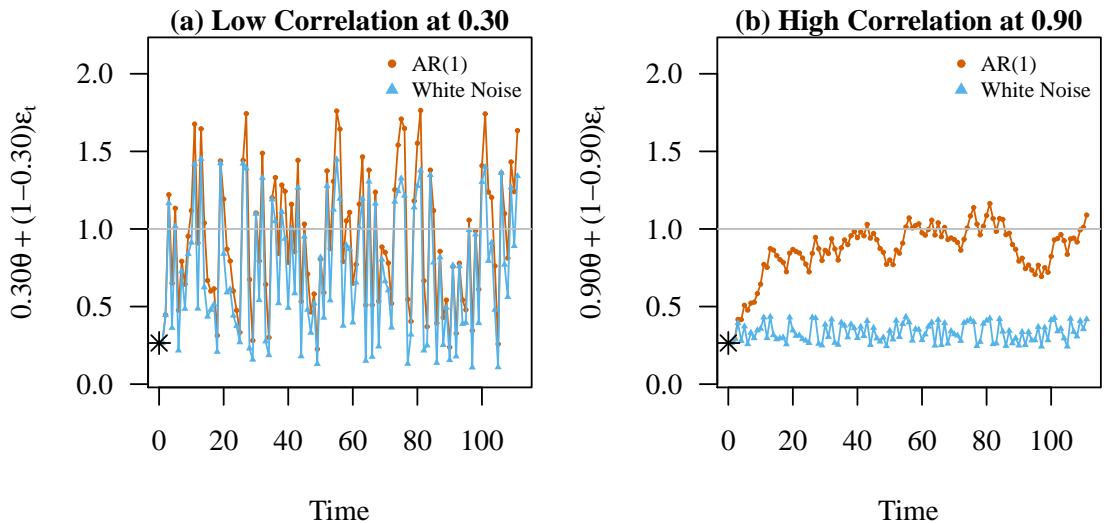}
\caption{AR(1) Model and White Noise at $\phi = 0.30$ (a) and $0.90$ (b) for $\boldsymbol{\theta}$ using $\delta=0.98$ and $\theta_0 = 0.265$.} 
\label{fig:3T:ThetaComp}
\end{figure}

We specify the following ``white noise'' process to incorporate correlation. First, $\theta^{*}_{i0} \sim U(-1,1)$ for all nodes $i$. Next, 
\begin{equation} \label{eqtn:DDCSBM:Theta:WhitNoise}
    \theta^{*}_{it} = \phi \cdot \theta^{*}_{i0} + (1 - \phi) \cdot \epsilon_{it} \text{ such that } \epsilon_{it} \sim U(-1,1) \text{ and } |\phi| < 1.
\end{equation} Then, $\theta_{it} = \delta*\theta^{*}_{it} + 1$ so that $\theta_{it} \in [1-\delta,1+\delta]$ for some $\delta \in (0,1)$. In Equation \ref{eqtn:DDCSBM:Theta:WhitNoise}, $\epsilon_t$ is scaled in order to achieve $\theta_{it} \in [1-\delta,1+\delta]$.
The reason for using a constant $\phi \cdot \theta^{*}_{i0}$ in Equation \ref{eqtn:DDCSBM:Theta:WhitNoise} rather than a more traditional AR(1), i.e., $\phi \cdot \theta^{*}_{i(t-1)}$, is that we want propensities to stay within a reasonable range of their starting value as $|\phi| \rightarrow 1$. After transformation when using an AR(1) model, we have the mean of the series as 1 with variance of $(1-\phi)^2 \sigma_{\epsilon}^2/ (1 - \phi^2)$. An issue is regardless of the starting value of a propensity, the AR(1) process tends toward the mean of 1 (after transformation). This implies, if $\theta^{*}_{i0} = -0.75$ and $\delta = 0.98$ so that $\theta_{i0} = \delta \cdot \theta^{*}_{i0} + 1 = 0.265$, then an AR(1) process with $\phi = 0.90$ for $\boldsymbol{\theta}_i$ settles around 1 rather than around $\theta_{i0} = 0.265$. We illustrate this tendency toward the mean of the process in Figure \ref{fig:3T:ThetaComp} using $\theta_{i0} = 0.265$ (shown using an asterisk) and $\delta=0.98$ at a relatively low and high correlation, $\phi = 0.30, 0.90$. In Figure \ref{fig:3T:ThetaComp}, we contrast an AR(1) model with our ``white noise'' model given in Equation \ref{eqtn:DDCSBM:Theta:WhitNoise} using the same $\epsilon$ values for both processes. Thus, low or high propensity to communicate nodes would be washed out if using a traditional AR(1) model. Therefore, scaling a ``white noise'' type process introduces a different kind of correlation based on the initial starting value of a propensity as can be seen in Figure \ref{fig:3T:ThetaComp}.

The default edge type is counts for the DDCSBM, but for any threshold, $b \in \mathbb{N}$, we can turn edge counts into a binary network. In this work, we use $b=1$ to obtain a binary network such that $y^{*}_{ijt} = 1$ if $y_{ijt} \geq b$. Settings of parameters in the simulation study include $K=3, ~\delta=0.98,$ $$~\boldsymbol{\pi} = \begin{pmatrix} 0.96 & 0.02 & 0.02\\ 0.02 & 0.96 & 0.02\\ 0.02 & 0.02 & 0.96 \end{pmatrix}, \text{ and } \boldsymbol{\omega} = \begin{pmatrix} 0.7 & 0.2 & 0.25\\ 0.2 & 0.6 & 0.3\\ 0.25 & 0.3 & 0.5 \end{pmatrix}.$$ Three communities are chosen rather than two for some additional variation. The $\theta_{it} \in [0.02, 1.98]$ are re-scaled within each community by the average propensity of each community respectively. Relatively low transition rates to other communities are selected for $\boldsymbol{\pi}$. Lastly, intra-communication was chosen to be much higher than inter-communications to help identify communities \citep{matias2017statistical}. An outline of the data generation from a DDCSBM is provided in Algorithm \ref{alg:DDCSBM}. Lastly, notation used to generate data from dynamic networks is summarized in Table \ref{tab:Notation:Models}.

\begin{algorithm}
    \caption{Data Generation from a DDCSBM.}
  \begin{algorithmic}[1]
    \item Set $n$, $T$, $K$, $\phi$, $\delta$, $\omega$, and $\pi$, and choose a binary or count network.
    \STATE Generate latent community assignments using $K$ and $\pi$.
    \begin{itemize}
        \item Sample initial latent community assignments, $Z_{i0} \sim \text{Multinomial}(\boldsymbol{\alpha} = \{1/K\}_{k=1}^{K})$.
        \item Generate $Z_{it} \sim \text{Multinomial}(\boldsymbol{\pi}_k)$ for $Z_{i(t-1)} = k$.
    \end{itemize}
    \STATE Generate propensities to communicate, $\boldsymbol{\Theta}$, using $\phi$ and $\delta$.
    \begin{itemize}
        \item Generate initial values $\{ \theta^{*}_{01}, \theta^{*}_{02}, \ldots, \theta^{*}_{0n} \} \sim U(-1,1)$.
        \item Generate $\boldsymbol{\theta^{*}_{i}}$ using $\theta^{*}_{it} = \phi \cdot \theta^{*}_{i0} + (1 - \phi) \cdot U(-1,1).$
        \item Recalculate $\boldsymbol{\theta_{i}} = \delta*\boldsymbol{\theta^{*}_{i}} + 1$ for all nodes $i$.
    \end{itemize}
    \STATE  \textbf{For $t$ in $1:T$ \{}
    \begin{itemize}
        \item[] \textbf{For $i$ in $1:n$ \{}
        \begin{itemize}
        \item Scale $\theta_{it}$ by the mean of propensities in community $k$: $\theta^{\star}_{it} = \theta_{it}/ \sum_{|k|} \theta_{jt}$.
        \item Draw $y_{ijt} \sim \text{Poisson}(p_{ijt} = \theta_{it}\theta_{jt}\omega_{Z_{it}Z_{jt}})$ (count) or $y*_{ijt} = \begin{cases} 0 & y_{ijt} = 0\\ 1 & y_{ijt} \geq 1 \end{cases} $ (binary).
        \end{itemize}
        \item[] \textbf{ \} END}
    \end{itemize}
     \textbf{ \} END}
  \end{algorithmic} \label{alg:DDCSBM}
\end{algorithm}

\begin{table}[H] \centering
\begin{tabular}{c|l} \toprule
\textbf{Symbol} & \textbf{Meaning} \\ \hline
$K$ & Number of communities (blocks)\\
$\beta_{IN}$ & Global DLSM parameter for popularity\\
$\beta_{OUT}$ & Global DLSM parameter for social activity\\
$\boldsymbol{r}$ & DLSM parameter for radius of communication of all nodes\\
$r_i$ & Radius of communication of node $i$\\
$\boldsymbol{\mathcal{X}}_t$ & Latent positions at time $t$ in a DLSM\\
$\sigma^2$ & DLSM parameter controlling spread of latent positions, $\boldsymbol{\mathcal{X}}_{1:T}$\\
$\phi$ & correlation via a time series\\
$\boldsymbol{\pi}$ & DDCSBM parameter for community transition rate\\
$\boldsymbol{\Theta}$ & DDCSBM parameter for propensity of communication for all nodes\\
$\theta_{it}$ & Propensity of communication for node $i$ at time $t$\\
$\boldsymbol{\omega}$ & Community structure matrix in a DDCSBM\\
\bottomrule
\end{tabular}
\caption{List of Notation Related to Data Generated from Dynamic Models.}
\label{tab:Notation:Models}
\end{table}


\section{Performance Evaluation of Monitoring Methods} \label{sec:simstudy}

Based on data generated from models explained in Section \ref{sec:models}, we conducted a comprehensive simulation study to evaluate the performance of monitoring network density, maximum degree, linear combinations of maximum degree and network density, as well as the scan statistic. A simulation study may allow for better understanding and evaluation of network monitoring approaches and types of changes detectable in dynamic network data. Performance evaluation is accomplished as follows. First, temporally-evolving network data are generated according to simulation study settings discussed in \ref{subsec:simstudy:settings}. Second, general network monitoring approaches are applied to such network data as was discussed in Section \ref{subsec:methods:monitoring}. Last, metrics for evaluating network monitoring output is elaborated on in Section \ref{subsec:simstudy:metrics}. To help facilitate understanding of performance evaluation output, effects of varying correlation and sparsity levels are briefly discussed in Section \ref{subsec:simstudy:CL}. 

In our simulation study, we considered two types of anomalies: anomalies in edge probabilities and expected degree.  All Monte Carlo simulations were performed using 100 nodes ($n=100$) and 110 time-points ($T=110$). For each type of anomaly, simulations were performed 200 times. Out of the 110 time-points, we set $T_1 = 50$, the first 50 time-points, as Phase I data where no anomalies are embedded. The latter 60 time-points ($t > 51$) are used as Phase II data with embedded anomalies in edge probabilities or expected degree for some duration. Performance evaluation from each type of anomaly is discussed respectively in Sections \ref{subsec:simstudy:anom:density} and \ref{subsec:simstudy:anom:maxdeg}.

\subsection{Simulation Study Settings} \label{subsec:simstudy:settings}

To evaluate performance of the methods based on summary statistics, temporally-evolving network data were simulated. Such simulated data take into account various amounts of correlation, sparsity, duration of anomaly, model, model parameters, and network types. Correlation is controlled by varying values of the VAR or white noise coefficient, $\phi$, and sparsity is controlled by varying values of average density, $\avgden$. In particular, the values considered are $$\phi \in \{ 0.1, 0.3, 0.5, 0.75, 0.9, 0.95, 0.99 \} \text{ and } \avgden \in \{ 0.03, 0.06, 0.09, 0.12, 0.15, 0.18, 0.21 \}.$$ 

\noindent Average density is fixed at 11\%, $\avgden = 0.11$, when varying $\phi$ in order to mimic realistic network densities of application data. Likewise, $\phi$ is fixed at 0.5 when varying $\avgden$ so that correlation across time is not too high nor too low. Duration of anomaly or change point length (CPL) represents consecutive time periods in which some kind of anomaly occurs, typically ranging from 5 time-points to 25 time-points throughout the study. Both binary and count network types are considered in DLSM and DDCSBM settings. 

In all our scenarios, negative correlation, e.g., $\phi = -0.5$, is not included. When exploring negative correlation in the context of dynamic networks, we found similar results with summary statistics on data generated using positive correlation. While performance is similar, the interpretation of such networks differs. In general, rather than nodes moving closer together or increasing the chance of a connection (form an edge), under negative correlation nodes may move farther apart or experience a decrease in the chance of a connection. For many brain studies and resulting dynamic networks applied on brain activity, correlation plays a big role in relating a behavior and the type of connections formed in the network \citep{hidalgo2009dynamic, zabelina2016dynamic}. In both application data examples of \cite{hidalgo2009dynamic} and \cite{zabelina2016dynamic} on brain activity, negatively correlated activities in the brain tend to decrease the prevalence of a behavior or task.

For all simulations, we used the model settings described at the end of Sections \ref{sec:models:DLSM} and \ref{sec:models:DDCSBM}. However, in order to achieve desired average densities, we must scale parameter(s) in network models using a scalar, $a_{\ell}$. A subscript $\ell$ is used to distinguish between a binary ($\ell=1$) and count ($\ell=2$) network. We chose to scale parameters $\sigma^2$ in a DLSM and $\boldsymbol{\omega}$ in a DDCSBM. Specific settings for $a_{\ell} \sigma^2$ in a DLSM and $a_{\ell} \boldsymbol{\omega}$ in DDCSBM are provided for ease of replicating results in our simulation study. When varying $\phi$ and controlling average network density at around 11\%, in a DLSM, $a_1 = 0.00014$ and $a_2 = 0.00042$ for $a_{\ell} \sigma^2,$ and in a DDCSBM, $a_1=0.16$ and $a_2=0.17$ for $a_{\ell} \cdot \boldsymbol{\omega}$. When varying $\avgden$ and fixing $\phi = 0.5$, specific settings of $a_{\ell}$ to appropriately scale $\sigma^2$ and $\boldsymbol{\omega}$ are given in Table \ref{tab:ADen:Settings}.

\begin{table}[H] \centering
\begin{tabular}{cc|ccccccc} \toprule
& &  \multicolumn{7}{c}{Average Network Density $\avgden$}\\
\cmidrule(lr){3-9}
\textbf{Model} & $\boldsymbol{a_{\ell}}$ &  0.21 & 0.18 & 0.15 & 0.12  & 0.09 & 0.06 & 0.03\\ \midrule
\multirow{2}{*}{DLSM} & $a_1$ & 0.0002 & 0.0002387 & 0.000292 & 0.000373 & 0.000493 & 0.000747 & 0.00153\\
 & $a_2$ & $3.5 \cdot a_1$ & $3.4 \cdot a_1$ & $3.4 \cdot a_1$ & $3.3 \cdot a_1$ & $3.3 \cdot a_1$ & $3.3 \cdot a_1$ & $3.3 \cdot a_1$\\ \midrule
\multirow{2}{*}{DDCSBM} & $a_1$ & 0.35 & 0.29 & 0.24 & 0.18 & 0.14 & 0.09 &  0.045\\
& $a_2$ & 0.32 & 0.265 &  0.22 & 0.17& 0.13& 0.085& 0.045\\
\bottomrule
\end{tabular}
\caption{Scalar Settings of $a_{\ell} \sigma^2$ and $a_{\ell} \boldsymbol{\omega}$ for Binary ($\ell = 1$) and Count ($\ell = 2$) Networks when Varying $\avgden$.}
\label{tab:ADen:Settings}
\end{table}

\subsection{Performance Evaluation Metrics} \label{subsec:simstudy:metrics}

Using the settings described in Section \ref{subsec:simstudy:settings}, network data were generated and network monitoring approaches were applied as discussed in Section \ref{subsec:methods:monitoring}. For each Monte Carlo simulation, the output of network monitoring approaches is a binary stream output, $\boldsymbol{A}$, where $A_t = 1$ if the method signals, and $A_t=0$ otherwise. Performance evaluation of output from network monitoring approaches is accomplished using two measures. One such measure is detection rate (DR), which is a binary measure of whether or not an anomaly was detected at all. If an anomaly is detected that outcome is assigned a 1 and 0 otherwise. DR provides a measure of the ability of a network monitoring approach to find anomalies. To further quantify this ability, a second measure utilized is AUC calculations from ROCs. In this context, we take advantage of confusion matrices. True labels of a confusion matrix are the time periods ($t$) within the duration of an anomaly, and predicted labels are alarms found from signals of network monitoring approaches. In a resulting ROC curve, both the true positive rate (TPR) and false positive rate (FPR) must range from [0,1]. Thus, we vary $q$ from -6 to 6, i.e., $q \in [-6,6]$, in control limits $CL = \hat{\mu} + q \hat{\sigma}$ and threshold $q$ for the scan statistic to achieve desired FPR and TPR in $[0,1]$. In essence, AUC reflects the number of times an anomaly is detectable. Both DR and AUC are calculated in each Monte Carlo simulation and later averaged over all Monte Carlo simulations.

\subsection{Effects of Correlation and Sparsity on Summary Statistics}
\label{subsec:simstudy:CL}

We explore the effect of correlation $(\phi)$ and sparsity $(\avgden)$ on means and standard deviations of summary statistics in Phase I data. Recall when monitoring summary statistics, control limits of a Shewhart individuals control chart are $\overline{x} \pm q s$. Choosing an appropriate $q$ was discussed in Section \ref{subsec:methods:Getq}, but learning the effects of correlation and sparsity on $\overline{x}$ and $s$ can aid understanding of the evaluation of network monitoring. Phase I data correspond to time-points less than 50, $t \leq 50$. Figures \ref{fig:Means:DLSM:B:Phi}-\ref{fig:SD:DDCSBM:B:Dens} are provided in Appendix \ref{subsec:AppA:MeansSD} displaying means and standard deviations of $W_t, ~D_t, ~M^{-}_t,$ and $M^{+}_t$ across varying amounts of correlation and sparsity. Examples across varying correlation, i.e., $\phi$, are shown for a binary DLSM and count DDCSBM settings and across varying sparsity, i.e., $\avgden$, are shown for count DLSM and binary DDCSBM. In DLSM settings, as correlation increases, the range of the means widens. In DDCSBM settings, this mainly affects means of maximum degree and subsequently the sum and difference statistics. The standard deviation is less affected by correlation overall. In terms of varying sparsity, actual values of the means vary while both range and values of the standard deviation change in DLSM and DDCSBM settings. In summary, the means more so than standard deviations of $W_t, ~D_t, ~M^{-}_t,$ and $M^{+}_t$ are affected by correlation and sparsity.

\subsection{Performance Evaluation with Anomalies in Edge Probabilities} \label{subsec:simstudy:anom:density}


We design a set of simulations to ultimately affect network density by manipulating $p_{ijt}$ (edge probabilities) in simulated network data. Such manipulation is accomplished via an odds ratio, denoted $OR$. For a DLSM, $y_{ijt} \sim \text{Bern}(p_{ijt})$. Hence, for a given $OR$, we compare $p_0= p_{ijt}$ and $p_1 = C_{ijt} \cdot p_{ijt}$ such that $OR = \left( \dfrac{1-p_0}{p_0} \right) / \left( \dfrac{1-p_1}{p_1} \right)$ and $C_{ijt} = \dfrac{OR}{(1-p_0 + OR \cdot p_0)}.$ Thus, as $OR$ increases, so does $C_{ijt}$. For a DDCSBM, $y_{ijt} \sim \text{Pois}(p_{ijt})$. Since $p_{ijt}$ is a rate, we directly compare $p_{ijt}$ and $OR \cdot p_{ijt}$. Such an anomaly implies a group of nodes all of a sudden increase (or decrease) communication. Specific scenarios are described in Table \ref{tab:1_Sim_NetDens}. Shift size refers to a medium or large change in resulting density. Let $N$ denote the number of nodes affected by a given change.

\begin{table}[H] \centering
\begin{tabular}{cccc} \toprule
Shift & Network & Number of  & $OR$ Change\\ 
Size & Type &  Anomalous Nodes ($N$) & from 1 to \\ \hline
Medium & DLSM & 33 & 4 \\ 
Medium & DDCSBM & 33 & 2.5 \\ 
Large & DLSM & 79 & 2.5 \\ 
Large & DDCSBM & 72 & 1.5 \\ \bottomrule
\end{tabular} 
\caption{Scenarios for Simulation Study Targeting Network Density via Odds Ratio.}
\label{tab:1_Sim_NetDens}
\end{table}

\begin{figure}[ht] \centering
\includegraphics[width=5in,angle=270]{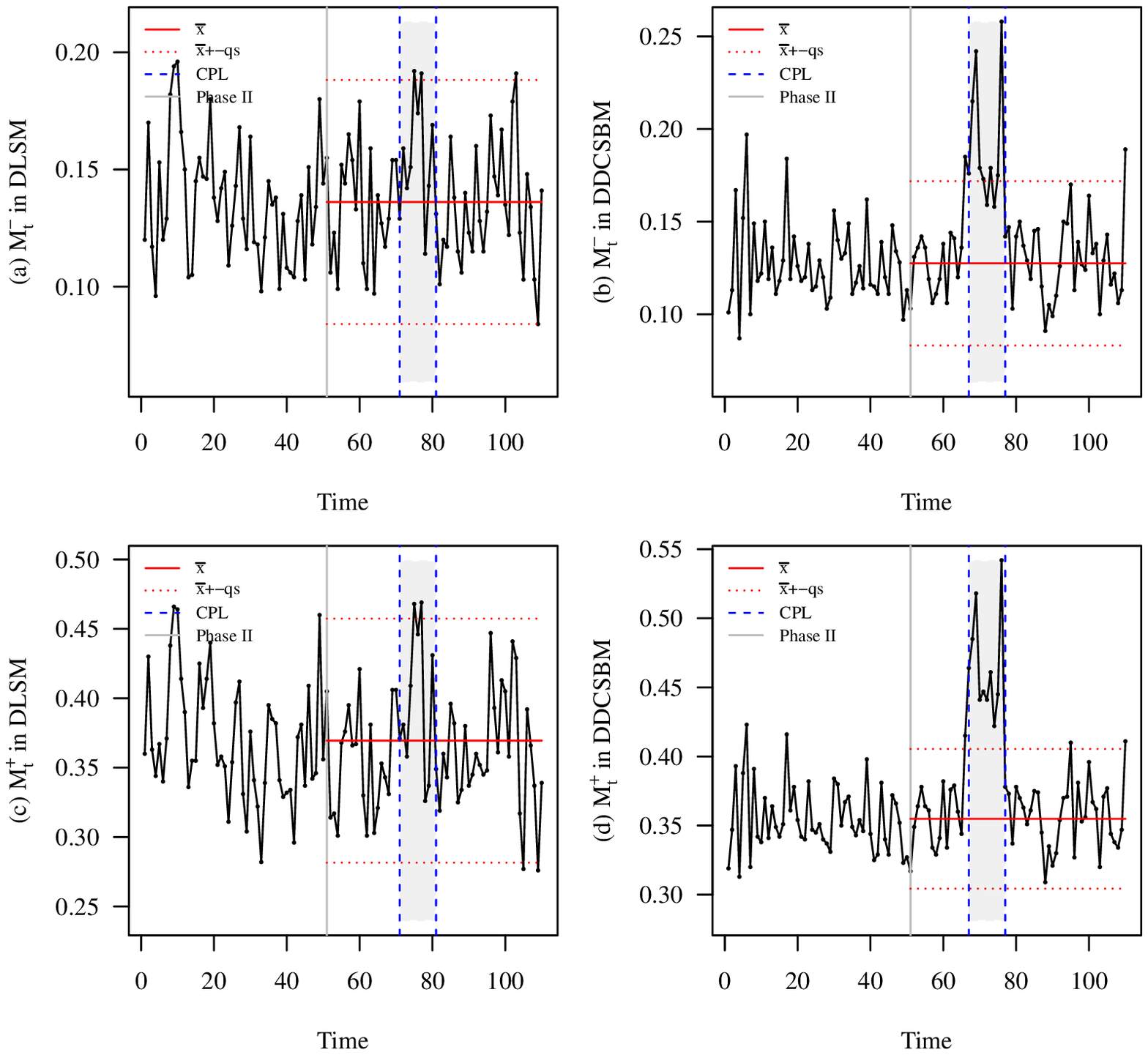}
\caption{Shewhart Individuals Control Charts of $M^{-}_t$ and $M^{+}_t$ for an Anomaly in Edge Probabilities in Binary Settings with $\phi=0.5$ and $\avgden=0.11$. Plots (a) and (c) show $M^{-}_t$ and $M^{+}_t$ in a DLSM setting, and plots (b) and (d) show $M^{-}_t$ and $M^{+}_t$ in a DDCSBM setting. Limits $q$ are determined setting empirical conditional false alarm probabilities at 0.03. }
\label{fig:Anom:NetDens}
\end{figure}

An example using $N=33$ anomalous nodes and OR change from 1 to 4 is shown in a binary DLSM and DDCSBM in Figure \ref{fig:Anom:NetDens} with $\phi = 0.5$, $\avgden=0.11$, and settings mentioned in Section \ref{subsec:simstudy:settings}. Shewhart control charts for this example are plotted for $M^{-}_t$ and $M^{+}_t$ with Phase I data for $t \leq 50$ in Figure \ref{fig:Anom:NetDens}. The solid red line indicates the mean from Phase I, the dashed red lines indicate $q$ standard deviations above or below the mean, the gray line separates Phase I and Phase II, and the blue dashed lines indicate the beginning and ending of the CPL. The number of standard deviations above the mean was determined by controlling the false alarm rates in the non-anomalous empirical data for $p = 0.03$ and $\phi = 0.5$ as was discussed in Section \ref{subsec:methods:Getq}. 

It is difficult to find meaningful changes to lower the odds ratio in the DLSM setting since manipulating $p_{ijt}$ (edge probabilities) closer to 0 appears to be affected by the randomness of latent positions. That is, the closer two latent positions are, the higher the edge probability becomes. It could be the case that two latent positions are so close together, that even scaling such edge probability yields a $p_{ijt} > 0.2$. Thus, only scenarios increasing the odds ratio are considered. 


Results of 200 simulations with $n=100$, $T=110$, and $CPL=5,10,$ or 15 are summarized below in Tables \ref{tab:ND:1:DLSM:TPR} -\ref{tab:ND:4:DDCSBM:TPR} using DR and in Appendix \ref{subsec:AppA:AUC:ND} Tables \ref{tab:ND:1:DLSM:AUC} -\ref{tab:ND:4:DDCSBM:AUC} using AUC. The values corresponding to the method which detects the planted anomaly best are in bold. Since anomalies planted affect edge probabilities (and ultimately density), it would be natural for the monitoring using $W_t$ to detect this change the best. Results are explained first by model and network type settings and then across trends in CPL, correlation, and sparsity. In binary DLSM settings, the best monitoring performance in terms of detection rates in the medium shift size case vary between $D_t, ~M^{-}_t,$ and $M^{+}_t$, while $W_t$ performs best in the large shift size case. 

In nearly all count DLSM settings, monitoring $W_t,~D_t, ~M^{-}_t,$ and $M^{+}_t$ all detect perfectly with a DR of 1. The worst performer tends to be $S^{*}_t$ in DLSM settings, but the DR of the scan statistic method is high in count DLSM settings with a large shift size. In DDCSBM settings, the best performancer is almost always $W_t$, which we would expect since the anomaly affected edge probabilities. In count DDCSBM settings, $M^{+}_t$ method does best, and in the large shift size setting, $S^{*}_t$ method also has a perfect detection rate. The shorter the duration, higher the correlation, and higher in sparsity, the worse most methods do in detecting an anomaly. These trends are observed more so in binary DLSM settings. From count to binary DDCSBM settings, recall that count data are transformed into binary using $y^{*}_{ijt} = 1$ if $y_{ijt} \geq 1$. In some cases, performance is better in count settings, but not in others. 

We make note the design of $S^*$ method is to detect if a change occurred rather than to identify the time-points said change occurred in \citep{priebe2005scan, zhao2018performance}. That is, when monitoring the scan statistic, the method is expected to signal at best only a few times if any anomaly occurred. In AUC results reported in  Tables \ref{tab:ND:1:DLSM:AUC} -\ref{tab:ND:4:DDCSBM:AUC} in Appendix \ref{subsec:AppA:AUC:ND}, the detection ability is mainly best for $M^{-}_t$ and $M^{+}_t$ methods in DLSM settings while the $W_t$ method does detect best in DDCSBM settings. The worst monitoring performance, as might be expected, is $S^{*}_t$ across all settings with a decrease in AUC of about 10\% to 40\% compared to other summary statistics. Across duration of anomaly (CPL) and correlation ($\phi$), there is little difference in AUC values. However, sparsity ($\avgden$) has some effect since the sparser the network, the more difficult it could be to detect an anomaly.

\begin{table}[H]
\caption{DR for DLSM with 33 Anomalous Nodes and $OR$ from 1 to 4.} \centering
\begin{tabular}{ccc rrrrr rrrrr} \toprule
\multicolumn{3}{c}{Settings} &  \multicolumn{5}{c}{Binary} & \multicolumn{5}{c}{Count}\\ 
 \cmidrule(lr){1-3} \cmidrule(lr){4-8} \cmidrule(lr){9-13}
CPL & $\phi$ & $\avgden$ &
$W_t$ & $D_{t}$ & $M_t^{-}$ & $M_t^{+}$ & $S_t^{*}$ & $W_t$ & $D_{t}$ & $M_t^{-}$ & $M_t^{+}$ & $S_t^{*}$\\ 
\cmidrule(lr){1-3} \cmidrule(lr){4-8} \cmidrule(lr){9-13}
5 & 0.5 & 0.11 & 0.320 & 0.365 & \textbf{0.400} & 0.365 & 0.210 & 0.980 & \textbf{1}& \textbf{1}& \textbf{1}& 0.545\\
10 & 0.5 & 0.11 & 0.560 & \textbf{0.605} & 0.595 & \textbf{0.605} & 0.405 & 0.995 & \textbf{1}& \textbf{1}& \textbf{1}& 0.560\\
15 & 0.5 & 0.11 & 0.700 & \textbf{0.775} & 0.730 & 0.745 & 0.460 & \textbf{1}& \textbf{1}& \textbf{1}& \textbf{1}& 0.545\\
\cmidrule(lr){1-3} \cmidrule(lr){4-8} \cmidrule(lr){9-13}
10 & 0.1 & 0.11 & 0.540 & \textbf{0.650} & 0.635 & 0.590 & 0.365 & \textbf{1}& \textbf{1}& \textbf{1}& \textbf{1}& 0.630\\
10 & 0.9 & 0.11 & 0.340 & 0.480 & \textbf{0.515} & 0.450 & 0.275 & 0.920 & \textbf{1}& \textbf{1}& \textbf{1}& 0.290\\
\cmidrule(lr){1-3} \cmidrule(lr){4-8} \cmidrule(lr){9-13}
10 & 0.5 & 0.03 & \textbf{0.530} & 0.470 & 0.430 & 0.520 & 0.350 & 0.900 & \textbf{0.970} & \textbf{0.970} & \textbf{0.970} & 0.630\\
10 & 0.5 & 0.21 & 0.530 & 0.725 & \textbf{0.780} & 0.680 & 0.340 & 0.990 & \textbf{1}& \textbf{1}& \textbf{1}& 0.850\\
\bottomrule
\end{tabular}
\label{tab:ND:1:DLSM:TPR}
\end{table}

\begin{table}[H]
\caption{DR for DLSM with 79 Anomalous Nodes and $OR$ from 1 to 2.5.} \centering
\begin{tabular}{ccc rrrrr rrrrr} \toprule
\multicolumn{3}{c}{Settings} &  \multicolumn{5}{c}{Binary} & \multicolumn{5}{c}{Count}\\ 
 \cmidrule(lr){1-3} \cmidrule(lr){4-8} \cmidrule(lr){9-13}
CPL & $\phi$ & $\avgden$ &
$W_t$ & $D_{t}$ & $M_t^{-}$ & $M_t^{+}$ & $S_t^{*}$ & $W_t$ & $D_{t}$ & $M_t^{-}$ & $M_t^{+}$ & $S_t^{*}$\\ 
\cmidrule(lr){1-3} \cmidrule(lr){4-8} \cmidrule(lr){9-13}
5 & 0.5 & 0.11 & \textbf{0.955} & 0.940 & 0.685 & 0.940 & 0.645 & \textbf{1}& \textbf{1}& \textbf{1}& \textbf{1}& 0.985\\
10 & 0.5 & 0.11 & \textbf{0.990} & 0.985 & 0.870 & \textbf{0.990} & 0.660 & \textbf{1}& \textbf{1}& \textbf{1}& \textbf{1}& 0.985\\
15 & 0.5 & 0.11 & \textbf{1}& 0.995 & 0.925 & 0.995 & 0.725 & \textbf{1}& \textbf{1}& \textbf{1}& \textbf{1}& 0.975\\
\cmidrule(lr){1-3} \cmidrule(lr){4-8} \cmidrule(lr){9-13}
10 & 0.1 & 0.11 & \textbf{0.995} & 0.990 & 0.840 & 0.990 & 0.715 & \textbf{1}& \textbf{1}& \textbf{1}& \textbf{1}& 0.990\\
10 & 0.9 & 0.11 & \textbf{0.880} & 0.860 & 0.740 & 0.870 & 0.360 & \textbf{1}& \textbf{1}& \textbf{1}& \textbf{1}& 0.930\\
\cmidrule(lr){1-3} \cmidrule(lr){4-8} \cmidrule(lr){9-13}
10 & 0.5 & 0.03 & \textbf{1}& 0.810 & 0.635 & 0.890 & 0.500 & \textbf{1}& \textbf{1}& 0.99 & \textbf{1}& 0.775\\
10 & 0.5 & 0.21 & \textbf{1}& \textbf{1}& 0.935 & \textbf{1}& 0.755 & \textbf{1}& \textbf{1}& \textbf{1}& \textbf{1}& 0.975\\
\bottomrule
\end{tabular}
\label{tab:ND:2:DDCSBM:TPR}
\end{table}

\begin{table}[H]
\caption{DR for DDCSBM with 33 Anomalous Nodes and $OR$ from 1 to 2.5.} \centering
\begin{tabular}{ccc rrrrr rrrrr} \toprule
\multicolumn{3}{c}{Settings} &  \multicolumn{5}{c}{Binary} & \multicolumn{5}{c}{Count}\\ 
 \cmidrule(lr){1-3} \cmidrule(lr){4-8} \cmidrule(lr){9-13}
CPL & $\phi$ & $\avgden$ &
$W_t$ & $D_{t}$ & $M_t^{-}$ & $M_t^{+}$ & $S_t^{*}$ & $W_t$ & $D_{t}$ & $M_t^{-}$ & $M_t^{+}$ & $S_t^{*}$\\ 
\cmidrule(lr){1-3} \cmidrule(lr){4-8} \cmidrule(lr){9-13}
5 & 0.5 & 0.11 & \textbf{1}& 0.825 & 0.605 & 0.925 & 0.695 & \textbf{1}& 0.915 & 0.755 & 0.970 & 0.760\\
10 & 0.5 & 0.11 & \textbf{1}& 0.945 & 0.820 & 0.995 & 0.730 & \textbf{1}& 0.985 & 0.930 & \textbf{1}& 0.795\\
15 & 0.5 & 0.11 & \textbf{1}& 0.980 & 0.905 & 0.995 & 0.790 & \textbf{1}& \textbf{1}& 0.985 & \textbf{1}& 0.785\\
\cmidrule(lr){1-3} \cmidrule(lr){4-8} \cmidrule(lr){9-13}
10 & 0.1 & 0.11 & \textbf{1}& 0.950 & 0.835 & 0.990 & 0.780 & \textbf{1}& 0.990 & 0.970 & \textbf{1}& 0.800\\
10 & 0.9 & 0.11 & \textbf{1}& 0.925 & 0.825 & 0.990 & 0.610 & \textbf{1}& 0.990 & 0.955 & 0.995 & 0.650\\
\cmidrule(lr){1-3} \cmidrule(lr){4-8} \cmidrule(lr){9-13}
10 & 0.5 & 0.03 & \textbf{0.99} & 0.780 & 0.665 & 0.855 & 0.630 & \textbf{0.985} & 0.820 & 0.745 & 0.880 & 0.540\\
10 & 0.5 & 0.21 & \textbf{1}& 0.985 & 0.830 & 0.995 & 0.960 & \textbf{1}& 0.995 & 0.980 & \textbf{1}& 0.995\\
\bottomrule
\end{tabular}
\label{tab:ND:3:DLSM:TPR}
\end{table}

\begin{table}[H]
\caption{DR for DDCSBM with 72 Anomalous Nodes and $OR$ from 1 to 1.5.} \centering
\begin{tabular}{ccc rrrrr rrrrr} \toprule
\multicolumn{3}{c}{Settings} &  \multicolumn{5}{c}{Binary} & \multicolumn{5}{c}{Count}\\ 
 \cmidrule(lr){1-3} \cmidrule(lr){4-8} \cmidrule(lr){9-13}
CPL & $\phi$ & $\avgden$ &
$W_t$ & $D_{t}$ & $M_t^{-}$ & $M_t^{+}$ & $S_t^{*}$ & $W_t$ & $D_{t}$ & $M_t^{-}$ & $M_t^{+}$ & $S_t^{*}$\\ 
\cmidrule(lr){1-3} \cmidrule(lr){4-8} \cmidrule(lr){9-13}
5 & 0.5 & 0.11 & \textbf{1}& 0.885 & 0.560 & 0.995 & 0.870 & \textbf{1}& 0.930 & 0.670 & 0.990 & 0.860\\
10 & 0.5 & 0.11 & \textbf{1}& 0.990 & 0.820 & \textbf{1}& 0.925 & \textbf{1}& 0.985 & 0.830 & \textbf{1}& 0.915\\
15 & 0.5 & 0.11 & \textbf{1}& \textbf{1}& 0.915 & \textbf{1}& 0.890 & \textbf{1}& 0.985 & 0.915 & \textbf{1}& 0.915\\
\cmidrule(lr){1-3} \cmidrule(lr){4-8} \cmidrule(lr){9-13}
10 & 0.1 & 0.11 & \textbf{1}& 0.990 & 0.835 & \textbf{1}& 0.865 & \textbf{1}& 0.990 & 0.910 & 0.995 & 0.920\\
10 & 0.9 & 0.11 & \textbf{1}& 0.985 & 0.800 & \textbf{1}& 0.870 & \textbf{1}& 0.985 & 0.890 & 0.995 & 0.820\\
\cmidrule(lr){1-3} \cmidrule(lr){4-8} \cmidrule(lr){9-13}
10 & 0.5 & 0.03 & \textbf{1}& 0.855 & 0.685 & 0.940 & 0.720 & \textbf{1}& 0.835 & 0.665 & 0.970 & 0.715\\
10 & 0.5 & 0.21 & \textbf{1}& 0.995 & 0.805 & \textbf{1}& \textbf{1}& \textbf{1}& \textbf{1}& 0.910 & \textbf{1}& \textbf{ 1 }\\
\bottomrule
\end{tabular}
\label{tab:ND:4:DDCSBM:TPR}
\end{table}

The scenarios considered thus far involved sustained anomalies in network density. Now, we compare such an anomaly with one that gradually increases network density over the duration of the anomaly or CPL. In the next set of scenarios, the odds ratio is gradually increased from 1 to 12 in a DLSM network and from 1 to 3.5 in a DDCSBM network for a sub-network of 39 nodes. When only varying CPL, CPL varies from 15, 20, to 25. For all other settings, CPL is 20.

DR results are reported in Tables \ref{tab:ND:7:DLSM:TPR} and \ref{tab:ND:8:DDCSBM:TPR} below with AUC results in Tables \ref{tab:ND:5:DLSM:AUC} and \ref{tab:ND:6:DDCSBM:AUC} in Appendix \ref{subsec:AppA:AUC:ND}. The values of the best monitoring performances are in bold, and in cases with a gradual change in the odds ratio, we see monitoring $W_t,~D_t, ~M^{-}_t,$ and $M^{+}_t$ all detect this embedded anomaly well. The scan statistic method detects less well in binary cases and rather well in count settings. AUC Results from Tables \ref{tab:ND:5:DLSM:AUC} and \ref{tab:ND:6:DDCSBM:AUC} in Appendix \ref{subsec:AppA:AUC:ND} are fairly similar to the sustained changes in odds ratio. Therefore, between sustained and gradual changes, monitoring $M^{+}_t$ does best in most DLSM settings while monitoring $W_t$ detects best in DDCSBM settings with respect to DR and AUC.

\begin{table}[H]
\caption{DR for DLSM with 39 Anomalous Nodes with $OR \in [1,12]$.} \centering
\begin{tabular}{ccc rrrrr rrrrr} \toprule
\multicolumn{3}{c}{Settings} &  \multicolumn{5}{c}{Binary} & \multicolumn{5}{c}{Count}\\ 
 \cmidrule(lr){1-3} \cmidrule(lr){4-8} \cmidrule(lr){9-13}
CPL & $\phi$ & $\avgden$ &
$W_t$ & $D_{t}$ & $M_t^{-}$ & $M_t^{+}$ & $S_t^{*}$ & $W_t$ & $D_{t}$ & $M_t^{-}$ & $M_t^{+}$ & $S_t^{*}$\\
\cmidrule(lr){1-3} \cmidrule(lr){4-8} \cmidrule(lr){9-13}
15 & 0.5 & 0.11 & 0.920 & \textbf{0.990} & 0.985 & \textbf{0.990} & 0.550 & \textbf{1}& \textbf{1}& \textbf{1}& \textbf{1}& 0.975\\
20 & 0.5 & 0.11 & 0.965 & \textbf{0.985} & \textbf{0.985} & \textbf{0.985} & 0.645 & \textbf{1}& \textbf{1}& \textbf{1}& \textbf{1}& 0.975\\
25 & 0.5 & 0.11 & 0.965 & \textbf{1}& \textbf{1}& \textbf{1}& 0.665 & \textbf{1}& \textbf{1}& \textbf{1}& \textbf{1}& 0.985\\
\cmidrule(lr){1-3} \cmidrule(lr){4-8} \cmidrule(lr){9-13}
20 & 0.1 & 0.11 & 0.980 & \textbf{0.995} & 0.990 & \textbf{0.995} & 0.675 & \textbf{1}& \textbf{1}& \textbf{1}& \textbf{1}& 0.965\\
20 & 0.9 & 0.11 & 0.735 & \textbf{0.955} & \textbf{0.955} & 0.910 & 0.520 & \textbf{1}& \textbf{1}& \textbf{1}& \textbf{1}& 0.735\\
\cmidrule(lr){1-3} \cmidrule(lr){4-8} \cmidrule(lr){9-13}
20 & 0.5 & 0.03 & 0.945 & 0.920 & 0.880 & \textbf{0.950} & 0.510 & \textbf{1}& \textbf{1}& \textbf{1}& \textbf{1}& 0.955\\
20 & 0.5 & 0.21 & 0.980 & \textbf{1}& \textbf{1}& \textbf{1}& 0.655 & \textbf{1}& \textbf{1}& \textbf{1}& \textbf{1}& 0.980\\
\bottomrule
\end{tabular} \label{tab:ND:7:DLSM:TPR}
\end{table}

\begin{table}[H]
\caption{DR for DDCSBM with 39 Anomalous Nodes with $OR \in [1,3.5]$.} \centering
\begin{tabular}{ccc rrrrr rrrrr} \toprule
\multicolumn{3}{c}{Settings} &  \multicolumn{5}{c}{Binary} & \multicolumn{5}{c}{Count}\\ 
\cmidrule(lr){1-3} \cmidrule(lr){4-8} \cmidrule(lr){9-13}
CPL & $\phi$ & $\avgden$ &
$W_t$ & $D_{t}$ & $M_t^{-}$ & $M_t^{+}$ & $S_t^{*}$ & $W_t$ & $D_{t}$ & $M_t^{-}$ & $M_t^{+}$ & $S_t^{*}$\\
\cmidrule(lr){1-3} \cmidrule(lr){4-8} \cmidrule(lr){9-13}
15 & 0.5 & 0.11 & \textbf{1}& \textbf{1}& \textbf{1}& \textbf{1}& 0.830 & \textbf{1}& \textbf{1}& \textbf{1}& \textbf{1}& 0.920\\
20 & 0.5 & 0.11 & \textbf{1}& \textbf{1}& 0.995 & \textbf{1}& 0.880 & \textbf{1}& \textbf{1}& \textbf{1}& \textbf{1}& 0.940\\
25 & 0.5 & 0.11 & \textbf{1}& \textbf{1}& \textbf{1}& \textbf{1}& 0.880 & \textbf{1}& \textbf{1}& \textbf{1}& \textbf{1}& 0.865\\
\cmidrule(lr){1-3} \cmidrule(lr){4-8} \cmidrule(lr){9-13}
20 & 0.1 & 0.11 & \textbf{1}& \textbf{1}& \textbf{1}& \textbf{1}& 0.910 & \textbf{1}& \textbf{1}& \textbf{1}& \textbf{1}& 0.915\\
20 & 0.9 & 0.11 & \textbf{1}& \textbf{1}& 0.995 & \textbf{1}& 0.710 & \textbf{1}& \textbf{1}& \textbf{1}& \textbf{1}& 0.815\\
\cmidrule(lr){1-3} \cmidrule(lr){4-8} \cmidrule(lr){9-13}
20 & 0.5 & 0.03 & \textbf{1}& 0.985 & 0.970 & \textbf{1}& 0.775 & \textbf{1}& 0.985 & 0.96 & \textbf{1}& 0.815\\
20 & 0.5 & 0.21 & \textbf{1}& \textbf{1}& 0.995 & \textbf{1}& 0.965 & \textbf{1}& \textbf{1}& \textbf{1}& \textbf{1}& 0.990\\
\bottomrule
\end{tabular} \label{tab:ND:8:DDCSBM:TPR}
\end{table}

\subsection{Performance Evaluation with Anomalies in Expected Degree} \label{subsec:simstudy:anom:maxdeg}

In attempts to manipulate maximum degree, we target effects of parameters in a DLSM and DDCSBM related to degree(s) of a node(s), which mainly affect expected degree. Such parameters are $r_{i}$ in a DLSM, which represents a radius of communication in the latent space for node $i$ that indirectly affects degree, and $\theta_{it}$ in a DDCSBM, which represents a propensity for node $i$ to communicate at time $t$ that directly affects degree. Recall, propensities of communication are a degree-correction for the stochastic block model. First note that having radii in a DLSM be time dependent neither drastically changes nor improves model fitting as discussed by \cite{SewellChen2015}. Second note that correlation via time in a DLSM is incorporated via latent positions and not $r_i$, whereas correlation via time for a DDCSBM is through degree parameters, $\Theta$. By manipulating degree parameters directly in a DDCSBM model, we cannot simply change the value of $\boldsymbol{\theta_{i}}$ at anomalous time points. Rather, $\theta_{it}$ is multiplied by some constant, $C$, throughout the anomalous time period. Table \ref{tab:2_Sim_MaxDeg} summarizes considered scenarios affecting a certain number of nodes and the settings for model parameters. For a DLSM, $r_i$ are directly manipulated for a binary (B) and count (C) network, while in a DDCSBM, $\theta_{it}$ are multiplied by some constant $C$. Shift size indicates a medium or large change in expected degrees.

\begin{table}[H] \centering
\begin{tabular}{|cccc|} \toprule
Shift & Number of & $r_{i} = 0.1$ & $C \cdot \Theta$\\ 
Size & Nodes ($N$) & to & $C=1$ to \\ \hline
Medium & 15 & 0.020(B);0.04(C) & 2.25 \\ 
Large & 35 & 0.015(B);0.0225(C) & \textbf{ 1 }.75 \\ \bottomrule
\end{tabular} 
\caption{Scenarios for Simulation Study Targeting Max Degree via Node Parameters.}
\label{tab:2_Sim_MaxDeg}
\end{table}

\begin{figure}[H] \centering
\includegraphics[width=2.2in,angle=270]{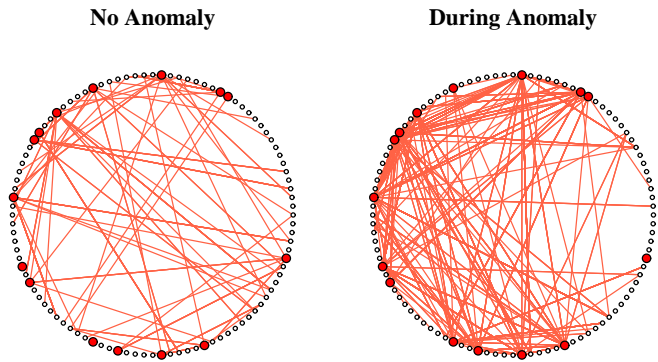}
\vspace{-0.3in}
\caption{Edges of 15 Nodes in Count DLSM Networks (a) Before and (b) During an Anomaly Targeting Max Degree. A subset of 15 nodes (large red vertices) have $r_i = 0.01$ increased to $r_i = 0.04$ for the anomaly. Only edges from those 15 nodes are displayed, and an increase in communication for those 15 nodes is observed.}
\label{fig:Anom:MaxDeg:DLSM}
\end{figure}

\begin{figure}[ht] \centering
\includegraphics[width=5in,angle=270]{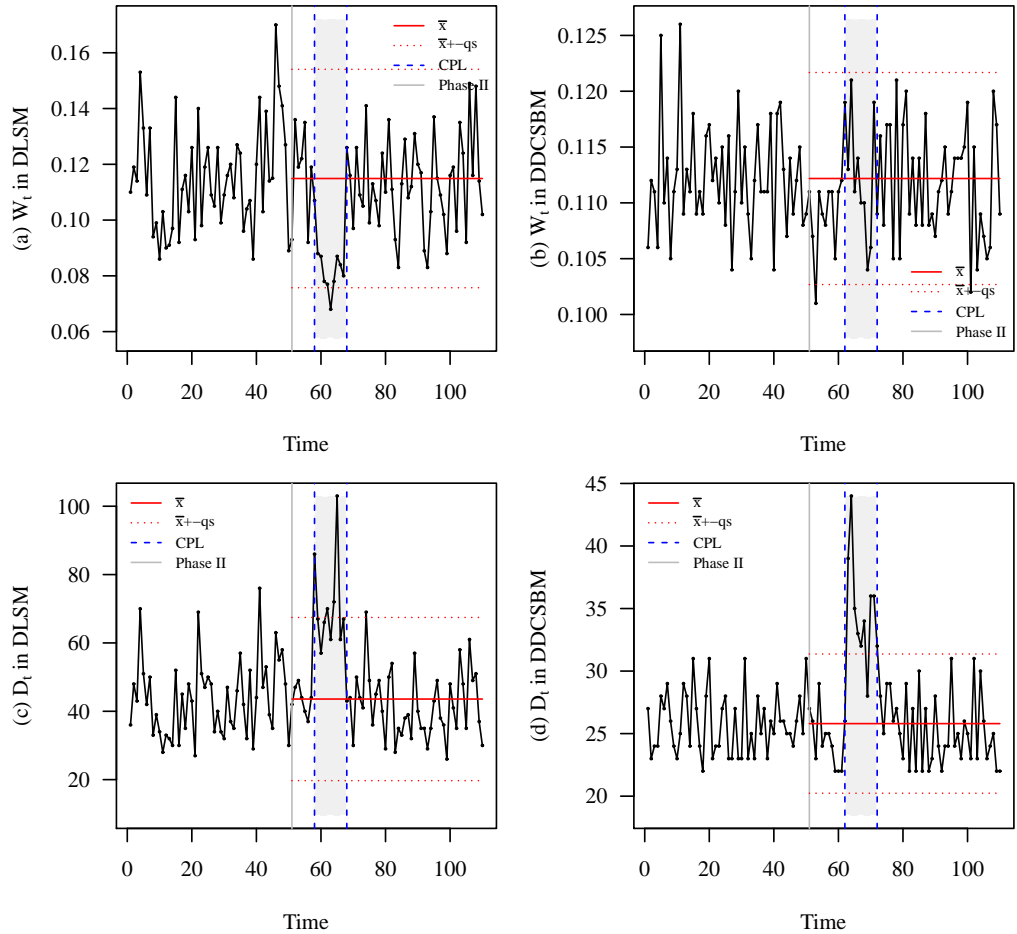}
\caption{Shewhart Individuals Control Charts of $W_t$ and $D_t$ for an Anomaly Targeting Expected Degree in Count Settings with $\phi=0.5$ and $\avgden=0.11$. Plots (a) and (c) show $W_t$ and $D_t$ in a DLSM setting, and plots (b) and (d) show $W_t$ and $D_t$ in a DDCSBM setting. Limits $q$ are determined setting empirical conditional false alarm probabilities at 0.03.}
\label{fig:Anom:MaxDeg}
\end{figure}

Recall, $N$ denotes the number of nodes affected by a given change. An example using $N=15$ with $r_i=0.04$ is shown in a count DLSM in Figure \ref{fig:Anom:MaxDeg:DLSM} with $\phi = 0.5$, $\avgden=0.11$, and settings mentioned in Section \ref{subsec:simstudy:settings}. In Figure \ref{fig:Anom:MaxDeg:DLSM}, the $N=15$ affected nodes are enlarged and colored red with only the associated edges to and from those nodes displayed in red as well. An increase in communication among the $N=15$ nodes can be viewed during an anomalous time within the change point interval. Note, while edge probabilities of these nodes are affected by manipulating expected degree during each time-point of the change point interval, not all time-points during the anomaly will have such an increase in communication as this example. Shewhart control charts for this example are plotted for $W_t$ and $D_t$ with Phase I data for $t \leq 50$ in Figure \ref{fig:Anom:MaxDeg}. In this example, the network density chart signals a decrease in network density in the DLSM setting, Figure \ref{fig:Anom:MaxDeg} (a). Figures \ref{fig:Anom:MaxDeg} (c) and (d) show $D_t$ signaling during the change point interval along with some false alarms. 

Results of 200 simulations with $n=100$, $T=110$, and $CPL=5,10,$ or 15 are summarized below in Tables \ref{tab:MD:1:DLSM:TPR} - \ref{tab:MD:4:DDCSBM:TPR} using DR and in Appendix \ref{subsec:AppA:AUC:MD} Tables \ref{tab:ND:1:DLSM:AUC} -\ref{tab:ND:4:DDCSBM:AUC} using AUC. The best value of the metric is in bold. Since anomalies planted affect expected degree (and desirably maximum degree), a natural best method would be $D_t$. As with anomalies in edge probabilities, results are explained first by model and network type settings and then across trends in CPL, correlation, and sparsity. In nearly all of DLSM and DDCSBM settings, use of $M^{-}_t$ yields the highest detection rates. In the medium shift size count DLSM settings, monitoring $D_t$ and $M^{+}_t$ detect just about as well. In medium shift size DDCSBM settings, $D_t, ~M^{-}_t,$ and $M^{+}_t$ methods all detect perfectly with a DR of 1. Detection rates across the board in the large shift size case (many nodes with a slight increase in expected degree) tend to be lower than the medium shift size case (some nodes with an increase in expected degree). This finding suggests detecting a change with some nodes affecting expected degree is easier than when many nodes affect expected degree. The worst monitoring performance in all scenarios is $W_t$, with the highest DR of 35\%. The second worst performing method is $S^{*}_t$, but it performs especially well in a count DLSM with a longer CPL and a count DDCSBM with a medium shift size. There is noticeably less effect from duration, correlation, and sparsity for most statistics. However, the scan statistic appears to be detect better with a longer CPL in DLSM settings. In general, most methods other than $W_t$ have high detection rates when planting an anomaly in expected degree. Unlike in the case where anomalies are embedded in edge probabilities, detection rates in DDCSBM count settings are better. 

In AUC results reported in  Tables \ref{tab:MD:1:DLSM:AUC} -\ref{tab:MD:4:DDCSBM:AUC} in Appendix \ref{subsec:AppA:AUC:MD}, the detection ability is mainly best for $M^{-}_t$ and $D_t$ methods in both DLSM and DDCSBM settings. The worst monitoring performance is $W_t$ with second worst being $S^{*}_t$ across all settings. In general, $W_t$ and $S^{*}_t$ methods suffer about a 30\% to 60\% loss in AUC compared to other summary statistics. Across duration of anomaly (CPL) and correlation ($\phi$), there is little difference in AUC values. It appears to be difficult to detect an anomaly when there is less communication (greater sparsity) in the network. For AUC values, DLSM count settings tend to have lower AUC than in DLSM binary settings. 

\begin{table}[H]
\caption{DR for DLSM from $r_i$ = 0.1 to $r_i= 0.020$(B); 0.04 (C) for $N=15$. } \centering
\begin{tabular}{ccc rrrrr rrrrr} \toprule
\multicolumn{3}{c}{Settings} &  \multicolumn{5}{c}{Binary} & \multicolumn{5}{c}{Count}\\ 
 \cmidrule(lr){1-3} \cmidrule(lr){4-8} \cmidrule(lr){9-13}
CPL & $\phi$ & $\avgden$ &
$W_t$ & $D_{t}$ & $M_t^{-}$ & $M_t^{+}$ & $S_t^{*}$ & $W_t$ & $D_{t}$ & $M_t^{-}$ & $M_t^{+}$ & $S_t^{*}$\\
\cmidrule(lr){1-3} \cmidrule(lr){4-8} \cmidrule(lr){9-13}
5 & 0.5 & 0.11 & 0.045 & \textbf{1}& \textbf{1}& 0.995 & 0.375 & 0.00 & 0.990 & \textbf{1}& 0.985 & 0.380\\
10 & 0.5 & 0.11 & 0.130 & \textbf{1}& \textbf{1}& 0.995 & 0.460 & 0.01 & \textbf{1}& \textbf{1}& \textbf{1}& 0.690\\
15 & 0.5 & 0.11 & 0.120 & \textbf{1}& \textbf{1}& \textbf{1}& 0.630 & 0.02 & \textbf{1}& \textbf{1}& \textbf{1}& 0.845\\
\cmidrule(lr){1-3} \cmidrule(lr){4-8} \cmidrule(lr){9-13}
10 & 0.1 & 0.11 & 0.115 & \textbf{1}& \textbf{1}& \textbf{1}& 0.530 & 0.01 & \textbf{1}& \textbf{1}& \textbf{1}& 0.610\\
10 & 0.9 & 0.11 & 0.055 & 0.990 & \textbf{1}& 0.930 & 0.500 & 0.01 & \textbf{0.995} & \textbf{0.995} & 0.960 & 0.685\\
\cmidrule(lr){1-3} \cmidrule(lr){4-8} \cmidrule(lr){9-13}
10 & 0.5 & 0.03 & 0.175 & 0.995 & \textbf{1}& 0.970 & 0.440 & 0.07 & 0.905 & \textbf{0.940} & 0.845 & 0.780\\
10 & 0.5 & 0.21 & 0.065 & \textbf{1}& \textbf{1}& \textbf{1}& 0.425 & 0.00 & \textbf{1}& \textbf{1}& \textbf{1}& 0.630\\
\bottomrule
\end{tabular} \label{tab:MD:1:DLSM:TPR}
\end{table}

\begin{table}[H]
\caption{DR for DLSM from $r_i$ = 0.1 to $r_i= 0.015$ (B); 0.0225 (C) for $N=35$. } \centering
\begin{tabular}{ccc rrrrr rrrrr} \toprule
\multicolumn{3}{c}{Settings} &  \multicolumn{5}{c}{Binary} & \multicolumn{5}{c}{Count}\\ 
 \cmidrule(lr){1-3} \cmidrule(lr){4-8} \cmidrule(lr){9-13}
CPL & $\phi$ & $\avgden$ &
$W_t$ & $D_{t}$ & $M_t^{-}$ & $M_t^{+}$ & $S_t^{*}$ & $W_t$ & $D_{t}$ & $M_t^{-}$ & $M_t^{+}$ & $S_t^{*}$\\ 
\cmidrule(lr){1-3} \cmidrule(lr){4-8} \cmidrule(lr){9-13}
5 & 0.5 & 0.11 & 0.095 & 0.970 & \textbf{1}& 0.845 & 0.345 & 0.065 & 0.940 & \textbf{0.995} & 0.860 & 0.465\\
10 & 0.5 & 0.11 & 0.130 & 0.980 & \textbf{1}& 0.920 & 0.530 & 0.075 & 0.995 & \textbf{1}& 0.960 & 0.655\\
15 & 0.5 & 0.11 & 0.220 & 0.995 & \textbf{1}& 0.965 & 0.620 & 0.120 & \textbf{1}& \textbf{1}& 0.980 & 0.945\\
\cmidrule(lr){1-3} \cmidrule(lr){4-8} \cmidrule(lr){9-13}
10 & 0.1 & 0.11 & 0.085 & \textbf{1}& \textbf{1}& 0.955 & 0.465 & 0.095 & 0.975 & \textbf{0.990} & 0.945 & 0.715\\
10 & 0.9 & 0.11 & 0.055 & 0.940 & \textbf{1}& 0.805 & 0.430 & 0.050 & 0.950 & \textbf{0.990} & 0.890 & 0.680\\
\cmidrule(lr){1-3} \cmidrule(lr){4-8} \cmidrule(lr){9-13}
10 & 0.5 & 0.03 & 0.205 & 0.930 & \textbf{0.950} & 0.865 & 0.495 & 0.165 & 0.745 & \textbf{0.770} & 0.705 & 0.680\\
10 & 0.5 & 0.21 & 0.060 & 0.995 & \textbf{1}& 0.865 & 0.375 & 0.030 & \textbf{1}& \textbf{1}& 0.985 & 0.600\\
\bottomrule
\end{tabular} \label{tab:MD:2:DLSM:TPR}
\end{table}

\begin{table}[H]
\caption{DR for DDCSBM from $C=1$ to $C=2.25$ in $C \cdot \Theta$ for $N=15$. } \centering
\begin{tabular}{ccc rrrrr rrrrr} \toprule
\multicolumn{3}{c}{Settings} &  \multicolumn{5}{c}{Binary} & \multicolumn{5}{c}{Count}\\ 
 \cmidrule(lr){1-3} \cmidrule(lr){4-8} \cmidrule(lr){9-13}
CPL & $\phi$ & $\avgden$ &
$W_t$ & $D_{t}$ & $M_t^{-}$ & $M_t^{+}$ & $S_t^{*}$ & $W_t$ & $D_{t}$ & $M_t^{-}$ & $M_t^{+}$ & $S_t^{*}$\\
\cmidrule(lr){1-3} \cmidrule(lr){4-8} \cmidrule(lr){9-13}
5 & 0.5 & 0.11 & 0.085 & \textbf{1}& \textbf{1}& \textbf{1}& 0.770 & 0.245 & \textbf{1}& \textbf{1}& \textbf{1}& 0.835\\
10 & 0.5 & 0.11 & 0.155 & \textbf{1}& \textbf{1}& \textbf{1}& 0.780 & 0.195 & \textbf{1}& \textbf{1}& \textbf{1}& 0.845\\
15 & 0.5 & 0.11 & 0.130 & \textbf{1}& \textbf{1}& \textbf{1}& 0.810 & 0.285 & \textbf{1}& \textbf{1}& \textbf{1}& 0.850\\
\cmidrule(lr){1-3} \cmidrule(lr){4-8} \cmidrule(lr){9-13}
10 & 0.1 & 0.11 & 0.115 & \textbf{1}& \textbf{1}& \textbf{1}& 0.770 & 0.285 & \textbf{1}& \textbf{1}& \textbf{1}& 0.840\\
10 & 0.9 & 0.11 & 0.140 & 0.990 & \textbf{0.995} & 0.990 & 0.735 & 0.185 & \textbf{1}& \textbf{1}& \textbf{1}& 0.800\\
\cmidrule(lr){1-3} \cmidrule(lr){4-8} \cmidrule(lr){9-13}
10 & 0.5 & 0.03 & 0.330 & \textbf{0.990} & \textbf{0.990} & 0.985 & 0.580 & 0.340 & 0.990 & \textbf{0.995} & 0.975 & 0.620\\
10 & 0.5 & 0.21 & 0.090 & \textbf{1}& \textbf{1}& \textbf{1}& 0.850 & 0.170 & \textbf{1}& \textbf{1}& \textbf{1}& 0.900\\
\bottomrule
\end{tabular} \label{tab:MD:3:DDCSBM:TPR}
\end{table}

\begin{table}[H]
\caption{DR for DDCSBM from $C=1$ to $C=1.75$ in $C \cdot \Theta$ for $N=35$. } \centering
\begin{tabular}{ccc rrrrr rrrrr} \toprule
\multicolumn{3}{c}{Settings} &  \multicolumn{5}{c}{Binary} & \multicolumn{5}{c}{Count}\\ 
 \cmidrule(lr){1-3} \cmidrule(lr){4-8} \cmidrule(lr){9-13}
CPL & $\phi$ & $\avgden$ &
$W_t$ & $D_{t}$ & $M_t^{-}$ & $M_t^{+}$ & $S_t^{*}$ & $W_t$ & $D_{t}$ & $M_t^{-}$ & $M_t^{+}$ & $S_t^{*}$\\
\cmidrule(lr){1-3} \cmidrule(lr){4-8} \cmidrule(lr){9-13}
5 & 0.5 & 0.11 & 0.145 & 0.950 & \textbf{0.965} & 0.935 & 0.585 & 0.255 & \textbf{0.985} & \textbf{0.985} & \textbf{0.985} & 0.625\\
10 & 0.5 & 0.11 & 0.215 & 0.975 & \textbf{0.990} & 0.965 & 0.625 & 0.285 & \textbf{0.990} & \textbf{0.990} & \textbf{0.990} & 0.635\\
15 & 0.5 & 0.11 & 0.245 & \textbf{1}& \textbf{1}& 0.985 & 0.650 & 0.350 & 0.990 & \textbf{0.995} & 0.990 & 0.715\\
\cmidrule(lr){1-3} \cmidrule(lr){4-8} \cmidrule(lr){9-13}
10 & 0.1 & 0.11 & 0.185 & \textbf{0.970} & \textbf{0.970} & 0.955 & 0.650 & 0.280 & \textbf{0.995} & \textbf{0.995} & 0.985 & 0.665\\
10 & 0.9 & 0.11 & 0.165 & 0.985 & \textbf{0.990} & 0.975 & 0.560 & 0.330 & 0.990 & \textbf{1}& 0.990 & 0.565\\
\cmidrule(lr){1-3} \cmidrule(lr){4-8} \cmidrule(lr){9-13}
10 & 0.5 & 0.03 & 0.330 & 0.845 & \textbf{0.855} & 0.815 & 0.530 & 0.395 & 0.835 & \textbf{0.860} & 0.850 & 0.570\\
10 & 0.5 & 0.21 & 0.075 & 0.980 & \textbf{0.990} & 0.965 & 0.630 & 0.285 & 0.995 & \textbf{1}& 0.995 & 0.690\\
\bottomrule
\end{tabular} \label{tab:MD:4:DDCSBM:TPR}
\end{table}

As before with anomalies within edge probabilities, we compare sustained changes in expected degree to a gradual increase of expected degree over the duration of the anomaly (CPL). In this last set of scenarios, $r_i$ is increased from $r_i = 1/n$ to $4/n$ in a DLSM setting and $C$ is increased from $C=1$ to 5 in a DDCSBM setting. 

DR results are reported in Tables \ref{tab:MD:7:DLSM:TPR} and \ref{tab:MD:8:DDCSBM:TPR} with AUC results in Tables \ref{tab:MD:5:DLSM:AUC} and \ref{tab:MD:6:DDCSBM:AUC} in Appendix \ref{subsec:AppA:AUC:MD}. Similar to the sustained change case, $M^{-}_t$ and $D_t$ methods detect this change the best in DLSM settings. However, monitoring $D_t, ~M^{-}_t,$ and $M^{+}_t$ all detect perfectly with a DR of 1 in DDCSBM settings. The worst monitoring performance is with $W_t$, and scan statistics detect more poorly in DLSM settings than in DDCSBM settings. One reason $S^{*}_t$ method might be performing poorly is because of window contamination as discussed in \cite{zhao2018aggregation}. Since the same subset of nodes are increasing slowly in expected degree, this effect can be captured in a given window and subsequent moving windows. Thus, the standardization process will have already included the anomaly, which makes detecting a gradual change much more difficult. In DDCSBM settings, the same possible window contamination does not affect detection rates as in DLSM settings. AUC results are in Tables \ref{tab:MD:5:DLSM:AUC} and \ref{tab:MD:6:DDCSBM:AUC} in Appendix \ref{subsec:AppA:AUC:MD}. The best monitoring performance with respect to AUC is $M^{-}_t$ for targeting anomalies that are aimed to affect maximum degree. As seen in the sustained case in DLSM settings, changes in radii are not as impactful as in count networks. In some instances, AUC increases, but in others, it decreases. For DDCSBM settings, count networks tend to have higher AUC than in binary networks, which suggests the loss of information in binary networks makes the change slightly harder to detect. 

\begin{table}[H]
\caption{DR for DLSM from $r_i \in [1/n, 4/n]$ for $N=20$. } \centering
\begin{tabular}{ccc rrrrr rrrrr} \toprule
\multicolumn{3}{c}{Settings} &  \multicolumn{5}{c}{Binary} & \multicolumn{5}{c}{Count}\\ 
 \cmidrule(lr){1-3} \cmidrule(lr){4-8} \cmidrule(lr){9-13}
CPL & $\phi$ & $\avgden$ &
$W_t$ & $D_{t}$ & $M_t^{-}$ & $M_t^{+}$ & $S_t^{*}$ & $W_t$ & $D_{t}$ & $M_t^{-}$ & $M_t^{+}$ & $S_t^{*}$\\ 
\cmidrule(lr){1-3} \cmidrule(lr){4-8} \cmidrule(lr){9-13}
15 & 0.5 & 0.11 & 0.065 & \textbf{1}& \textbf{1}& 0.890 & 0.250 & 0.110 & 0.955 & \textbf{0.985} & 0.910 & 0.345\\
20 & 0.5 & 0.11 & 0.095 & \textbf{1}& \textbf{1}& 0.910 & 0.305 & 0.170 & 0.975 & \textbf{0.995} & 0.950 & 0.410\\
25 & 0.5 & 0.11 & 0.135 & \textbf{1}& \textbf{1}& 0.970 & 0.300 & 0.175 & \textbf{0.995} & \textbf{0.995} & 0.975 & 0.465\\
\cmidrule(lr){1-3} \cmidrule(lr){4-8} \cmidrule(lr){9-13}
20 & 0.1 & 0.11 & 0.090 & \textbf{1}& \textbf{1}& 0.945 & 0.345 & 0.185 & 0.995 & \textbf{1}& 0.975 & 0.400\\
20 & 0.9 & 0.11 & 0.060 & 0.98 & \textbf{1}& 0.730 & 0.320 & 0.100 & 0.990 & \textbf{0.995} & 0.935 & 0.485\\
\cmidrule(lr){1-3} \cmidrule(lr){4-8} \cmidrule(lr){9-13}
20 & 0.5 & 0.03 & 0.150 & \textbf{1}& \textbf{1}& 0.985 & 0.440 & 0.245 & 0.790 & \textbf{0.795} & 0.720 & 0.490\\
20 & 0.5 & 0.21 & 0.050 & \textbf{1}& \textbf{1}& 0.735 & 0.165 & 0.180 & \textbf{0.995} & \textbf{0.995} & 0.985 & 0.365\\
\bottomrule
\end{tabular} \label{tab:MD:7:DLSM:TPR}
\end{table}

\begin{table}[H]
\caption{DR for DDCSBM from $C \in [1,5]$ in $C \cdot \Theta$ for $N=20$. } \centering
\begin{tabular}{ccc rrrrr rrrrr} \toprule
\multicolumn{3}{c}{Settings} &  \multicolumn{5}{c}{Binary} & \multicolumn{5}{c}{Count}\\ 
\cmidrule(lr){1-3} \cmidrule(lr){4-8} \cmidrule(lr){9-13}
CPL & $\phi$ & $\avgden$ &
$W_t$ & $D_{t}$ & $M_t^{-}$ & $M_t^{+}$ & $S_t^{*}$ & $W_t$ & $D_{t}$ & $M_t^{-}$ & $M_t^{+}$ & $S_t^{*}$\\
\cmidrule(lr){1-3} \cmidrule(lr){4-8} \cmidrule(lr){9-13}
15 & 0.5 & 0.11 & 0.065 & \textbf{1}& \textbf{1}& \textbf{1}& 0.825 & 0.180 & \textbf{1}& \textbf{1}& \textbf{1}& 0.865\\
20 & 0.5 & 0.11 & 0.135 & \textbf{1}& \textbf{1}& \textbf{1}& 0.825 & 0.170 & \textbf{1}& \textbf{1}& \textbf{1}& 0.865\\
25 & 0.5 & 0.11 & 0.110 & \textbf{1}& \textbf{1}& \textbf{1}& 0.770 & 0.195 & \textbf{1}& \textbf{1}& \textbf{1}& 0.825\\
\cmidrule(lr){1-3} \cmidrule(lr){4-8} \cmidrule(lr){9-13}
20 & 0.1 & 0.11 & 0.105 & \textbf{1}& \textbf{1}& \textbf{1}& 0.775 & 0.155 & \textbf{1}& \textbf{1}& \textbf{1}& 0.855\\
20 & 0.9 & 0.11 & 0.120 & \textbf{1}& \textbf{1}& \textbf{1}& 0.685 & 0.205 & \textbf{1}& \textbf{1}& \textbf{1}& 0.815\\
\cmidrule(lr){1-3} \cmidrule(lr){4-8} \cmidrule(lr){9-13}
20 & 0.5 & 0.03 & 0.265 & \textbf{1}& \textbf{1}& \textbf{1}& 0.710 & 0.310 & \textbf{1}& \textbf{1}& \textbf{1}& 0.760\\
20 & 0.5 & 0.21 & 0.080 & \textbf{1}& \textbf{1}& \textbf{1}& 0.700 & 0.110 & \textbf{1}& \textbf{1}& \textbf{1}& 0.830\\
\bottomrule
\end{tabular} \label{tab:MD:8:DDCSBM:TPR}
\end{table}

\subsection{Overall summary of results}
In the light of the results from our performance evaluation study, we make the following general observations:

\begin{enumerate}
    \item Summary statistics like network density, maximum degree, and their linear combinations can be valuable and effective monitoring tools for detecting anomalous changes in time-evolving networks.
    In particular, such summary statistics can be much more powerful than more complicated and computationally expensive monitoring techniques like the scan statistic of \cite{priebe2005scan}.
    This remarkable fact is demonstrated throughout our study and establishes the value of using summary statistics in network monitoring.
    
    \item Network density ($W_t$) is effective in detecting changes in the odds ratio (Tables 5--10), but ineffective in detecting changes in individual node behavior (Tables 12--17).
    This is consistent with what one would expect, as changes in individual node behavior do not significantly affect the overall network density.
    
    \item Maximum degree  ($D_{t}$) is effective in detecting changes in individual node behavior (Tables 12--17), which is expected.
    In addition, maximum degree is also effective in detecting changes in odds ratio (Tables 5--10), often performing close to or better than network density.
    This makes maximum degree a versatile summary statistic for network monitoring.
    
    \item The linear combinations ($M_t^{-}$ and $M_t^{+}$) can combine the strengths of network density and maximum degree.
    For example, when the anomaly corresponds to an increase in the odds ratio, there is an increase in both network density and maximum degree.  In such cases, the detection rates of $M_t^{+}$ are often higher than maximum degree (Tables 5--10).
    Furthermore, when the anomaly consists of change in individual node behavior, the detection rates of $M_t^{-}$ and $M_t^{+}$ are much higher than network density (Tables 12--17).
    
    \item Another way to combine the strengths of network density and maximum degree would be to consider ($W_t$,$D_{t}$) as a bivariate summary statistic and employ bivariate process monitoring methods.
    To accomplish this, we need to consider the covariance between $W_t$ and $D_{t}$ and update the calibration of $\bar{x}, s,$  and $q$ accordingly to construct bivariate control limits.
    This seems to be a very promising approach that we plan to explore in future work.
    
\end{enumerate}

\section{Conclusion} \label{sec:conclusion}

Anomaly detection in temporally-evolving networks is an active area of research, but often subject to a specific network model. In this work, we explored network monitoring approaches on calculated summary statistics using a comprehensive simulation study. Performance evaluations of summary statistics, density, maximum degree, and linear combinations of density and maximum degree, were compared to that of the scan statistic. To introduce interesting complexities, temporally-evolving network models, DLSMs and DDCSBMs, were used to incorporate correlation over time. This correlation better models phenomena in time-varying networks as opposed to independent snapshots over time. 

In evaluating performance, metrics such as detection rates and area under a receiver-operating curve suggest that simple, relatively easy-to-compute summary statistics can outperform the more sophisticated, difficult-to-implement scan method. Albeit, the measures of success analyzed may not be best suited for documenting the advantages of a scan statistic. The scan statistic is, by construct, vulnerable to missing gradual changes in networks, i.e., window contamination. The types of planted anomalies in our simulations resulted from intentional changes in  edge probabilities and expected degree. Specifically, adjustments were made to the odds ratios of edge probabilities and model parameters governing expected degree. While use of $W_t$ performs better to detect anomalies resulting from changes in  edge probabilities, $W_t$ method performed the worst with anomalies concerning expected degree. Use of maximum degree ($D_t$), however,  does fairly well in both scenarios, yet use of linear combinations of $W_t$  and $D_t$, $M^{-}_t$ and $M^{+}_t$, perform the best in both scenarios. 

To summarize, this paper demonstrates that monitoring summary statistics has clear advantages.  They are simple to calculate,  easy to interpret, and able to catch several types of anomalies.  Based on results from a detailed simulation study, summary statistics showed effective in detecting anomalies under varying conditions pertaining to the following: anomaly duration, correlation, sparsity, network types, and network models. Admittingly, summary statistics will not detect some anomalies that do not impact the statistics directly; e.g., extreme-node-switching, where two nodes that have, say, the maximum degree and minimum degrees in a network at time $t$, swap at time $t+1$.  To detect such an anomaly would require detailed modeling efforts, whereas the \nonpara\ approach presented here with summary statistics saves time and fosters consistency across efforts in detecting anomalies.  However, we might improve the effectiveness of monitoring summary statistics by considering multivariate, rather than univariate analytic approaches; i.e., at time $t$, assessing density ($W_t$), maximum degree ($D_t$), difference ($M_t^{-}$), and/or sum ($M_t^{+}$) jointly. In future work, generalizing univariate methods to multivariate methods should be considered.

Lastly, efforts in this paper concentrated on performance evaluation assessment of summary statistics by using simulated network data. These network monitoring methods can be easily applied to non-synthesized data as a first step in anomaly detection. A few issues arise when utilizing a process monitoring application. One issue is determining the number of time-points for Phase I data since control limits need to be estimated accurately. Another issue is being reasonably sure that no anomalies occurred in Phase I. Any anomalies in Phase I data would be incorrectly incorporated as expected behavior. After control limits are established from Phase I data, monitoring begins in Phase II. If any anomalies are detected using network monitoring methods, then further investigation is needed to find a root cause. Thus, network monitoring methods can provide preliminary analysis on any anomalies in the data.

\newpage
\appendix
\section{Simulations with No Anomaly Additional Figures} \label{sec:appxA:NonAnomFig}

\subsection{Supplemental Figures for Section \ref{subsec:methods:Getq}} \label{subsec:AppA:FPR}

\begin{figure}[H] \centering
\includegraphics[width=3.57in,angle=270]{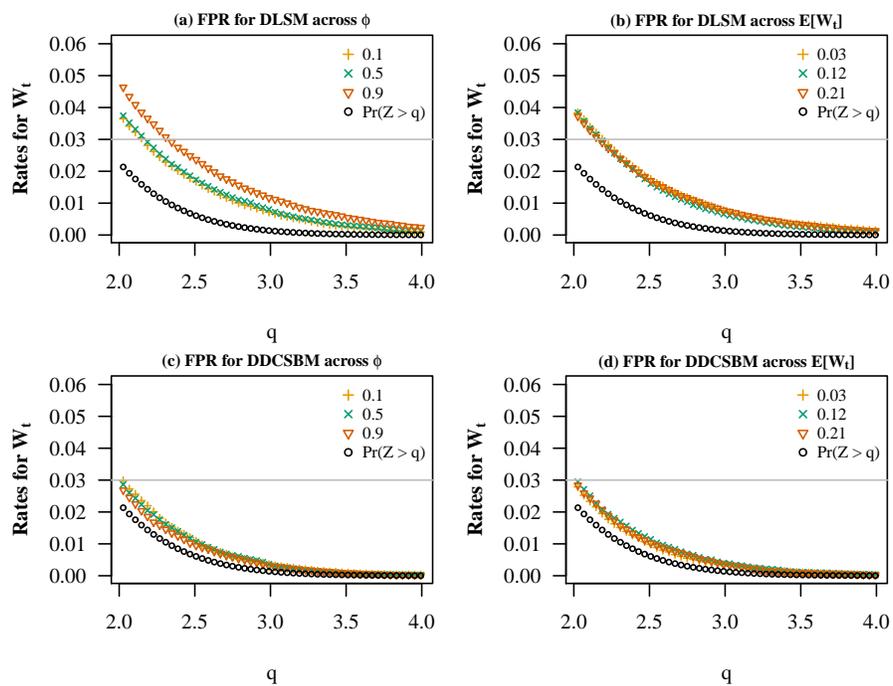}
\caption{Plot of False Alarm Rates Monitoring $W_t$ Across Varying Correlation [(a) and (c)] and Sparsity [(b) and (d)] in Count DLSM and DDCSBM Settings.}
\label{fig:FPR:Dens:Ct}
\end{figure}

\begin{figure}[H] \centering
\includegraphics[width=3.57in,angle=270]{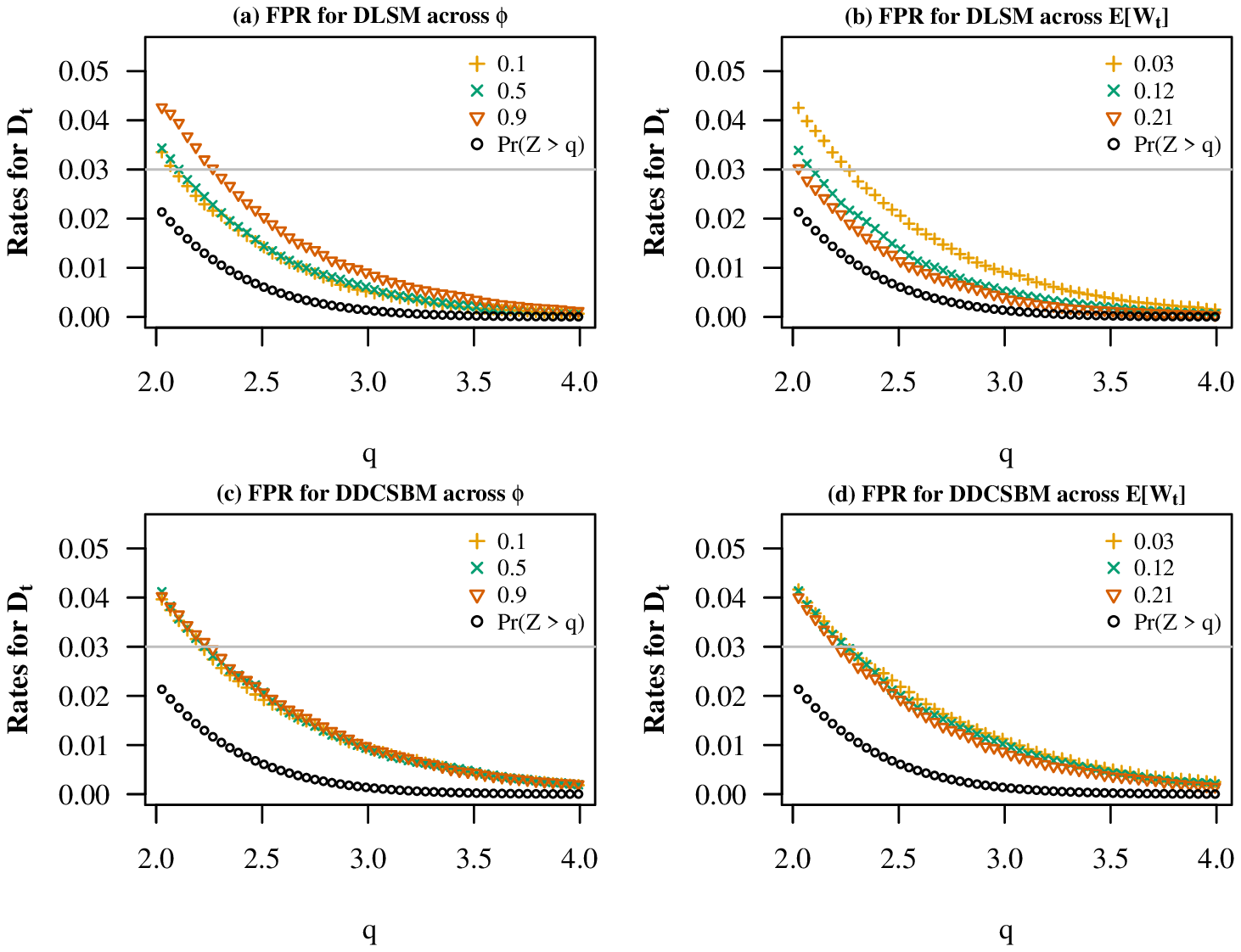}
\caption{Plot of False Alarm Rates Monitoring $D_t$ Across Varying Correlation [(a) and (c)] and Sparsity [(b) and (d)] Values in Binary DLSM and DDCSBM Settings.}
\label{fig:FPR:Max:Bin}
\end{figure}

\begin{figure}[H] \centering
\includegraphics[width=3.57in,angle=270]{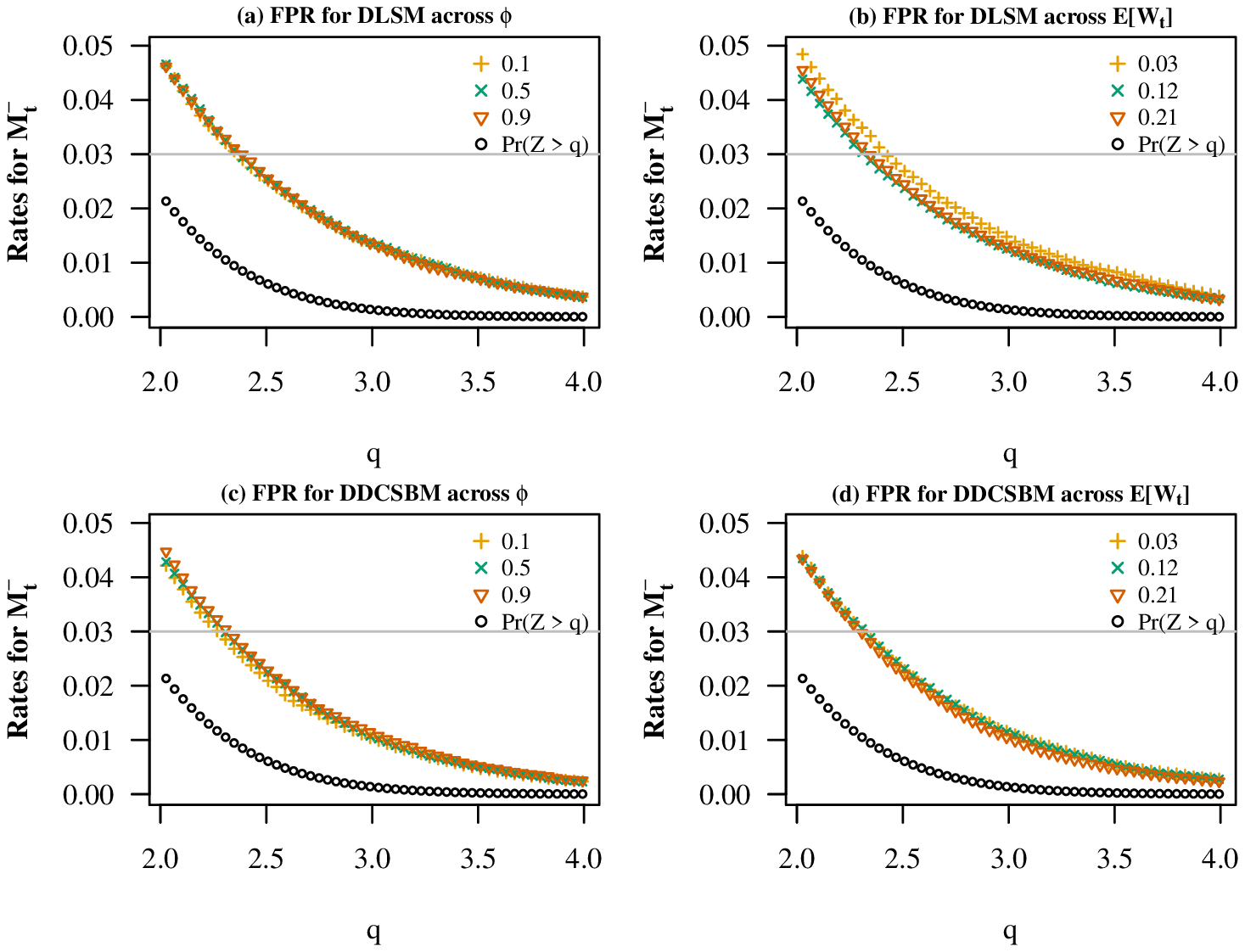}
\caption{Plot of False Alarm Rates Monitoring $M^{-}_t$ Across Varying Correlation [(a) and (c)] and Sparsity [(b) and (d)] Values in Count DLSM and DDCSBM Settings.}
\label{fig:FPR:Diff:Ct}
\end{figure}

\begin{figure}[H] \centering
\includegraphics[width=3.57in,angle=270]{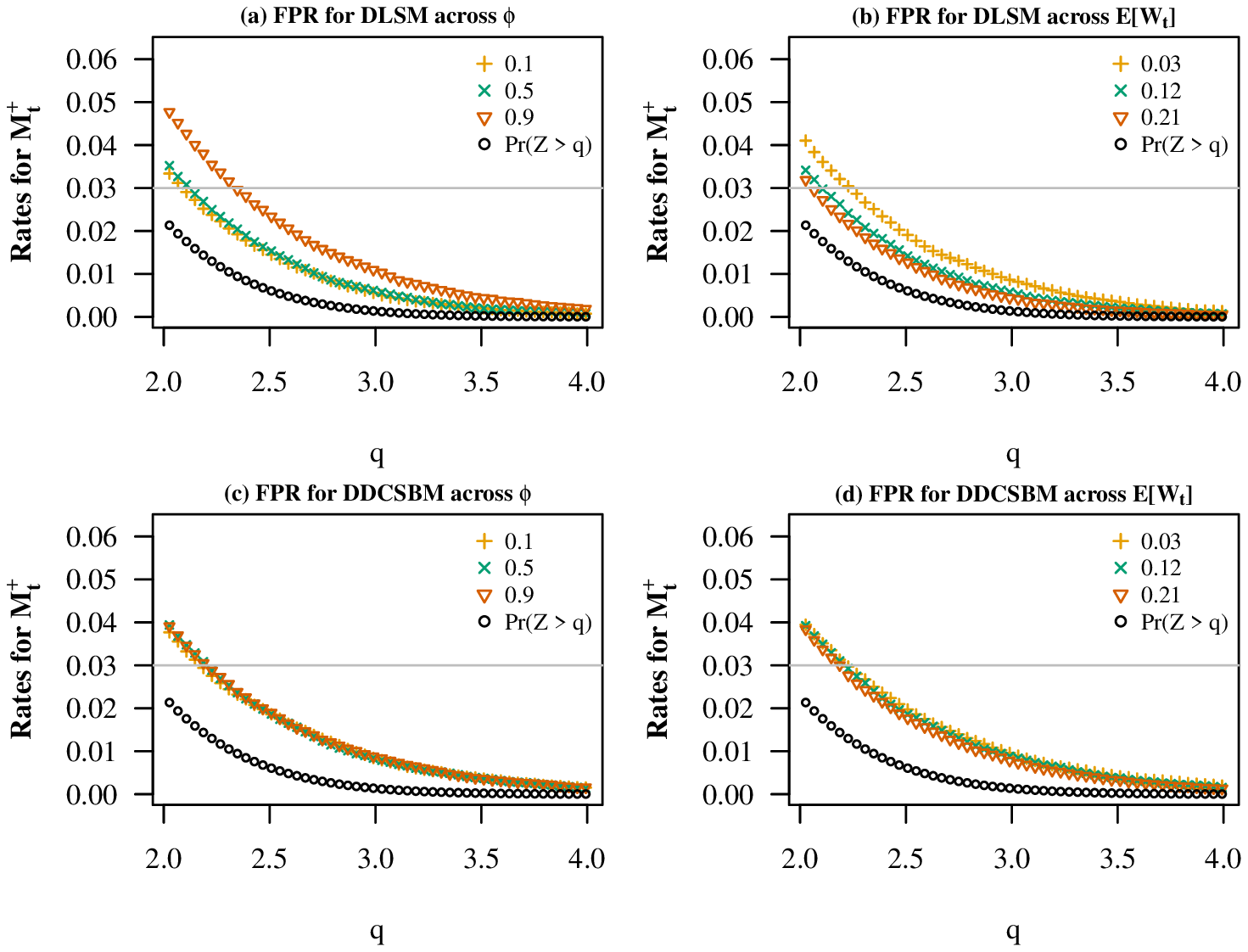}
\caption{Plot of False Alarm Rates Monitoring $M^{+}_t$ Across Varying Correlation [(a) and (c)] and Sparsity [(b) and (d)] in Binary DLSM and DDCSBM Settings.}
\label{fig:FPR:Sum:Bin}
\end{figure}


\subsection{Supplemental Figures for Section \ref{subsec:simstudy:CL}} \label{subsec:AppA:MeansSD}

\begin{figure}[H] \centering
\includegraphics[width=3.81in,angle=270]{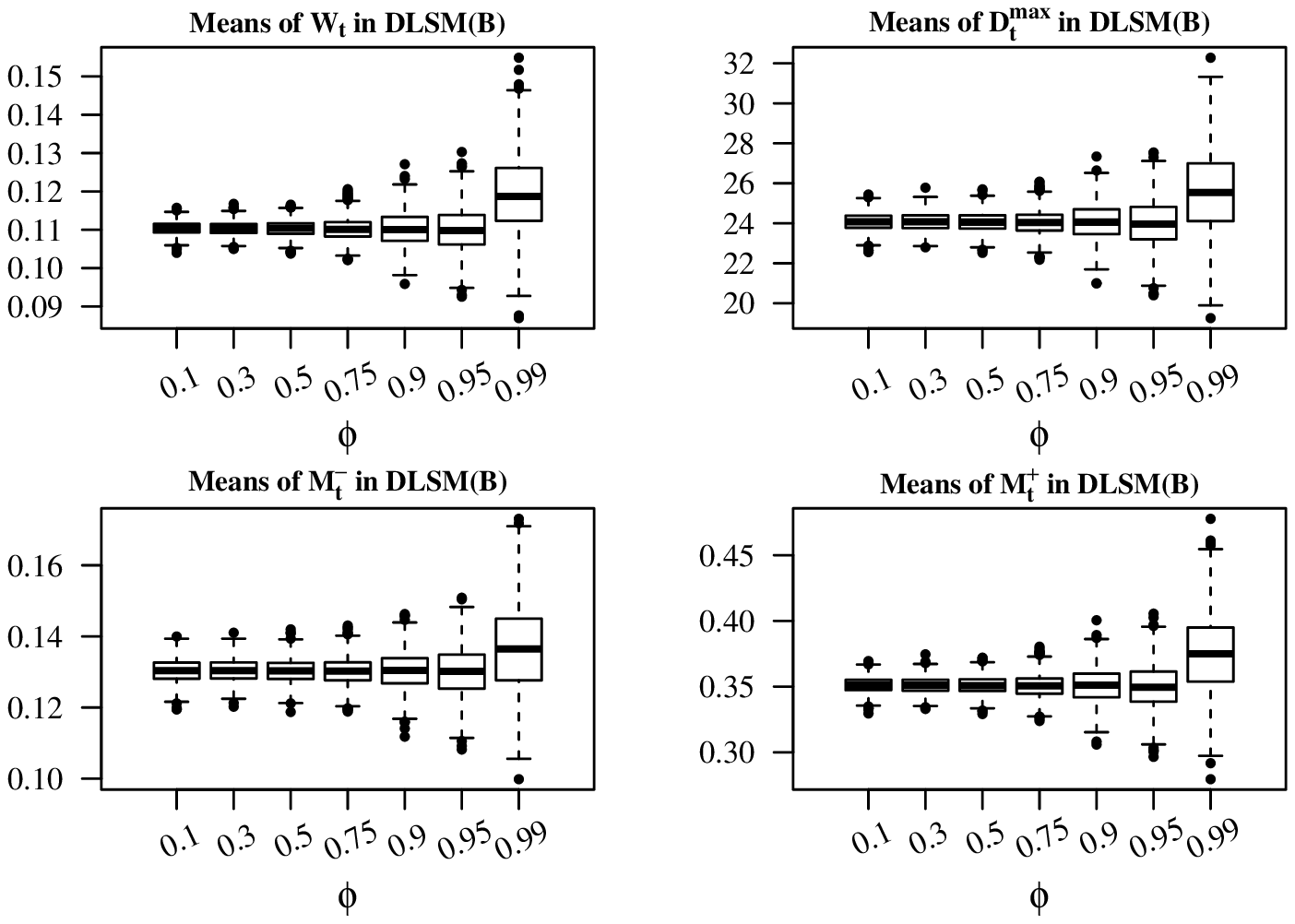}
\caption{Plot of Means in a DLSM Binary (B) Setting Across Varying Correlation.}
\label{fig:Means:DLSM:B:Phi}
\end{figure}

\begin{figure}[H] \centering
\includegraphics[width=3.81in,angle=270]{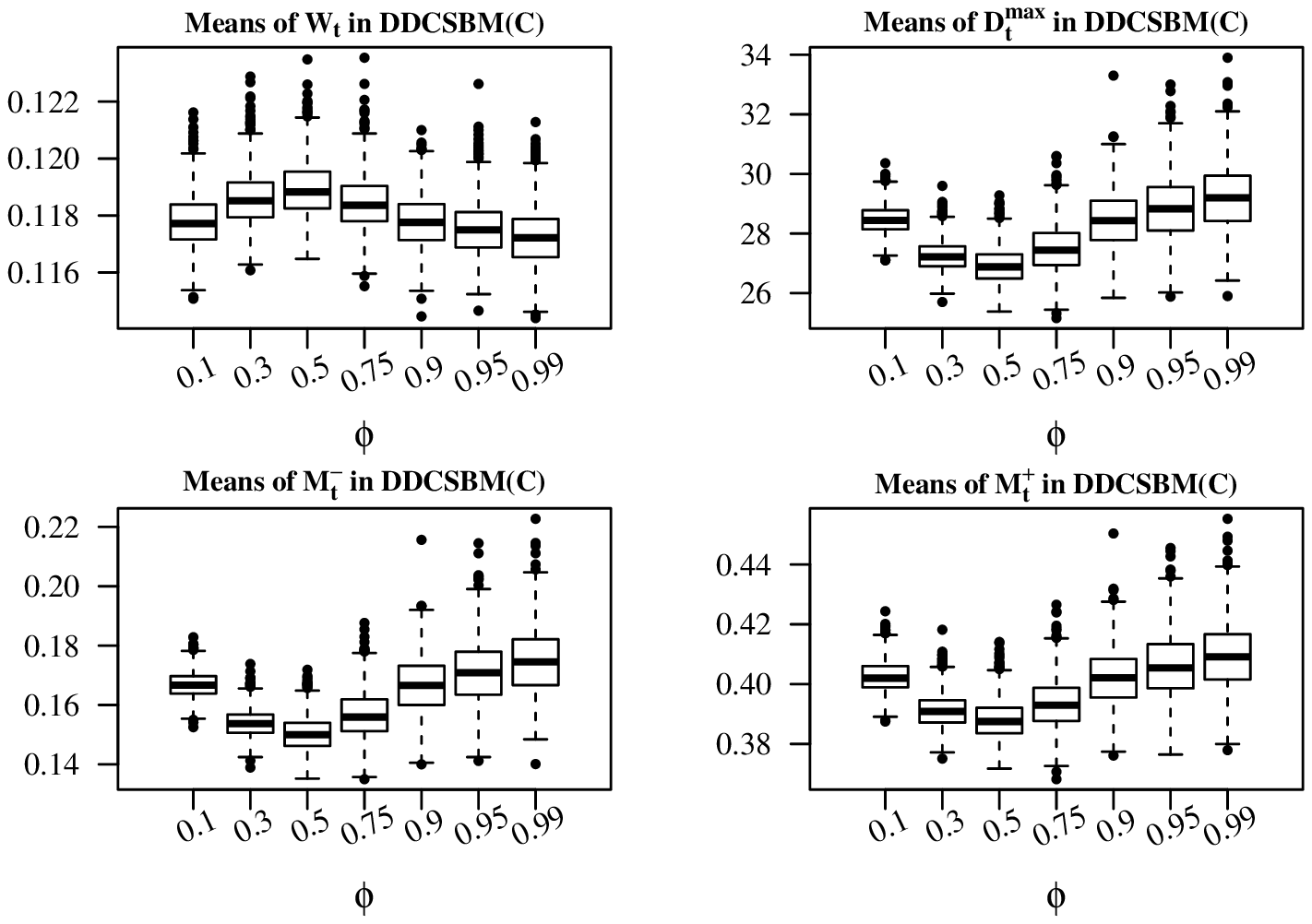}
\caption{Plot of Means in a DDCSBM Count (C) Setting Across Varying Correlation.}
\label{fig:Means:DDCSBM:C:Phi}
\end{figure}

\begin{figure}[H] \centering
\includegraphics[width=3.81in,angle=270]{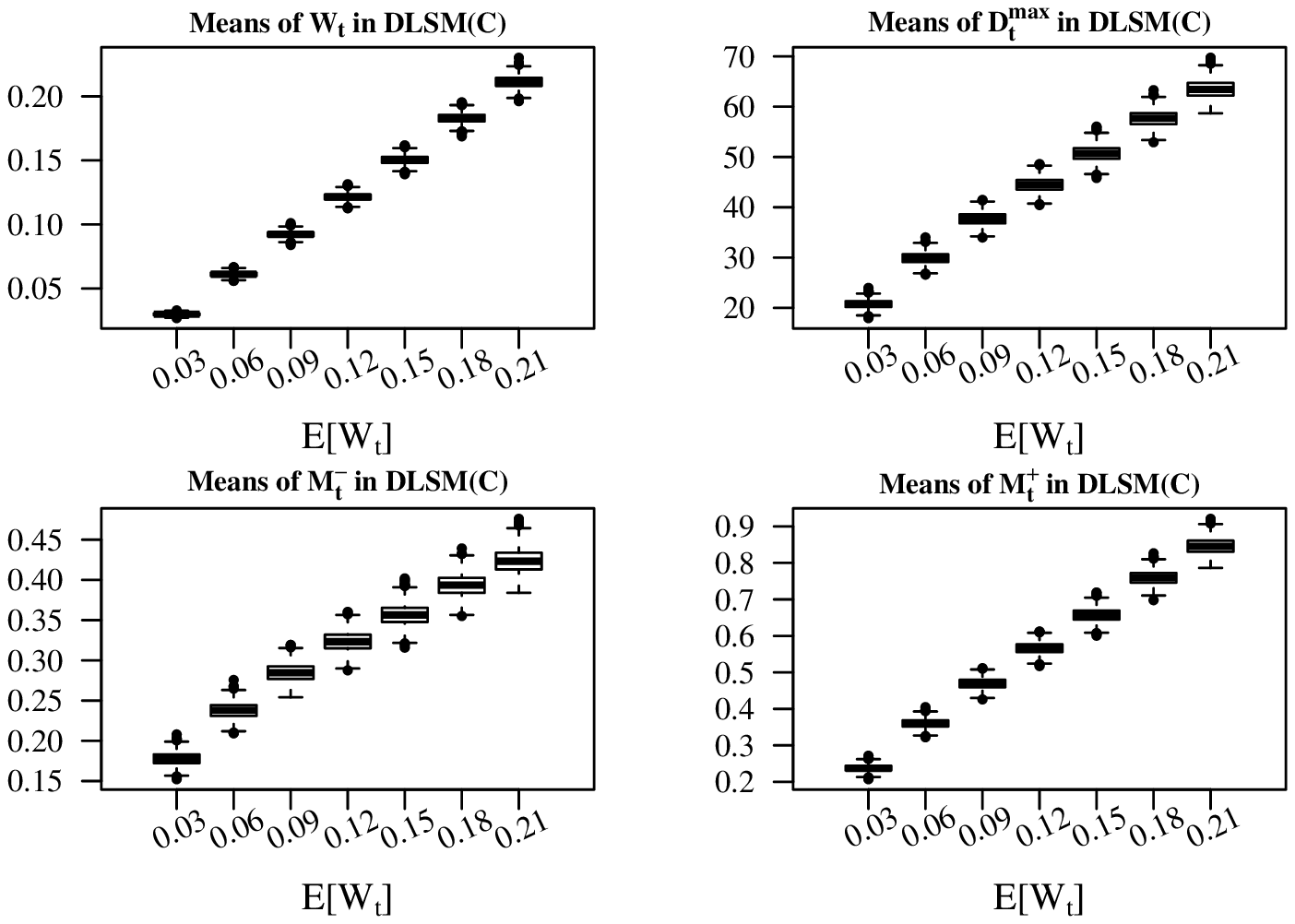}
\caption{Plot of Means in a DLSM Count (C) Setting Across Varying Sparsity.}
\label{fig:Means:DLSM:C:Dens}
\end{figure}

\begin{figure}[H] \centering
\includegraphics[width=3.81in,angle=270]{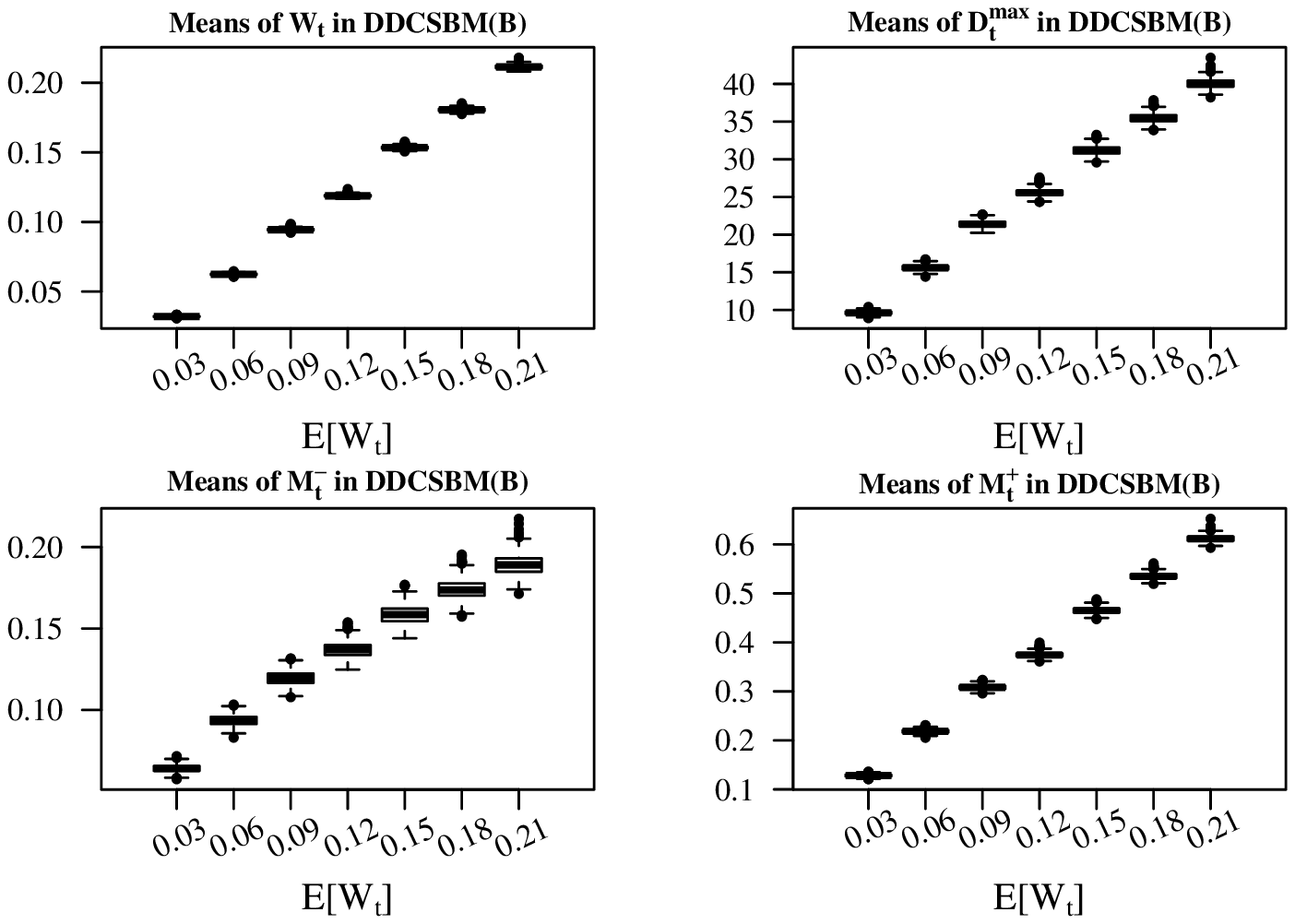}
\caption{Plot of Means in a DDCSBM Binary (B) Setting Across Varying Correlation.}
\label{fig:Means:DDCSBM:B:Dens}
\end{figure}
\begin{figure}[H] \centering
\includegraphics[width=3.81in,angle=270]{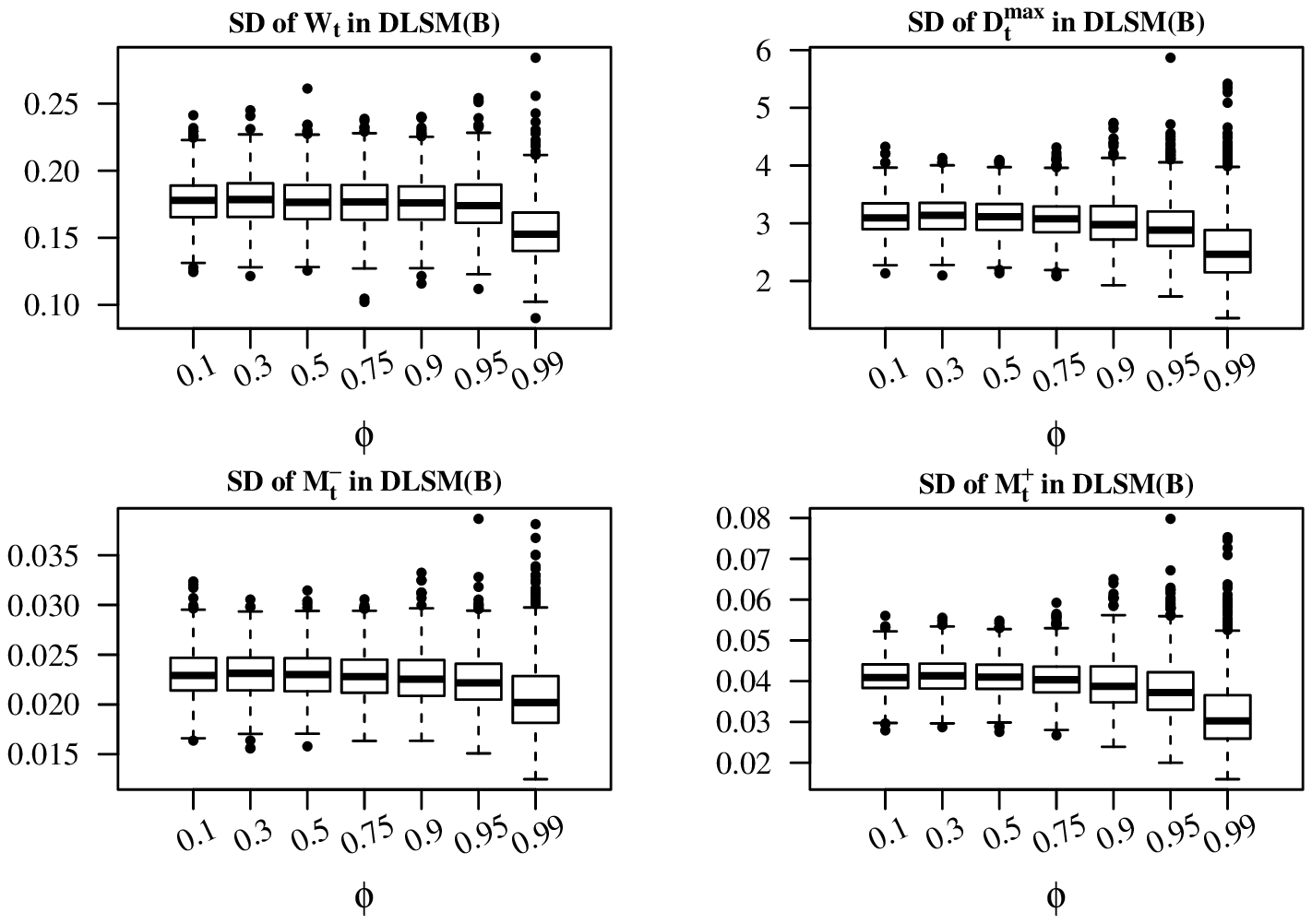}
\caption{Plot of SD in a DLSM Binary (B) Setting Across Varying Correlation.}
\label{fig:SD:DLSM:B:Phi}
\end{figure}

\begin{figure}[H] \centering
\includegraphics[width=3.81in,angle=270]{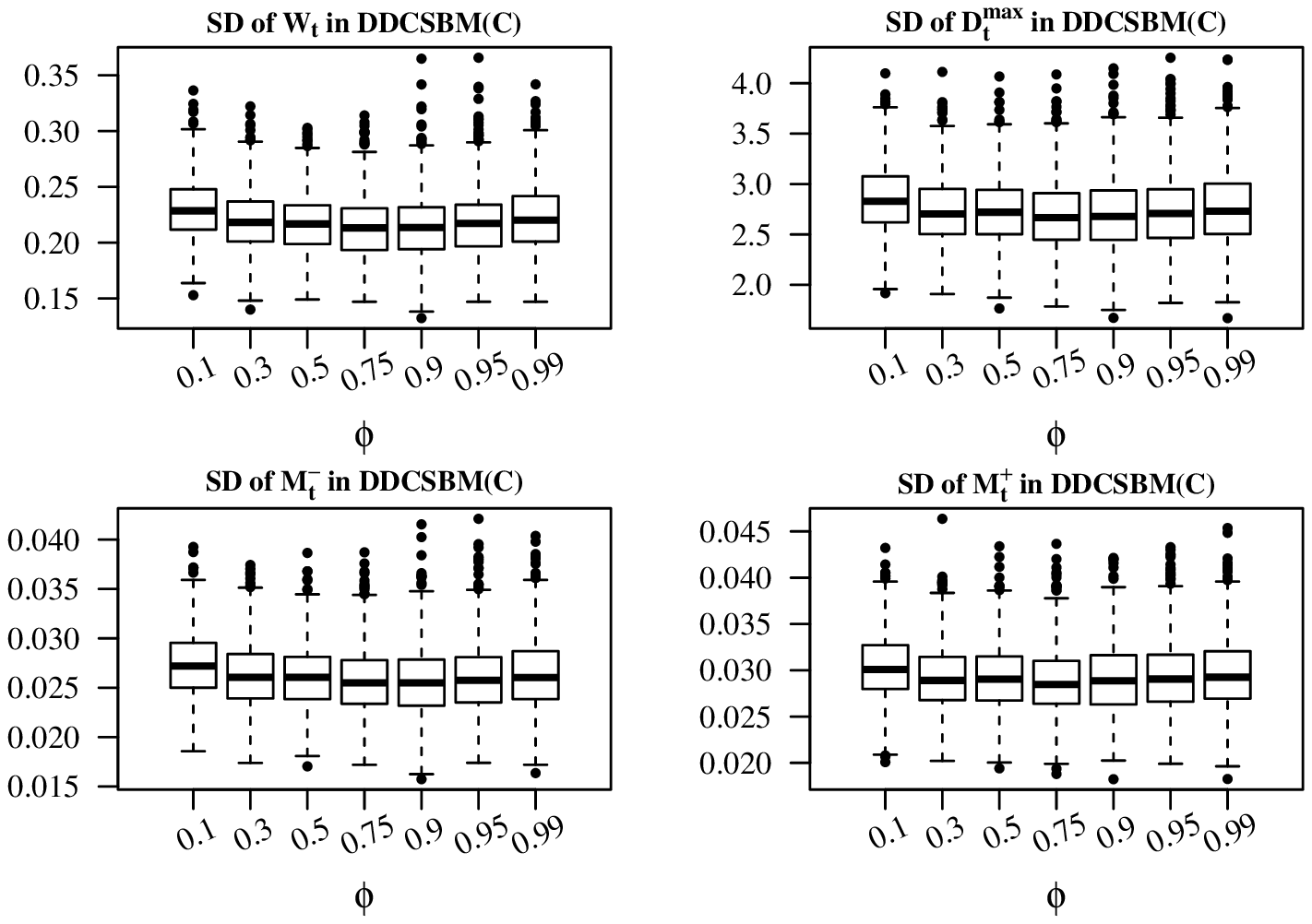}
\caption{Plot of SD in a DDCSBM Count (C) Setting Across Varying Correlation.}
\label{fig:SD:DDCSBM:C:Phi}
\end{figure}

\begin{figure}[H] \centering
\includegraphics[width=3.81in,angle=270]{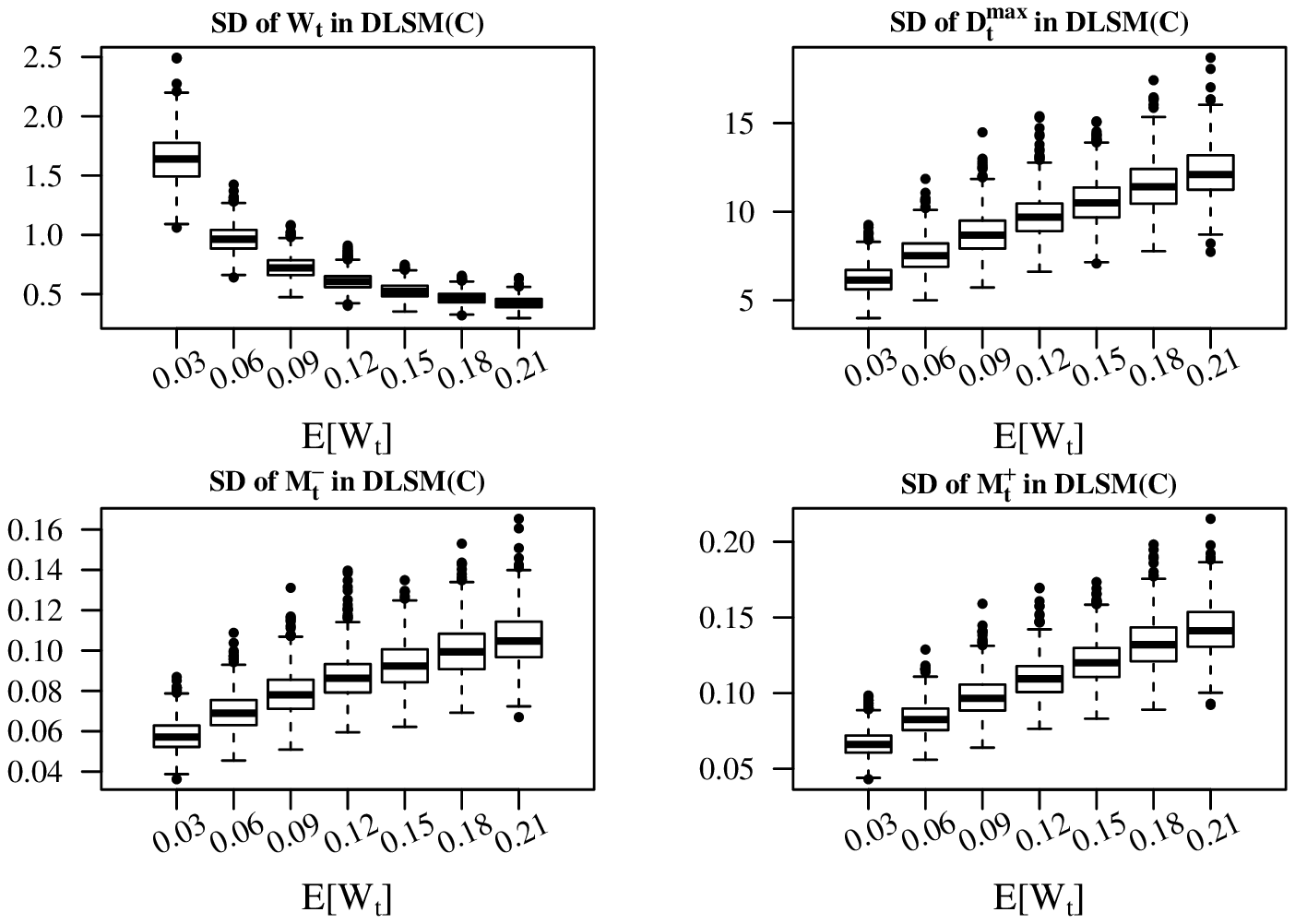}
\caption{Plot of SD in a DLSM Count (C) Setting Across Varying Sparsity.}
\label{fig:SD:DLSM:C:Dens}
\end{figure}

\begin{figure}[H] \centering
\includegraphics[width=3.81in,angle=270]{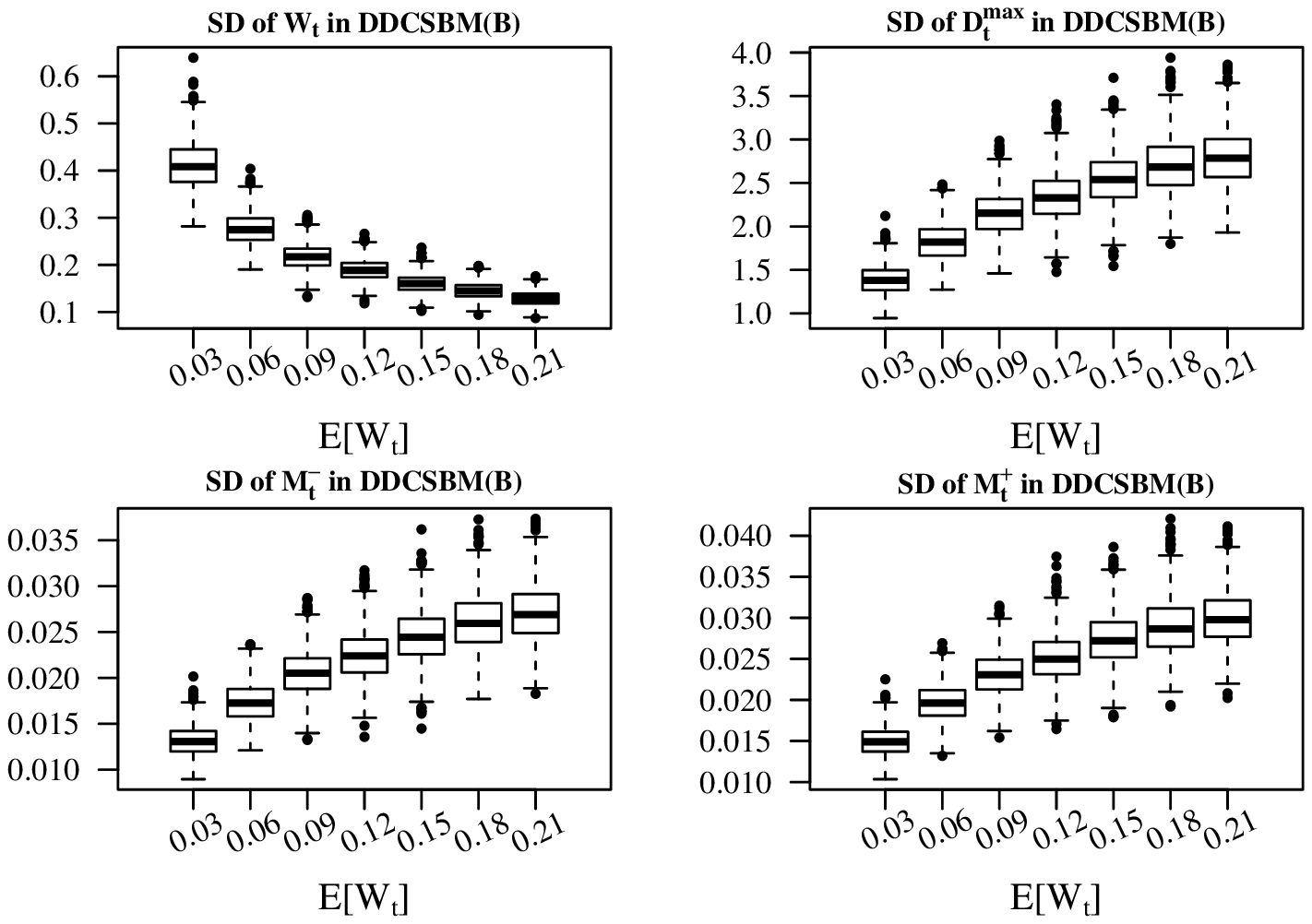}
\caption{Plot of SD in a DDCSBM Binary (B) Setting Across Varying Correlation.}
\label{fig:SD:DDCSBM:B:Dens}
\end{figure}

\newpage
\subsection{AUC Results for Section \ref{subsec:simstudy:anom:density}} \label{subsec:AppA:AUC:ND}

Results of 200 simulations with $n=100$, $T=110$, and $CPL=5,10,$ or 15 are summarized in Tables \ref{tab:ND:1:DLSM:AUC} -\ref{tab:ND:4:DDCSBM:AUC} using AUC. The method which detects the embedded anomaly best are in bold. 

\begin{table}[H]
\caption{AUC for DLSM with 33 Anomalous Nodes and $OR$ from 1 to 4.} \centering
\begin{tabular}{ccc rrrrr rrrrr} \toprule
\multicolumn{3}{c}{Settings} &  \multicolumn{5}{c}{Binary} & \multicolumn{5}{c}{Count}\\ 
 \cmidrule(lr){1-3} \cmidrule(lr){4-8} \cmidrule(lr){9-13}
CPL & $\phi$ & $\avgden$ &
$W_t$ & $D_{t}$ & $M_t^{-}$ & $M_t^{+}$ & $S_t^{*}$ & $W_t$ & $D_{t}$ & $M_t^{-}$ & $M_t^{+}$ & $S_t^{*}$\\ 
\cmidrule(lr){1-3} \cmidrule(lr){4-8} \cmidrule(lr){9-13}
5 & 0.5 & 0.11 & 0.649 & 0.667 & 0.648 & \textbf{0.670} & 0.598 & 0.937 & 0.992 & \textbf{0.994} & 0.990 & 0.877\\
10 & 0.5 & 0.11 & 0.641 & 0.664 & 0.652 & \textbf{0.665} & 0.584 & 0.933 & 0.994 & \textbf{0.995} & 0.991 & 0.768\\
15 & 0.5 & 0.11 & 0.658 & 0.665 & 0.645 & \textbf{0.671} & 0.547 & 0.935 & 0.993 & \textbf{0.994} & 0.991 & 0.681\\
\cmidrule(lr){1-3} \cmidrule(lr){4-8} \cmidrule(lr){9-13}
10 & 0.1 & 0.11 & 0.652 & 0.667 & 0.648 & \textbf{0.672} & 0.576 & 0.933 & 0.993 & \textbf{0.995} & 0.990 & 0.759\\
10 & 0.9 & 0.11 & 0.695 & 0.697 & 0.668 & \textbf{0.705} & 0.546 & 0.936 & 0.993 & \textbf{0.994} & 0.991 & 0.667\\
\cmidrule(lr){1-3} \cmidrule(lr){4-8} \cmidrule(lr){9-13}
10 & 0.5 & 0.03 & \textbf{0.645} & 0.615 & 0.595 & 0.629 & 0.554 & 0.727 & 0.751 & 0.745 & \textbf{0.755} & 0.618\\
10 & 0.5 & 0.21 & 0.655 & 0.720 & \textbf{0.723} & 0.707 & 0.570 & 0.871 & 0.937 & 0.932 & \textbf{0.938} & 0.706\\
\bottomrule
\end{tabular}
\label{tab:ND:1:DLSM:AUC}
\end{table}

\begin{table}[H]
\caption{AUC for DLSM with 79 Anomalous Nodes and $OR$ from 1 to 2.5.} \centering
\begin{tabular}{ccc rrrrr rrrrr} \toprule
\multicolumn{3}{c}{Settings} &  \multicolumn{5}{c}{Binary} & \multicolumn{5}{c}{Count}\\ 
 \cmidrule(lr){1-3} \cmidrule(lr){4-8} \cmidrule(lr){9-13}
CPL & $\phi$ & $\avgden$ &
$W_t$ & $D_{t}$ & $M_t^{-}$ & $M_t^{+}$ & $S_t^{*}$ & $W_t$ & $D_{t}$ & $M_t^{-}$ & $M_t^{+}$ & $S_t^{*}$\\ 
\cmidrule(lr){1-3} \cmidrule(lr){4-8} \cmidrule(lr){9-13}
5 & 0.5 & 0.11 & \textbf{0.935} & 0.883 & 0.798 & 0.909 & 0.792 & \textbf{1} & \textbf{1} & 0.997 & \textbf{1} & 0.932\\
10 & 0.5 & 0.11 & \textbf{0.929} & 0.870 & 0.781 & 0.899 & 0.704 & \textbf{1} & \textbf{1} & 0.998 & \textbf{1} & 0.790\\
15 & 0.5 & 0.11 & \textbf{0.928} & 0.868 & 0.773 & 0.898 & 0.614 & \textbf{1} & \textbf{1} & 0.997 & \textbf{1} & 0.649\\
\cmidrule(lr){1-3} \cmidrule(lr){4-8} \cmidrule(lr){9-13}
10 & 0.1 & 0.11 & \textbf{0.925} & 0.864 & 0.773 & 0.894 & 0.702 & \textbf{1} & \textbf{1} & 0.997 & \textbf{1} & 0.787\\
10 & 0.9 & 0.11 & \textbf{0.942} & 0.882 & 0.780 & 0.914 & 0.601 & \textbf{1} & \textbf{1} & 0.997 & \textbf{1} & 0.719\\
\cmidrule(lr){1-3} \cmidrule(lr){4-8} \cmidrule(lr){9-13}
10 & 0.5 & 0.03 & \textbf{0.900} & 0.779 & 0.698 & 0.826 & 0.653 & \textbf{0.952} & 0.926 & 0.914 & 0.934 & 0.706\\
10 & 0.5 & 0.21 & \textbf{0.944} & 0.929 & 0.840 & 0.941 & 0.730 & \textbf{0.999} & 0.992 & 0.978 & 0.996 & 0.771\\
\bottomrule
\end{tabular}
\label{tab:ND:2:DLSM:AUC}
\end{table}

\begin{table}[H]
\caption{AUC for DDCSBM with 33 Anomalous Nodes and $OR$ from 1 to 2.5.} \centering
\begin{tabular}{ccc rrrrr rrrrr} \toprule
\multicolumn{3}{c}{Settings} &  \multicolumn{5}{c}{Binary} & \multicolumn{5}{c}{Count}\\ 
 \cmidrule(lr){1-3} \cmidrule(lr){4-8} \cmidrule(lr){9-13}
CPL & $\phi$ & $\avgden$ &
$W_t$ & $D_{t}$ & $M_t^{-}$ & $M_t^{+}$ & $S_t^{*}$ & $W_t$ & $D_{t}$ & $M_t^{-}$ & $M_t^{+}$ & $S_t^{*}$\\
 \cmidrule(lr){1-3} \cmidrule(lr){4-8} \cmidrule(lr){9-13}
5 & 0.5 & 0.11 & \textbf{0.982} & 0.861 & 0.772 & 0.915 & 0.822 & \textbf{0.987} & 0.898 & 0.831 & 0.936 & 0.820\\
10 & 0.5 & 0.11 & \textbf{0.980} & 0.853 & 0.763 & 0.909 & 0.702 & \textbf{0.988} & 0.899 & 0.833 & 0.937 & 0.705\\
15 & 0.5 & 0.11 & \textbf{0.977} & 0.867 & 0.781 & 0.916 & 0.588 & \textbf{0.986} & 0.899 & 0.835 & 0.936 & 0.608\\
 \cmidrule(lr){1-3} \cmidrule(lr){4-8} \cmidrule(lr){9-13}
10 & 0.1 & 0.11 & \textbf{0.973} & 0.868 & 0.789 & 0.916 & 0.716 & \textbf{0.982} & 0.903 & 0.847 & 0.937 & 0.704\\
10 & 0.9 & 0.11 & \textbf{0.975} & 0.880 & 0.804 & 0.925 & 0.664 & \textbf{0.983} & 0.911 & 0.855 & 0.943 & 0.664\\
 \cmidrule(lr){1-3} \cmidrule(lr){4-8} \cmidrule(lr){9-13}
10 & 0.5 & 0.03 & \textbf{0.884} & 0.752 & 0.691 & 0.805 & 0.642 & \textbf{0.897} & 0.763 & 0.705 & 0.814 & 0.656\\
10 & 0.5 & 0.21 & \textbf{0.992} & 0.887 & 0.777 & 0.943 & 0.717 & \textbf{0.998} & 0.952 & 0.901 & 0.976 & 0.716\\
\bottomrule
\end{tabular}
\label{tab:ND:3:DDCSBM:AUC}
\end{table}

\begin{table}[H]
\caption{AUC for DDCSBM with 72 Anomalous Nodes and $OR$ from 1 to 1.5.} \centering
\begin{tabular}{ccc rrrrr rrrrr} \toprule
\multicolumn{3}{c}{Settings} &  \multicolumn{5}{c}{Binary} & \multicolumn{5}{c}{Count}\\ 
 \cmidrule(lr){1-3} \cmidrule(lr){4-8} \cmidrule(lr){9-13}
CPL & $\phi$ & $\avgden$ &
$W_t$ & $D_{t}$ & $M_t^{-}$ & $M_t^{+}$ & $S_t^{*}$ & $W_t$ & $D_{t}$ & $M_t^{-}$ & $M_t^{+}$ & $S_t^{*}$\\
\cmidrule(lr){1-3} \cmidrule(lr){4-8} \cmidrule(lr){9-13}
5 & 0.5 & 0.11 & \textbf{1} & 0.923 & 0.776 & 0.977 & 0.870 & \textbf{1} & 0.935 & 0.812 & 0.978 & 0.873\\
10 & 0.5 & 0.11 & \textbf{1} & 0.925 & 0.781 & 0.976 & 0.729 & \textbf{1} & 0.928 & 0.806 & 0.975 & 0.732\\
15 & 0.5 & 0.11 & \textbf{1} & 0.924 & 0.776 & 0.976 & 0.616 & \textbf{1} & 0.929 & 0.808 & 0.975 & 0.621\\
\cmidrule(lr){1-3} \cmidrule(lr){4-8} \cmidrule(lr){9-13}
10 & 0.1 & 0.11 & \textbf{1} & 0.930 & 0.801 & 0.978 & 0.757 & \textbf{1} & 0.932 & 0.823 & 0.975 & 0.769\\
10 & 0.9 & 0.11 & \textbf{1} & 0.931 & 0.799 & 0.977 & 0.700 & \textbf{1} & 0.941 & 0.837 & 0.980 & 0.704\\
\cmidrule(lr){1-3} \cmidrule(lr){4-8} \cmidrule(lr){9-13}
10 & 0.5 & 0.03 & \textbf{0.981} & 0.807 & 0.708 & 0.878 & 0.698 & \textbf{0.976} & 0.809 & 0.718 & 0.874 & 0.693\\
10 & 0.5 & 0.21 & \textbf{1} & 0.950 & 0.768 & 0.992 & 0.727 & \textbf{1} & 0.966 & 0.858 & 0.993 & 0.728\\
\bottomrule
\end{tabular} \label{tab:ND:4:DDCSBM:AUC}
\end{table}

AUC results are reported in Tables \ref{tab:ND:5:DLSM:AUC} and \ref{tab:ND:6:DDCSBM:AUC} below for the gradual change in odds ratio.

\begin{table}[H]
\caption{AUC for DLSM with 39 Anomalous Nodes with $OR \in [1,12]$.} \centering
\begin{tabular}{ccc rrrrr rrrrr} \toprule
\multicolumn{3}{c}{Settings} &  \multicolumn{5}{c}{Binary} & \multicolumn{5}{c}{Count}\\ 
 \cmidrule(lr){1-3} \cmidrule(lr){4-8} \cmidrule(lr){9-13}
CPL & $\phi$ & $\avgden$ &
$W_t$ & $D_{t}$ & $M_t^{-}$ & $M_t^{+}$ & $S_t^{*}$ & $W_t$ & $D_{t}$ & $M_t^{-}$ & $M_t^{+}$ & $S_t^{*}$\\
\cmidrule(lr){1-3} \cmidrule(lr){4-8} \cmidrule(lr){9-13}
15 & 0.5 & 0.11 & 0.786 & 0.814 & 0.794 & \textbf{0.817} & 0.617 & 0.941 & 0.947 & 0.942 & \textbf{0.949} & 0.793\\
20 & 0.5 & 0.11 & 0.782 & 0.809 & 0.790 & \textbf{0.812} & 0.600 & 0.937 & 0.947 & 0.943 & \textbf{0.949} & 0.758\\
25 & 0.5 & 0.11 & 0.779 & 0.805 & 0.786 & \textbf{0.808} & 0.560 & 0.936 & 0.942 & 0.938 & \textbf{0.945} & 0.722\\
\cmidrule(lr){1-3} \cmidrule(lr){4-8} \cmidrule(lr){9-13}
20 & 0.1 & 0.11 & 0.784 & 0.812 & 0.793 & \textbf{0.815} & 0.604 & 0.935 & 0.943 & 0.939 & \textbf{0.945} & 0.754\\
20 & 0.9 & 0.11 & 0.782 & 0.810 & 0.792 & \textbf{0.812} & 0.560 & 0.940 & 0.948 & 0.943 & \textbf{0.950} & 0.715\\
\cmidrule(lr){1-3} \cmidrule(lr){4-8} \cmidrule(lr){9-13}
20 & 0.5 & 0.03 & \textbf{0.760} & 0.719 & 0.684 & 0.741 & 0.576 & 0.871 & 0.882 & \textbf{0.878} & 0.886 & 0.715\\
20 & 0.5 & 0.21 & 0.791 & \textbf{0.862} & 0.861 & 0.848 & 0.605 & 0.952 & 0.963 & 0.959 & \textbf{0.964} & 0.767\\
\bottomrule
\end{tabular} \label{tab:ND:5:DLSM:AUC}
\end{table}

\begin{table}[H]
\caption{AUC for DDCSBM with 39 Anomalous Nodes with $OR \in [1,3.5]$.} \centering
\begin{tabular}{ccc rrrrr rrrrr} \toprule
\multicolumn{3}{c}{Settings} &  \multicolumn{5}{c}{Binary} & \multicolumn{5}{c}{Count}\\ 
 \cmidrule(lr){1-3} \cmidrule(lr){4-8} \cmidrule(lr){9-13}
CPL & $\phi$ & $\avgden$ &
$W_t$ & $D_{t}$ & $M_t^{-}$ & $M_t^{+}$ & $S_t^{*}$ & $W_t$ & $D_{t}$ & $M_t^{-}$ & $M_t^{+}$ & $S_t^{*}$\\
\cmidrule(lr){1-3} \cmidrule(lr){4-8} \cmidrule(lr){9-13}
15 & 0.5 & 0.11 & \textbf{0.949} & 0.869 & 0.803 & 0.902 & 0.763 & \textbf{0.952} & 0.879 & 0.829 & 0.905 & 0.771\\
20 & 0.5 & 0.11 & \textbf{0.947} & 0.865 & 0.795 & 0.900 & 0.715 & \textbf{0.949} & 0.872 & 0.821 & 0.899 & 0.725\\
25 & 0.5 & 0.11 & \textbf{0.944} & 0.857 & 0.786 & 0.893 & 0.681 & \textbf{0.944} & 0.873 & 0.823 & 0.900 & 0.692\\
\cmidrule(lr){1-3} \cmidrule(lr){4-8} \cmidrule(lr){9-13}
20 & 0.1 & 0.11 & \textbf{0.938} & 0.860 & 0.799 & 0.892 & 0.717 & \textbf{0.945} & 0.878 & 0.836 & 0.901 & 0.730\\
20 & 0.9 & 0.11 & \textbf{0.940} & 0.868 & 0.809 & 0.897 & 0.677 & \textbf{0.946} & 0.887 & 0.847 & 0.909 & 0.697\\
\cmidrule(lr){1-3} \cmidrule(lr){4-8} \cmidrule(lr){9-13}
20 & 0.5 & 0.03 & \textbf{0.888} & 0.787 & 0.733 & 0.828 & 0.670 & \textbf{0.885} & 0.793 & 0.743 & 0.830 & 0.677\\
20 & 0.5 & 0.21 & \textbf{0.959} & 0.885 & 0.803 & 0.920 & 0.741 & \textbf{0.962} & 0.902 & 0.858 & 0.925 & 0.759\\
\bottomrule
\end{tabular} \label{tab:ND:6:DDCSBM:AUC}
\end{table}

\newpage
\subsection{AUC Results for Section \ref{subsec:simstudy:anom:maxdeg}} \label{subsec:AppA:AUC:MD}

Results of 200 simulations with  with $n=100$, $T=110$, and $CPL=5,10,$ or 15 are summarized in Tables \ref{tab:MD:1:DLSM:AUC} - \ref{tab:MD:4:DDCSBM:AUC} using AUC. 

\begin{table}[H]
\caption{AUC for DLSM from $r_i$ = 0.1 to $r_i= 0.020$(B); 0.04 (C) for $N=15$. } \centering
\begin{tabular}{ccc rrrrr rrrrr} \toprule
\multicolumn{3}{c}{Settings} &  \multicolumn{5}{c}{Binary} & \multicolumn{5}{c}{Count}\\ 
 \cmidrule(lr){1-3} \cmidrule(lr){4-8} \cmidrule(lr){9-13}
CPL & $\phi$ & $\avgden$ &
$W_t$ & $D_{t}$ & $M_t^{-}$ & $M_t^{+}$ & $S_t^{*}$ & $W_t$ & $D_{t}$ & $M_t^{-}$ & $M_t^{+}$ & $S_t^{*}$\\
\cmidrule(lr){1-3} \cmidrule(lr){4-8} \cmidrule(lr){9-13}
5 & 0.5 & 0.11 & 0.324 & 0.931 & \textbf{0.982} & 0.853 & 0.508 & 0.069 & 0.874 & \textbf{0.931} & 0.801 & 0.439\\
10 & 0.5 & 0.11 & 0.337 & 0.931 & \textbf{0.980} & 0.859 & 0.496 & 0.068 & 0.877 & \textbf{0.934} & 0.805 & 0.480\\
15 & 0.5 & 0.11 & 0.323 & 0.929 & \textbf{0.981} & 0.853 & 0.494 & 0.069 & 0.875 & \textbf{0.933} & 0.805 & 0.524\\
\cmidrule(lr){1-3} \cmidrule(lr){4-8} \cmidrule(lr){9-13}
10 & 0.1 & 0.11 & 0.327 & 0.930 & \textbf{0.981} & 0.856 & 0.505 & 0.068 & 0.882 & \textbf{0.939} & 0.811 & 0.473\\
10 & 0.9 & 0.11 & 0.322 & 0.926 & \textbf{0.980} & 0.850 & 0.507 & 0.063 & 0.879 & \textbf{0.937} & 0.806 & 0.501\\
\cmidrule(lr){1-3} \cmidrule(lr){4-8} \cmidrule(lr){9-13}
10 & 0.5 & 0.03 & 0.395 & 0.805 & \textbf{0.855} & 0.760 & 0.510 & 0.195 & 0.663 & \textbf{0.705} & 0.625 & 0.563\\
10 & 0.5 & 0.21 & 0.277 & 0.953 & \textbf{0.997} & 0.849 & 0.494 & 0.027 & 0.942 & \textbf{0.983} & 0.864 & 0.447\\
\bottomrule
\end{tabular} \label{tab:MD:1:DLSM:AUC}
\end{table}

\begin{table}[H]
\caption{AUC for DLSM from $r_i$ = 0.1 to $r_i= 0.015$ (B); 0.0225 (C) for $N=35$. } \centering
\begin{tabular}{ccc rrrrr rrrrr} \toprule
\multicolumn{3}{c}{Settings} &  \multicolumn{5}{c}{Binary} & \multicolumn{5}{c}{Count}\\ 
 \cmidrule(lr){1-3} \cmidrule(lr){4-8} \cmidrule(lr){9-13}
CPL & $\phi$ & $\avgden$ &
$W_t$ & $D_{t}$ & $M_t^{-}$ & $M_t^{+}$ & $S_t^{*}$ & $W_t$ & $D_{t}$ & $M_t^{-}$ & $M_t^{+}$ & $S_t^{*}$\\ 
\cmidrule(lr){1-3} \cmidrule(lr){4-8} \cmidrule(lr){9-13}
5 & 0.5 & 0.11 & 0.348 & 0.827 & \textbf{0.924} & 0.733 & 0.485 & 0.194 & 0.795 & \textbf{0.860} & 0.729 & 0.474\\
10 & 0.5 & 0.11 & 0.364 & 0.839 & \textbf{0.929} & 0.749 & 0.501 & 0.209 & 0.812 & \textbf{0.871} & 0.747 & 0.517\\
15 & 0.5 & 0.11 & 0.369 & 0.835 & \textbf{0.927} & 0.747 & 0.491 & 0.209 & 0.805 & \textbf{0.867} & 0.740 & 0.574\\
\cmidrule(lr){1-3} \cmidrule(lr){4-8} \cmidrule(lr){9-13}
10 & 0.1 & 0.11 & 0.362 & 0.835 & \textbf{0.928} & 0.745 & 0.499 & 0.209 & 0.805 & \textbf{0.866} & 0.741 & 0.511\\
10 & 0.9 & 0.11 & 0.362 & 0.845 & \textbf{0.934} & 0.756 & 0.478 & 0.193 & 0.803 & \textbf{0.866} & 0.737 & 0.521\\
\cmidrule(lr){1-3} \cmidrule(lr){4-8} \cmidrule(lr){9-13}
10 & 0.5 & 0.03 & 0.413 & 0.718 & \textbf{0.771} & 0.674 & 0.502 & 0.307 & 0.623 & \textbf{0.653} & 0.595 & 0.530\\
10 & 0.5 & 0.21 & 0.311 & 0.869 & \textbf{0.980} & 0.734 & 0.502 & 0.134 & 0.862 & \textbf{0.933} & 0.774 & 0.484\\
\bottomrule
\end{tabular} \label{tab:MD:2:DLSM:AUC}
\end{table}

\begin{table}[H]
\caption{AUC for DDCSBM from $C=1$ to $C=2.25$ in $C \cdot \Theta$ for $N=15$. } \centering
\begin{tabular}{ccc rrrrr rrrrr} \toprule
\multicolumn{3}{c}{Settings} &  \multicolumn{5}{c}{Binary} & \multicolumn{5}{c}{Count}\\ 
 \cmidrule(lr){1-3} \cmidrule(lr){4-8} \cmidrule(lr){9-13}
CPL & $\phi$ & $\avgden$ &
$W_t$ & $D_{t}$ & $M_t^{-}$ & $M_t^{+}$ & $S_t^{*}$ & $W_t$ & $D_{t}$ & $M_t^{-}$ & $M_t^{+}$ & $S_t^{*}$\\
\cmidrule(lr){1-3} \cmidrule(lr){4-8} \cmidrule(lr){9-13}
5 & 0.5 & 0.11 & 0.349 & 0.920 & \textbf{0.931} & 0.907 & 0.512 & 0.438 & 0.932 & \textbf{0.938} & 0.925 & 0.525\\
10 & 0.5 & 0.11 & 0.351 & 0.929 & \textbf{0.939} & 0.914 & 0.483 & 0.422 & 0.938 & \textbf{0.945} & 0.929 & 0.496\\
15 & 0.5 & 0.11 & 0.337 & 0.928 & \textbf{0.939} & 0.914 & 0.469 & 0.434 & 0.944 & \textbf{0.951} & 0.936 & 0.482\\
\cmidrule(lr){1-3} \cmidrule(lr){4-8} \cmidrule(lr){9-13}
10 & 0.1 & 0.11 & 0.343 & 0.932 & \textbf{0.940} & 0.919 & 0.502 & 0.434 & 0.944 & \textbf{0.949} & 0.938 & 0.503\\
10 & 0.9 & 0.11 & 0.330 & 0.930 & \textbf{0.941} & 0.915 & 0.476 & 0.433 & 0.942 & \textbf{0.947} & 0.935 & 0.483\\
\cmidrule(lr){1-3} \cmidrule(lr){4-8} \cmidrule(lr){9-13}
10 & 0.5 & 0.03 & 0.473 & 0.779 & \textbf{0.798} & 0.761 & 0.508 & 0.481 & 0.780 & \textbf{0.798} & 0.762 & 0.517\\
10 & 0.5 & 0.21 & 0.199 & 0.952 & \textbf{0.960} & 0.938 & 0.468 & 0.381 & 0.963 & \textbf{0.966} & 0.960 & 0.483\\
\bottomrule
\end{tabular} \label{tab:MD:3:DDCSBM:AUC}
\end{table}

\begin{table}[H]
\caption{AUC for DDCSBM from $C=1$ to $C=1.75$ in $C \cdot \Theta$ for $N=35$. } \centering
\begin{tabular}{ccc rrrrr rrrrr} \toprule
\multicolumn{3}{c}{Settings} &  \multicolumn{5}{c}{Binary} & \multicolumn{5}{c}{Count}\\ 
 \cmidrule(lr){1-3} \cmidrule(lr){4-8} \cmidrule(lr){9-13}
CPL & $\phi$ & $\avgden$ &
$W_t$ & $D_{t}$ & $M_t^{-}$ & $M_t^{+}$ & $S_t^{*}$ & $W_t$ & $D_{t}$ & $M_t^{-}$ & $M_t^{+}$ & $S_t^{*}$\\
\cmidrule(lr){1-3} \cmidrule(lr){4-8} \cmidrule(lr){9-13}
5 & 0.5 & 0.11 & 0.416 & 0.799 & \textbf{0.818} & 0.777 & 0.537 & 0.465 & 0.814 & \textbf{0.828} & 0.798 & 0.548\\
10 & 0.5 & 0.11 & 0.403 & 0.802 & \textbf{0.823} & 0.777 & 0.506 & 0.453 & 0.817 & \textbf{0.832} & 0.800 & 0.506\\
15 & 0.5 & 0.11 & 0.407 & 0.801 & \textbf{0.821} & 0.777 & 0.492 & 0.455 & 0.821 & \textbf{0.836} & 0.804 & 0.491\\
\cmidrule(lr){1-3} \cmidrule(lr){4-8} \cmidrule(lr){9-13}
10 & 0.1 & 0.11 & 0.406 & 0.811 & \textbf{0.831} & 0.788 & 0.518 & 0.452 & 0.833 & \textbf{0.847} & 0.817 & 0.524\\
10 & 0.9 & 0.11 & 0.398 & 0.827 & \textbf{0.849} & 0.802 & 0.493 & 0.464 & 0.843 & \textbf{0.857} & 0.826 & 0.507\\
\cmidrule(lr){1-3} \cmidrule(lr){4-8} \cmidrule(lr){9-13}
10 & 0.5 & 0.03 & 0.474 & 0.662 & \textbf{0.679} & 0.646 & 0.514 & 0.484 & 0.666 & \textbf{0.681} & 0.651 & 0.507\\
10 & 0.5 & 0.21 & 0.292 & 0.845 & \textbf{0.871} & 0.809 & 0.490 & 0.416 & 0.886 & \textbf{0.898} & 0.870 & 0.522\\
\bottomrule
\end{tabular} \label{tab:MD:4:DDCSBM:AUC}
\end{table}

Results using AUC are shown in Tables \ref{tab:MD:5:DLSM:AUC} and \ref{tab:MD:6:DDCSBM:AUC} below for a gradual change in expected degree. 

\begin{table}[H]
\caption{AUC for DLSM from $r_i \in [1/n, 4/n]$ for $N=20$. } \centering
\begin{tabular}{ccc rrrrr rrrrr} \toprule
\multicolumn{3}{c}{Settings} &  \multicolumn{5}{c}{Binary} & \multicolumn{5}{c}{Count}\\ 
 \cmidrule(lr){1-3} \cmidrule(lr){4-8} \cmidrule(lr){9-13}
CPL & $\phi$ & $\avgden$ &
$W_t$ & $D_{t}$ & $M_t^{-}$ & $M_t^{+}$ & $S_t^{*}$ & $W_t$ & $D_{t}$ & $M_t^{-}$ & $M_t^{+}$ & $S_t^{*}$\\ 
\cmidrule(lr){1-3} \cmidrule(lr){4-8} \cmidrule(lr){9-13}
15 & 0.5 & 0.11 & 0.156 & 0.835 & \textbf{0.951} & 0.636 & 0.394 & 0.260 & 0.776 & \textbf{0.820} & 0.726 & 0.473\\
20 & 0.5 & 0.11 & 0.155 & 0.838 & \textbf{0.950} & 0.633 & 0.356 & 0.256 & 0.770 & \textbf{0.814} & 0.719 & 0.447\\
25 & 0.5 & 0.11 & 0.163 & 0.836 & \textbf{0.945} & 0.640 & 0.318 & 0.260 & 0.772 & \textbf{0.816} & 0.724 & 0.430\\
\cmidrule(lr){1-3} \cmidrule(lr){4-8} \cmidrule(lr){9-13}
20 & 0.1 & 0.11 & 0.160 & 0.839 & \textbf{0.949} & 0.638 & 0.351 & 0.263 & 0.772 & \textbf{0.816} & 0.724 & 0.448\\
20 & 0.9 & 0.11 & 0.153 & 0.835 & \textbf{0.946} & 0.634 & 0.359 & 0.254 & 0.770 & \textbf{0.813} & 0.720 & 0.462\\
\cmidrule(lr){1-3} \cmidrule(lr){4-8} \cmidrule(lr){9-13}
20 & 0.5 & 0.03 & 0.217 & 0.840 & \textbf{0.887} & 0.775 & 0.455 & 0.350 & 0.611 & \textbf{0.635} & 0.590 & 0.475\\
20 & 0.5 & 0.21 & 0.131 & 0.704 & \textbf{0.965} & 0.467 & 0.293 & 0.213 & 0.827 & \textbf{0.873} & 0.765 & 0.411\\
\bottomrule
\end{tabular} \label{tab:MD:5:DLSM:AUC}
\end{table}

\begin{table}[H]
\caption{AUC for DDCSBM from $C \in [1,5]$ in $C \cdot \Theta$ for $N=35$. } \centering
\begin{tabular}{ccc rrrrr rrrrr} \toprule
\multicolumn{3}{c}{Settings} &  \multicolumn{5}{c}{Binary} & \multicolumn{5}{c}{Count}\\ 
\cmidrule(lr){1-3} \cmidrule(lr){4-8} \cmidrule(lr){9-13}
CPL & $\phi$ & $\avgden$ &
$W_t$ & $D_{t}$ & $M_t^{-}$ & $M_t^{+}$ & $S_t^{*}$ & $W_t$ & $D_{t}$ & $M_t^{-}$ & $M_t^{+}$ & $S_t^{*}$\\
\cmidrule(lr){1-3} \cmidrule(lr){4-8} \cmidrule(lr){9-13}
15 & 0.5 & 0.11 & 0.219 & 0.900 & \textbf{0.906} & 0.891 & 0.593 & 0.354 & 0.902 & \textbf{0.905} & 0.898 & 0.653\\
20 & 0.5 & 0.11 & 0.228 & 0.904 & \textbf{0.910} & 0.897 & 0.521 & 0.347 & 0.910 & \textbf{0.913} & 0.906 & 0.580\\
25 & 0.5 & 0.11 & 0.225 & 0.908 & \textbf{0.913} & 0.900 & 0.464 & 0.348 & 0.912 & \textbf{0.916} & 0.907 & 0.524\\
\cmidrule(lr){1-3} \cmidrule(lr){4-8} \cmidrule(lr){9-13}
20 & 0.1 & 0.11 & 0.216 & 0.902 & \textbf{0.907} & 0.893 & 0.508 & 0.336 & 0.913 & \textbf{0.917} & 0.909 & 0.583\\
20 & 0.9 & 0.11 & 0.219 & 0.906 & \textbf{0.912} & 0.897 & 0.531 & 0.336 & 0.908 & \textbf{0.911} & 0.904 & 0.579\\
\cmidrule(lr){1-3} \cmidrule(lr){4-8} \cmidrule(lr){9-13}
20 & 0.5 & 0.03 & 0.420 & 0.832 & \textbf{0.846} & 0.818 & 0.569 & 0.446 & 0.838 & \textbf{0.849} & 0.826 & 0.594\\
20 & 0.5 & 0.21 & 0.132 & 0.922 & \textbf{0.928} & 0.911 & 0.441 & 0.249 & 0.925 & \textbf{0.928} & 0.921 & 0.548\\
\bottomrule
\end{tabular} \label{tab:MD:6:DDCSBM:AUC}
\end{table}

\newpage
\bibliographystyle{jasa}
\bibliography{networks}


\end{document}